\documentclass{aastex631}
\usepackage{comment}
\usepackage{empheq}

\newcommand{\Add}[1]{\textcolor{black}{#1}}

\shorttitle{Superflares and Prominence Eruptions on Young Solar-type Star}
\shortauthors{Namekata et al.}
\graphicspath{{./}{../analysis-2022/figures/}{figures/}}

\begin{document}

\title{Multiwavelength Campaign Observations of a Young Solar-type Star, EK Draconis. I. Discovery of Prominence Eruptions Associated with Superflares}

\author[0000-0002-1297-9485]{Kosuke Namekata}
\affiliation{ALMA Project, National Astronomical Observatory of Japan, NINS, Osawa, Mitaka, Tokyo, 181-8588, Japan}
\email{kosuke.namekata@astro.nao.ac.jp, namekata@kusastro.kyoto-u.ac.jp}

\author[0000-0003-4452-0588]{Vladimir S. Airapetian}
\affiliation{Sellers Exoplanetary Environments Collaboration, NASA Goddard Space Flight Center, Greenbelt, MD, USA}
\affiliation{Department of Physics, American University, Washington, DC, USA}

\author[0000-0001-7624-9222]{Pascal Petit}
\affiliation{Institut de Recherche en Astrophysique et Plan\'{e}tologie, Universit\'{e} de Toulouse, CNRS, CNES, 14 avenue \'{E}douard Belin, 31400 Toulouse, France}

\author[0000-0003-0332-0811]{Hiroyuki Maehara}
\affil{Okayama Branch Office, Subaru Telescope, National Astronomical Observatory of Japan, NINS, Kamogata, Asakuchi, Okayama 719-0232, Japan}

\author[0000-0002-5978-057X]{Kai Ikuta}
\affil{Department of Multidisciplinary Sciences, The University of Tokyo, 3-8-1 Komaba, Meguro, Tokyo 153-8902, Japan}

\author[0000-0003-3085-304X]{Shun Inoue}
\affil{Department of Physics, Kyoto University, Sakyo, Kyoto 606-8502, Japan}

\author[0000-0002-0412-0849]{Yuta Notsu}
\affil{Laboratory for Atmospheric and Space Physics, University of Colorado Boulder, 3665 Discovery Drive, Boulder, CO 80303, USA}
\affil{National Solar Observatory, 3665 Discovery Drive, Boulder, CO 80303, USA}

\author[0000-0002-8090-3570]{Rishi R. Paudel}
\affil{NASA Goddard Space Flight Center, 8800 Greenbelt Road, Greenbelt, MD 20771, USA}
\affil{University of Maryland, Baltimore County, 1000 Hilltop Circle, Baltimore, MD 21250, USA}

\author{Zaven Arzoumanian}
\affil{NASA Goddard Space Flight Center, Greenbelt, MD, USA}

\author[0009-0001-5099-8070]{Antoaneta A. Avramova-Boncheva}
\affil{The Institute of Astronomy and National Astronomical Observatory,  Bulgarian Academy of Sciences, 72 Tsarigradsko Chaussee Blvd.,1784 Sofia, Bulgaria}

\author[0000-0001-7115-2819]{Keith Gendreau}
\affil{NASA Goddard Space Flight Center, Greenbelt, MD, USA}

\author[0000-0003-2490-4779]{Sandra V. Jeffers}
\affil{Max Planck Institute for Solar System Research, Justus-von-Liebig-weg 3, 37077 G\"ottingen, Germany}

\author[0000-0001-5522-8887]{Stephen Marsden}
\affil{Centre for Astrophysics, University of Southern Queensland, Toowoomba, Queensland 4350, Australia}

\author[0000-0002-4996-6901]{Julien Morin}
\affil{LUPM, Universit\'e de Montpellier, CNRS, Place Eug\`ene Bataillon, F-34095 Montpellier, France}

\author[0000-0003-1978-9809]{Coralie Neiner}
\affil{LESIA, Paris Observatory, PSL University, CNRS, Sorbonne University, Universit\'e Paris Cit\'e, 5 place Jules Janssen, 92195 Meudon, France}

\author[0000-0001-5371-2675]{Aline A. Vidotto}
\affil{Leiden Observatory, Leiden University, PO Box 9513, 2300 RA Leiden, The Netherlands}

\author{Kazunari Shibata}
\affil{Kwasan Observatory, Kyoto University, Yamashina, Kyoto 607-8471, Japan}
\affil{School of Science and Engineering, Doshisha University, Kyotanabe, Kyoto 610-0321, Japan}

\begin{abstract}

Young solar-type stars frequently produce superflares, serving as a unique window into the young Sun-Earth environments. 
Large solar flares are closely linked to coronal mass ejections (CMEs) associated with filament/prominence eruptions, but its observational evidence for stellar superflares remains scarce. 
Here, we present a 12-day multi-wavelength campaign observation of young solar-type star EK Draconis (G1.5V, 50-120 Myr age) utilizing TESS, NICER, and Seimei telescope. 
The star has previously exhibited blueshifted H$\alpha$ absorptions as evidence for a filament eruption associated with a superflare. 
Our simultaneous optical and X-ray observations identified three superflares of $1.5\times10^{33}$ -- $1.2\times10^{34}$ erg. 
We report the first discovery of two prominence eruptions on a solar-type star, observed as blueshifted H$\alpha$ emissions at speed of 690 and 430 km s$^{-1}$ and masses of $1.1\times10^{19}$ and $3.2\times10^{17}$ g, respectively.
The faster, massive event shows a candidate of post-flare X-ray dimming with the amplitude of up to $\sim$10 \%.
Several observational aspects consistently point to the occurrence of a fast CME associated with this event.
The comparative analysis of the estimated length scales of flare loops, prominences, possible dimming region, and starspots provides the overall picture of the eruptive phenomena.
Furthermore, the energy partition of the observed superflares in the optical and X-ray bands is consistent with flares from the Sun, M-dwarfs, and close binaries, yielding the unified empirical relations.
These discoveries provide profound implications of impact of these eruptive events on the early Venus, Earth and Mars and young exoplanets.

\end{abstract}

\keywords{Stellar flares (1603); Stellar coronal mass ejections (1881); Optical flares (1166); Stellar x-ray flares (1637); Flare stars (540); G dwarf stars (556); Solar analogs (1941)}


\section{Introduction}\label{sec:1}

Solar and stellar flares are the most powerful explosive phenomena in atmospheres of cool stars observed in the wide range of wavelengths, from radio to X-ray bands \citep[see][for reviews]{2011LRSP....8....6S,2017LRSP...14....2B}.
They are thought to be caused by the conversion of magnetic energy stored in stellar coronae into kinetic and thermal energy via magnetic reconnection. 
These events on the Sun are frequently accompanied by coronal mass ejections (CMEs) of hundreds of millions of tons of hot coronal material propagating into the interplanetary space at a few hundreds to a few thousands of km s$^{-1}$. 
X-rays and Extreme Ultraviolet (EUV) emission from large solar flares and associated fast CMEs often have severe impacts on the planetary magnetospheres and ionospheres \citep[see][for reviews]{2020IJAsB..19..136A,2021LRSP...18....4T,2022LRSP...19....2C}.

The largest flare energy observed on our Sun is about 10$^{32}$ erg \citep[e.g.,][]{2012ApJ...759...71E}, while larger flares with the energy over 10$^{33}$ erg referred to as ``superflares" have never been reported in the modern solar observations \citep[e.g.,][]{2013A&A...549A..66A}. However, recent studies have identified multiple possible extreme solar energetic events with energies up to 10$^{34}$ ergs have occurred in the past millennia using cosmogenic radionuclides ($^{14}$C, $^{10}$Be, $^{36}$Cl) recorded in tree rings and ice cores \citep[e.g.,][]{2019esps.book.....M,2023LRSP...20....2U}.
Furthermore, recent observations of solar-type (G-type main-sequence) stars have provided important insights into the rate of occurrence of superflares from the young solar-type stars and the analogs of the present-day Sun.
Vigorous observational studies of stellar flares over the last 30 yrs revealed that young rapidly-rotating solar-type stars show frequent superflares at a rate of 1 event per day \cite[age of $\sim$100 Myr;][]{1999ApJ...513L..53A,2000ApJ...541..396A,2022ApJ...926L...5N,2022PASJ...74.1295Y,2022A&A...661A.148C}, while mature slowly-rotating solar-type stars show superflares at much lower occurrence rates of about 1 event per a few thousand years \citep[age of several Gyr;][]{2012Natur.485..478M,2013ApJS..209....5S,2019ApJ...876...58N,2020arXiv201102117O}.
These data suggest that superflares from the young Sun could have had significant impact on the magnetospheric and atmospheric environments of the young Venus, Earth, and Mars, on a time scale of 300 Myr that is controlled
the stellar rotation rate \citep{2015A&A...577L...3T}.
Also, the detection of superflares on the slowly-rotating stars imply a possibility that superflares can occur on the present-day, mature Sun. Such studies are critical for understanding the frequency and impact of these events on our civilization and its infrastructures.
Thus, superflares on solar-type stars have been highlighted from the context of solar physics \citep[e.g.,][]{2013PASJ...65...49S,2013A&A...549A..66A,2022LRSP...19....2C}, planetary science \citep[e.g.,][]{2018AsBio..18..739F,2019LNP...955.....L,2020IJAsB..19..136A}, biochemistry \citep[e.g.,][]{2023Life...13.1103K}, and historical and geophysical studies \citep[e.g.,][]{2012Natur.486..240M,2023LRSP...20....2U,2017ApJ...850L..31H}.

High magnetic activity of young solar-type stars is characterized not only by stellar superflares but also by strong average surface magnetic fields \citep[e.g.,][]{2014MNRAS.441.2361V,2020A&A...635A.142K} and giant starspots covering a substantial portion of the hemisphere (up to tens of percent) \citep[e.g.,][]{2017MNRAS.465.2076W,2017PASJ...69...41M,2019ApJ...871..187N,2020ApJ...891..103N,2019ApJ...876...58N,2021ApJ...907...89H,2020arXiv201102117O,2022PASJ...74.1295Y}. The magnetic energy stored in active regions associated with starspots power X-ray and EUV bright (up to 10$^{30}$ erg s$^{-1}$), hot (up to 10 MK) and dense (10$^{10-11}$ cm$^{-3}$) coronae and massive winds \citep[e.g.,][]{1995A&A...301..201G,1997ApJ...479..416G,2003ApJ...598.1387P,2004A&ARv..12...71G,2011ApJ...743...48W,2020ApJ...901...70T,2021A&A...656A.111S,2021ApJ...916...96A,2022ApJ...927..179T,2023ApJ...945..147N}. 
Observations of the nearby young solar-type stars in various phases of evolution have provided insights into the time history of our Sun and its heliosphere, which have had a profound influence on the young Earth over the span of millions to billions of years \citep[e.g.,][]{1997ApJ...483..947G,2007LRSP....4....3G,2019A&A...624L..10J,2020arXiv201102117O}.
According to recent geophysical and fossil data, life on Earth is thought to have originated between 4.4 and 3.8 billion years ago \citep{2023FrASS...995701W}. This time period is known as the ``Hadean Eon", during which the Earth would have been subjected to intense XUV radiation and frequent solar flares and CMEs from the young Sun. 
The XUV radiation from the young Sun could have triggered photoionization processes and contributed to the atmospheric escape of the early Earth \citep{2017ApJ...836L...3A,2021MNRAS.505.2941Y}. 
Moreover, the powerful energetic particles associated with flares/CMEs could have induced chemical reactions, resulting in the formation of greenhouse gases and prebiotic chemistry in the Earth's primitive atmosphere \citep{2016NatGe...9..452A,2023Life...13.1103K}. 
These processes may intimately contribute to the formation and disruption of habitable worlds, conditions crucial for the emergence and sustainment of life.
As there is little direct evidence of the strong XUV flare radiation and energetic CMEs of the young Sun from the present-day Solar system, we need to infer them from observations of young solar-type stars at ages of a few tens to hundreds of millions of years characteristic of the Hadean period.

Several nearby solar-type stars have been considered as likely proxies for the young Sun \citep[e.g.,][]{1997ApJ...483..947G,2007LRSP....4....3G,2019A&A...624L..10J}. These include $\kappa^1$ Ceti (G5V, $\sim$ 600 Myr old), EK Draconis (EK Dra, G1.5V, $\sim$ 125 Myr, \citealt{2017MNRAS.465.2076W,2021MNRAS.502.3343S}), and DS Tucanae A (DS Tuc A, G6V, $\sim$ 45 Myr old).
\cite{1999ApJ...513L..53A,2000ApJ...541..396A} reported on X-ray flare frequencies for $\kappa^1$ Ceti and EK Dra, using data from the \textit{Extreme Ultraviolet Explorer} (\textit{EUVE}). They examined the contribution of flares to coronal heating and investigated the relationship between flares and quiescent X-ray emissions. 
Later, \cite{2023ApJ...944..163H} reported X-ray superflares on $\kappa^1$ Ceti using the \textit{Neutron star Interior Composition ExploreR} (\textit{NICER}).
\cite{2015AJ....150....7A} detailed UV spectroscopic observations of the decay phase of a superflare on EK Dra, captured by the \textit{Hubble Space Telescope}. They concluded that Si IV and O IV UV emission lines forming at 8$\times$10$^5$ K redshifted at 30--40 km s$^{-1}$ are the signatures of the post-flare loop downflows.
\cite{2019AA...622A.210G} conducted multi-wavelength observations of a superflare on the active solar-type star HII 345 (G8V), using the Kepler Space Telescope (K2 mission, 30 min cadence) and \textit{XMM-Newton}. They estimated the energy partition between X-ray and white-light flares and derived the length scale of the flare loop from X-ray data analysis, although its time delay between the peaks in the optical and X-ray bands was not clearly resolved.
More recently, \cite{2022A&A...666A.198P} reported two superflares on DS Tuc A, an exoplanet-hosting star, in X-ray and Near-Ultraviolet (NUV) wavelengths using \textit{XMM-Newton}. The X-ray emission peaks are notably delayed compared to NUV peaks, showing the relationship between heating processes associated with the Neupert effect as observed in solar flare events \citep{1968ApJ...153L..59N,1993SoPh..146..177D}. Specifically, these delays between peaks inform us about the impulsive sources that transport energetic particle energy flux from the solar/stellar coronae and deposit heating in lower chromospheric layers, which respond back in the form of evaporated flows filling coronal loops with hot and dense X-ray emitting plasma.
While these studies have uncovered many critical aspects of flaring atmospheres of magnetically active stars, multi-wavelength observations remain relatively scarce. The \textit{Kepler Space Telescope} established a robust relationship between rotation (age) and the frequency of white-light flares (WLFs) for solar-type stars \citep[][]{2012Natur.485..478M,2013ApJS..209....5S,2019ApJ...876...58N,2020arXiv201102117O}. 
Application of the extensive \textit{Kepler}'s statistics to assess impact of flares on exoplanetary environments requires characterizion of the radiative flux from superflares on young solar-type stars in the wide range of wavelengths: from XUV to optical.

Recent optical spectroscopic observations of superflares on young solar-type stars provided indications of stellar CMEs.
\cite{2022NatAs...6..241N} reported the first detection of the optical H$\alpha$ spectra of a superflare of $\sim 10^{33}$ erg on the young solar-type star EK Dra (the flare is labeled as ``E4" in the following sections).
The H$\alpha$ spectra show a blueshifted absorption as evidence of a gigantic stellar filament eruption (10$^{4}$ K plasma) with a velocity of 510 km s$^{-1}$ and mass of 10$^{18}$ g (approximately 10 times larger than the largest solar CMEs).
In the standard solar flare model, photospheric motions produce a stressed magnetic flux rope connected to a filament, which becomes unstable and is ejected into the solar corona, initiating a flare through magnetic reconnection \citep{2011LRSP....8....6S,2019ApJ...880...97L}. These prominence signatures relate to the formation of magnetic flux ropes in the solar and stellar corona, a phenomenon widely researched in solar studies.
On the Sun, filament or prominence eruptions (cores of CMEs) are often associated with interplanetary CMEs, the sources of space weather \citep{2022LRSP...19....2C}.
By directly comparing with solar and Sun-as-a-star observations, \cite{2022NatAs...6..241N} suggested that a massive high-energy stellar CME is probably associated with the eruption of an extended filament, affecting the young planetary systems and stellar mass loss.
On the other hand, \cite{2022ApJ...926L...5N} discovered another $10^{34}$ erg-class superflare on EK Dra (labeled as ``E5") which did not show significant blueshifted H$\alpha$ spectra, casting a doubt in the one-to-one relationship between filament eruptions/CMEs and superflares.
Alternatively, \cite{2022ApJ...926L...5N} suggested that the superflare could have occurred near the stellar limb and possibly is radiated from flare loops \citep{2018ApJ...859..143H}, resulting in no blue/redshifts due to the location.
The above observational studies demonstrate that time-resolved optical spectroscopic observations are the promising -- and currently the only approach to studying CMEs on young solar-type stars, which are also beneficial for understanding other chromospheric heating phenomena \citep[cf.][for solar and M-dwarf flares]{1962BAICz..13...37S,1984SoPh...93..105I,1987ApJ...322..999C,2013ApJS..207...15K,2020PASJ...72...68N,2020ApJ...895....6G,2022ApJ...933..209N,2022ApJ...939...98O,2023ApJ...945...61N}. However, to date, only two such observational events have been reported in the literature. To further our understanding of the impact on young planetary environments, we need to significantly increase the number of samples.

The search for signatures of stellar CMEs on other types of stars, such as M/K-dwarfs and active close binaries, has been relatively successful, by using the strategy based on solar CME observations (see, recent reviews by \citealt{2017IAUS..328..243O,2022SerAJ.205....1L,2022arXiv221105506N}). 
This is thanks to the long history of ground- and space-based observations, which is a result of their frequent flares and faint stellar quiescent background \citep[e.g.,][]{1984ApJS...54..375P,1989SoPh..121..299P,1992ApJS...78..565H}. 
Representatively, spectroscopic observations  revealed many flare events associated with blueshifted emission line profiles, called ``blue asymmetries", in optical Hydrogen Balmer lines \citep{1990A&A...238..249H,1992AJ....104.1161E,1994A&A...285..489G,2006A&A...452..987C,2008A&A...487..293F,2011A&A...534A.133F,2018A&A...615A..14F,2016A&A...590A..11V,2019A&A...623A..49V,2017A&A...597A.101F,2018PASJ...70...62H,2020arXiv201200786K,2020A&A...637A..13M,2020MNRAS.499.5047M,2020PASJ..tmp..253M,2023ApJ...948....9I}, UV lines \citep{2007PASP..119...67H,2014MNRAS.443..898L}, and low-temperature X-ray lines \citep{2019NatAs...3..742A} during flares on M dwarfs, close binaries, and subgiant/giant stars.
These signatures represent the direct evidence of plasma motions in line-of-sight (LOS) direction and the detected number is very large ($>$100).
Although their relatively low velocities of 100--300 km s$^{-1}$ in most cases make it difficult to distinguish whether they originate from prominence eruptions/CMEs or flare-related surface phenomena, some of them are characterized as prominence eruptions leading to CMEs by the faster velocities than the stellar surface escape velocities \citep{1990A&A...238..249H,2016A&A...590A..11V,2019A&A...623A..49V,2023ApJ...948....9I}.
Another promising method to indentify stellar CMEs is based on the detection of the XUV dimming observed during solar CMEs forming as a direct result of evacuation and outward propagation of materials from the solar corona, which can last for hours \citep{2016ApJ...830...20M,2022ApJ...928..154J}.
\cite{2021NatAs...5..697V} reported on possible post-flare X-ray and EUV dimming in cool dwarfs by using archive dataset, and later \cite{2022ApJ...936..170L} found post-flare dimming in coronal Far Ultraviolet (FUV) lines from $\epsilon$ Eridani (K2V).
Furthermore, \cite{2020ApJ...905...23Z} reported on a possible type-IV radio burst after a flare on Proxima Centauri (M5.5V), although no detections of type-II burst have been reported to date \citep{2018ApJ...856...39C,2018ApJ...862..113C,2019ApJ...871..214V}.
These eruptive events on cool dwarfs have been in general highlighted as a possible driver of mass and angular momentum loss evolution of the low-mass stars \citep[e.g.,][]{2015ApJ...809...79O,2017ApJ...840..114C,2021ApJ...915...37W} and as a threat to the exoplanet habitability \citep[e.g.,][]{2010AsBio..10..751S,2019AsBio..19...64T,2019ApJ...881..114Y,2020IJAsB..19..136A,2021NatAs...5..298C}.
Although there is still a lack of conclusive observational evidence, these observations provide a hint for potential approaches that could further be extended to the search of eruptions from solar-type stars, as discussed in our study.

\startlongtable 
\begin{deluxetable*}{lccc}
   \tablecaption{Observation log}
   \tablehead{
     Telescope & \colhead{UT date} & \colhead{Total Obs.} & \colhead{Exp. Time}  \\
 (Data Type) & \colhead{} & \colhead{} & \colhead{(sec)}  
     }
   \startdata
    \hline 
   3.8m Seimei & 2022 Apr 10 & 8.0 (hr)$^\dagger$ & 60,120  \\
   (Optical spec.) & 2022 Apr 11 & 3.0 (hr) & 60,120    \\
    & 2022 Apr 12 & 4.2 (hr) & 60,120    \\
    & 2022 Apr 15 & 2.6 (hr) & 60,120    \\
    & 2022 Apr 16 & 9.6 (hr) & 60,120   \\
    & 2022 Apr 17 & 9.2 (hr) & 60,120    \\
    & 2022 Apr 18 & 4.8 (hr) & 60,120    \\
    & 2022 Apr 19 & 8.1 & 60,120    \\
    & 2022 Apr 20 & 7.3 (hr) & 60,120    \\
     \hline 
   \textit{TESS} &   \multicolumn{2}{c}{2022 Mar 26 -- Apr 22} & 600 \\
   (Optical photo.) &   \multicolumn{2}{c}{(Sector 50)}  &   \\
   \hline 
   \textit{NICER}$^{\ast}$ &  2022 Apr 10  &  2192 (sec)$^\S$  & --  \\
   (Soft X-ray) &  2022 Apr 11 &  1361 (sec) & --  \\
    & 2022 Apr 12 & 687 (sec) & -- \\
    & 2022 Apr 13 & 3989 (sec) & -- \\
    & 2022 Apr 14 & 2666 (sec) & -- \\
    & 2022 Apr 15 & 3311 (sec) & -- \\
    & 2022 Apr 16 & 8355 (sec) & -- \\
    & 2022 Apr 17 & 9948 (sec) & -- \\
    & 2022 Apr 18 & 560 (sec) & -- \\
    & 2022 Apr 20 & 4436 (sec) & -- \\
    & 2022 Apr 21 & 3357 (sec) & -- \\
   \enddata
   \tablecomments{
$^\dagger$These are calculated by multiplying the number of frames times the each exposure time (plus readout time $\sim$17 sec).
$^{\ast}$The Observation ID is 5202490102--5202490113.
$^\S$The exposure was obtained from \url{https://heasarc.gsfc.nasa.gov/db-perl/W3Browse/w3table.pl?tablehead=name\%3Dnicermastr&Action=More+Options}.
}
   \label{tables:1:obs_log}
 \end{deluxetable*}

Given the above context, the aim of this study is threefold: (1) to understand the radiative energy distribution and evolution of flares among white-light, H$\alpha$, and X-ray wavelength; (2) to study the diversity of H$\alpha$ spectra from superflares on young solar-type stars by increasing the sample size; and (3) to provide a comprehensive understanding of stellar CMEs through multi-wavelength observations. To achieve these goals, we conducted the optical and X-ray monitoring observations of the young solar-type star, EK Dra, over a 12-day period from 2022 April 10 to 2022 April 21 (refer to Table \ref{tables:1:obs_log}). 
This period coincides with the optical photometric observations performed by the \textit{Transiting Exoplanet Survey Satellite} (\textit{TESS}, \citealt{2015JATIS...1a4003R}, Section \ref{sec:2-2}).
The high-time-cadence optical spectroscopic observations were conducted with 3.8m Seimei telescope in Okayama Observatory, Japan (Section \ref{sec:2-3}).
The X-ray monitoring observations were also performed by \textit{NICER} onboard International Space Station (ISS) (Section \ref{sec:2-4}).
This paper is organized as follows. Section \ref{sec:2} outlines the observational details, while Section \ref{sec:3} describes the analyses and results. In Section \ref{sec:4}, we investigate the radiative properties of superflares and their length scales. Section \ref{sec:5} examines the characteristics of prominence eruptions and a potential X-ray dimming. Section \ref{sec:6} integrates the information from Sections \ref{sec:4} and \ref{sec:5} to address our main questions and present our conclusions. 
In Section \ref{sec:future}, we outline our future studies, which include modeling prominence eruptions and starspots, as discussed in our forthcoming paper II (Namekata et al. in preparation).




\section{Observation and Data Reduction}\label{sec:2}

\subsection{EK Draconis}\label{sec:2-1}

EK Dra (HD 129333, G1.5V) is a young solar-type star at an age of 50--125 Myr. 
Its effective temperature of 5560--5700 K, radius of 0.94 $R_{\odot}$, and mass of 0.95 $M_{\odot}$ make this star one of the best proxies for an infant Sun at the time of Hadean period on Earth  \citep{2017MNRAS.465.2076W,2021MNRAS.502.3343S}.
It is a rapidly rotating star with the rotation period of 2.77 days \citep{1999ApJ...513L..53A,2015AJ....150....7A,2022NatAs...6..241N}, and it exhibits a high level of magnetic activity in the form of the hot and dense corona \citep[e.g.,][]{2005A&A...432..671S,2012ApJ...745...25L,2023ApJ...945..147N}, strong surface magnetic field and large starspots \citep[e.g.,][]{2005AN....326..283B,2017MNRAS.465.2076W,2018A&A...620A.162J} and frequent  superflares \citep[e.g.,][]{1999ApJ...513L..53A,2000ApJ...541..396A,2015AJ....150....7A,2022NatAs...6..241N,2022ApJ...926L...5N}.
Although a faint low-mass companion is reported $\sim$20 AU away in the projected distance from the G-type primary star (here we call the primary G dwarf EK Dra), these two stars are not believed to be magnetically connected \citep{2017MNRAS.465.2076W}, and the G-dwarf is the main source of stellar superflares and associated eruptive events \citep[cf. Section ``Low mass companion" in Supplementary Information of][]{2022NatAs...6..241N}.

\subsection{TESS}\label{sec:2-2}

The \textit{TESS} observed EK Dra (TIC 159613900) in its Sector 50 from 26 March to 22 April 2022\footnote{\url{https://archive.stsci.edu/missions/tess/doc/tess_drn/tess_sector_50_drn72_v02.pdf}}. \textit{TESS} conducts photometric observations using four cameras with the \textit{TESS} filter in the optical band of 6000--10000 {\AA}. Each sector is observed for approximately 27 days \citep{2015JATIS...1a4003R}. During this period, EK Dra was observed with a 10-minute time cadence mode.
The data were obtained from the Multimission Archive at the Space Telescope (MAST) archive\footnote{\url{https://mast.stsci.edu/portal/Mashup/Clients/Mast/Portal.html}}. We used \textsf{eleanor} for photometric analysis to extract light curves from TESS full-frame image (FFI) data \citep{2019PASP..131i4502F}. The \textsf{eleanor} yields ``raw" flux, ``corrected" flux that is regressed against instrumental effects, and ``PCA (Principal Component Analysis)" flux that has common systematics between targets on the same camera removed.
We employed the ``raw" flux for flare analysis due to the relatively high noise level of the ``corrected" and ``PCA" fluxes. For the analysis of rotational modulations, only the long-term rotational modulations in the ``raw" flux were calibrated using a smoothed ``PCA" flux.

\subsection{3.8-m Seimei Telescope}\label{sec:2-3}

\begin{figure}
\epsscale{0.5}
\plotone{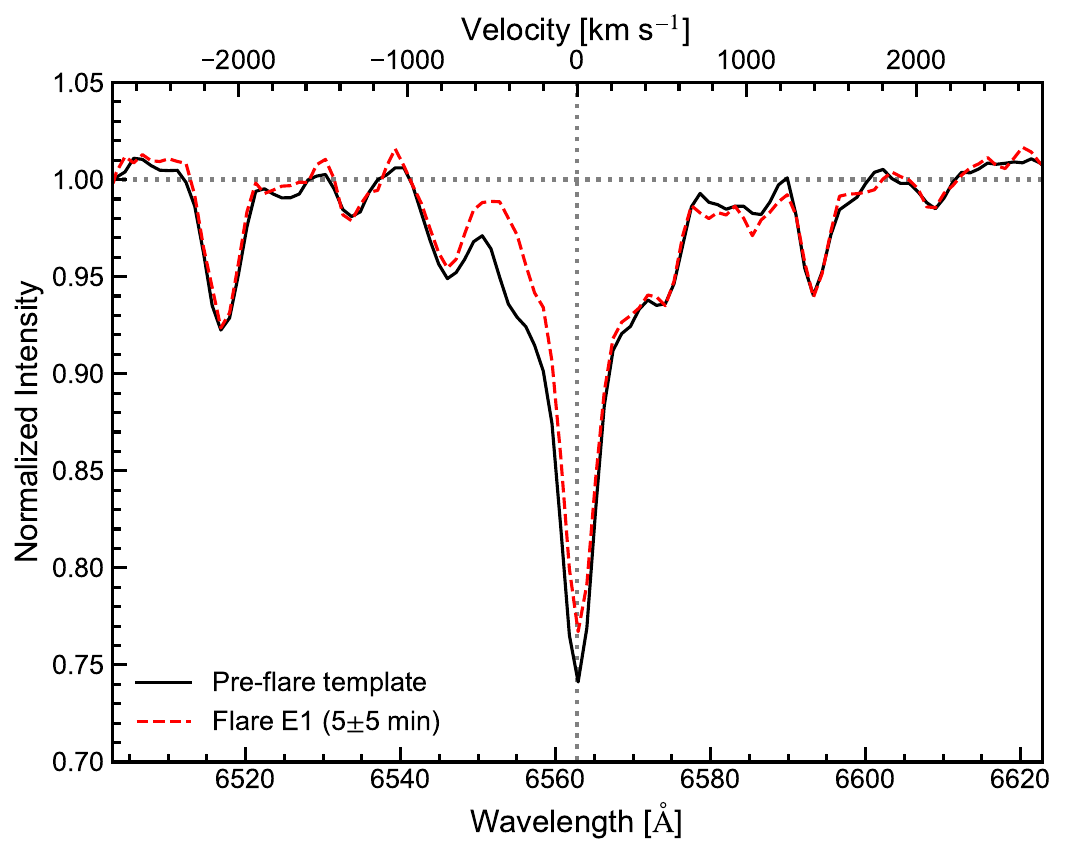}
\caption{Examples of H$\alpha$ spectra of EK Dra on 10 Apr 2022. A black solid line is the pre-flare template for flare E1, and a red dashed line is the flaring spectrum of flare E1.}
\label{fig:0}
\end{figure}

We conducted optical spectroscopic monitoring observations using the Kyoto Okayama Optical Low-dispersion Spectrograph with optical-fiber Integral Field Unit (KOOLS-IFU) \citep{2019PASJ...71..102M}, installed on the 3.8-m Seimei Telescope \citep{2020PASJ...72...48K} at Okayama Observatory of Kyoto University. The KOOLS-IFU is an optical spectrograph that provides a spectral resolution of R $\sim$ 2000 and covers a wavelength range from 5800-8000 {\AA}. The campaign ran from 2022 April 10 to 2022 April 20, with the exposure time set at either 60 or 120 seconds, depending on weather conditions.
We followed the data reduction procedure outlined in \cite{2020PASJ...72...68N,2022NatAs...6..241N,2022ApJ...926L...5N} using \textsf{IRAF}\footnote{IRAF is distributed by the National Optical Astronomy Observatories, which are operated by the Association of Universities for Research in Astronomy, Inc., under cooperate agreement with the National Science Foundation.} and \textsf{PyRAF}\footnote{PyRAF is part of the stscipython package of astronomical data analysis tools and is a product of the Science Software Branch at the Space Telescope Science Institute.} packages. This produced a wavelength-calibrated and continuum-normalized 1D spectrum for each frame. The wavelength of the spectra was calibrated to account for the radial velocity shift caused by Earth's motion and the radial velocity of EK Dra \citep[--20.7 km s$^{-1}$;][]{2018A&A...616A...2L}.
Figure \ref{fig:0} shows examples of the obtained H$\alpha$ spectrum of EK Dra, which is typical G-dwarf spectra with absorption (even during flares).
We computed the H$\alpha$ equivalent width (hereafter ``EW", which refers to H$\alpha$ emission integrated for 6562.8--10 {\AA} $\sim$ 6562.8+10 {\AA}), and used this value to plot the H$\alpha$ light curves.
Note that we defined an EW in terms of an emission profile to be negative. In this literature, on the other hand, we often refer pre-flare subtracted equivalent width as $\Delta$EW = -- (EW -- EW$_{\rm preflare}$), where an emission is positive value.

\subsection{NICER}\label{sec:2-4}

\textit{NICER} is a soft X-ray instrument (0.2 -- 12 keV) onboard the International Space Station (ISS) \citep{2016SPIE.9905E..1HG}. 
\textit{NICER} monitored EK Dra from 2022 April 10 to 2022 April 21 (Observation ID: 5202490102 -- 5202490113)\footnote{Observations also took place on 2022 April 9 (ID: 5202490101) and 2022 April 19 (ID: 5202490111), but with exposure times of 9 and 0 seconds respectively, they were not included in our analysis.}. 
These observations were scheduled concurrently with those from the 3.8m Seimei telescope and \textit{TESS}.
We downloaded the data from NASA's website\footnote{\url{https://heasarc.gsfc.nasa.gov/db-perl/W3Browse/w3table.pl?tablehead=name\%3Dnicermastr\&Action=More+Options}}. 
The exposure time for each ISS orbit (about 90 minutes) varied from a few minutes to a few tens of minutes, depending on the dates.


We processed the datasets using the NICER calibration version CALDB XTI(20221001) and the \textsf{nicerl2} tool within HEASoft version 6.31 (and NICERDAS version V010c). \textit{NICER}'s X-ray Timing Instrument (XTI) consists of 56 X-ray modules. In our data processing with \textsf{nicerl2}, we eliminated six modules (id = 11, 14, 20, 22, 34, 60) that were either inactive or noisy. We performed barycentric correction for each event file using \textsf{barycor}.

The XTI has a large collection area between 0.2--12 keV (approximately 1900 cm$^{-2}$ at 1.5 keV). We used \textsf{xselect} and \textsf{lcurve} to produce a light curve with a 2-minute time cadence and an energy band of 0.5 -- 3 keV. As \textit{NICER} is not an imaging instrument, there is some background contamination in the higher and lower energy bands. Thus, we chose 0.5 -- 3 keV as a conservative energy range where the variation in background is relatively negligible.

We extracted the time-resolved X-ray spectra using \textsf{xselect}. The background files were created with \textsf{nibackgen3c50} \citep{2022AJ....163..130R}. We performed spectral model fitting with \textsf{xspec}. We will explain the details of the model in Section \ref{sec:3:x}.

\section{Analysis and Results}\label{sec:3}

\subsection{Light Curves}\label{sec:3:lc}

\begin{figure}
\epsscale{1.0}
\plotone{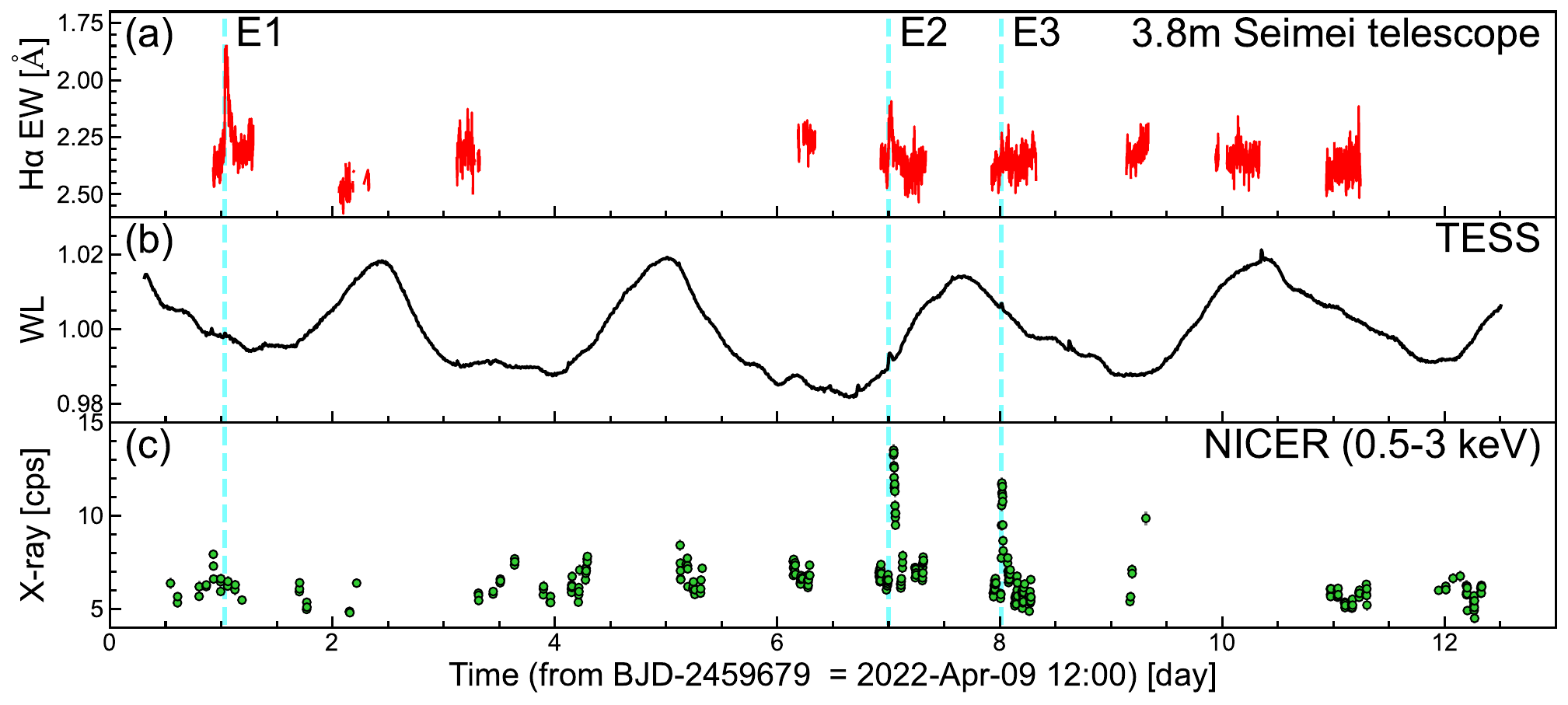}
\caption{Overall light curves of EK Dra for (a) H$\alpha$ EW, (b) \textit{TESS}'s optical white light, and (c) \textit{NICER}'s X-ray count rate (0.5-3 keV), taken from 2022 April 9 to 2022 April 21. H$\alpha$ EW is calculated within the wavelength range of 6562.8$\pm$10 {\AA}. Note that a negative (or negative-direction) EW value corresponds to emission by definition. \textit{TESS}'s optical white light fluxes are normalized to the average flux. The vertical dashed lines represent detected superflares, labeled as ``E1", ``E2", and ``E3".}
\label{fig:1}
\end{figure}

\begin{deluxetable*}{lccccccccc}
\tablecaption{The list of superflares on EK Dra in this study and past observations.}
\tablewidth{0pt}
\tablehead{
\colhead{} & \colhead{$GOES$} & \colhead{$E_{\rm X,bol}$} & \colhead{$E_{\rm WL,bol}$} & \colhead{$E_{\rm H\alpha}$} & \colhead{$t_{\rm WL}$} & \colhead{$\tau_{\rm WL}$} & \colhead{$t_{\rm H\alpha}$} &  Ref.  \\
\colhead{}& \colhead{} & \colhead{(0.1--100 keV)} & \colhead{} & \colhead{} & \colhead{(FWHM)} & \colhead{(e-decay)} & \colhead{(FWHM)} & &   \\
\colhead{} & \colhead{} & \colhead{[10$^{33}$ erg]} & \colhead{[10$^{33}$ erg]} & \colhead{[10$^{31}$ erg]} & \colhead{[min]} & \colhead{[min]} & \colhead{[min]}  &   
}
\startdata
2022 Apr 10 (E1) & -- & --  & 1.5$_{\pm 0.1}$ & 48.7$_{\pm 1.7}$(9.1$_{\pm 2.6}$)$^{\ast}$ & 20 & 18 & 48 & (1)  \\
2022 Apr 16 (E2) & X4000-5400$^{(\ae)}$ & $12.3_{-0.5}^{+0.3}$-$16.7_{-0.7}^{+0.4}$$^{(\ae)}$  & 12.2$_{\pm 0.2}$ & 17.7$_{\pm 0.7}$ & 40 & 20 &  54 & (1) \\
2022 Apr 17 (E3) & X3200 & 1.14$_{-0.05}^{+0.06}$ & 3.4$_{\pm 0.1}$  & 2.9$_{\pm 0.3}$ & 30  & 9.6 & 23 & (1) \\
\hline
2020 Apr 05 (E4) & -- & --  & 2.0$_{\pm 0.1}$ & 1.7$_{\pm 0.1}$$^{(\ddag)}$ & 6.0 & 5.4  & 7.8$^{(\ddag)}$ & (2) \\
2020 Mar 14 (E5) & -- & --  & 26$_{\pm 3}$ & 40$_{\pm 4}$ & 130 & 26  & 130 & (3) \\
\enddata
\tablecomments{
``GOES" is the goes X-ray class. $E_{\rm X,bol}$ is the bolometric X-ray flare energy in 0.1--100 keV. 
Note that the value of ``GOES" and $E_{\rm X,bol}$ are calibrated for the light curve extrapolation.
$t_{\rm WL}$ and $t_{\rm H\alpha}$ is the FWHM duration of the WL and H$\alpha$ flare. 
$\tau_{\rm WL}$ is the e-folding decay time of the flare.
$^{\ae}$The energy and flux values are already corrected for light curve extrapolations. 
$^{\ast}$The value in bracket is the radiation energy of the central component without the blueshifted component. 
$^{\ddag}$The flare emission on April 5 in \cite{2022NatAs...6..241N} is very short-lived and its light curve is a combination of blueshift absorption and flaring emission, so the H$\alpha$ flare duration is expected to be underestimated. 
References: (1) This study, (2) \cite{2022NatAs...6..241N}, (3) \cite{2022ApJ...926L...5N}.
}
\label{tab:flare-basic}
\end{deluxetable*}

\begin{figure}
\epsscale{0.5}
\plotone{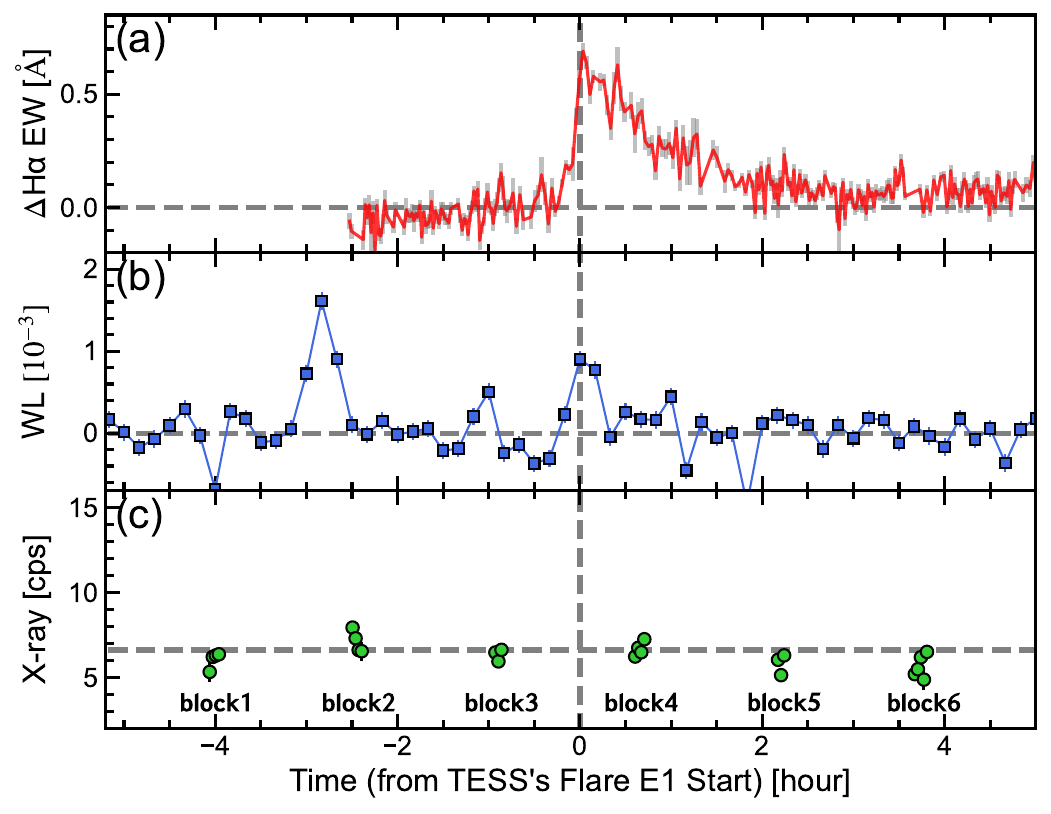}
\plotone{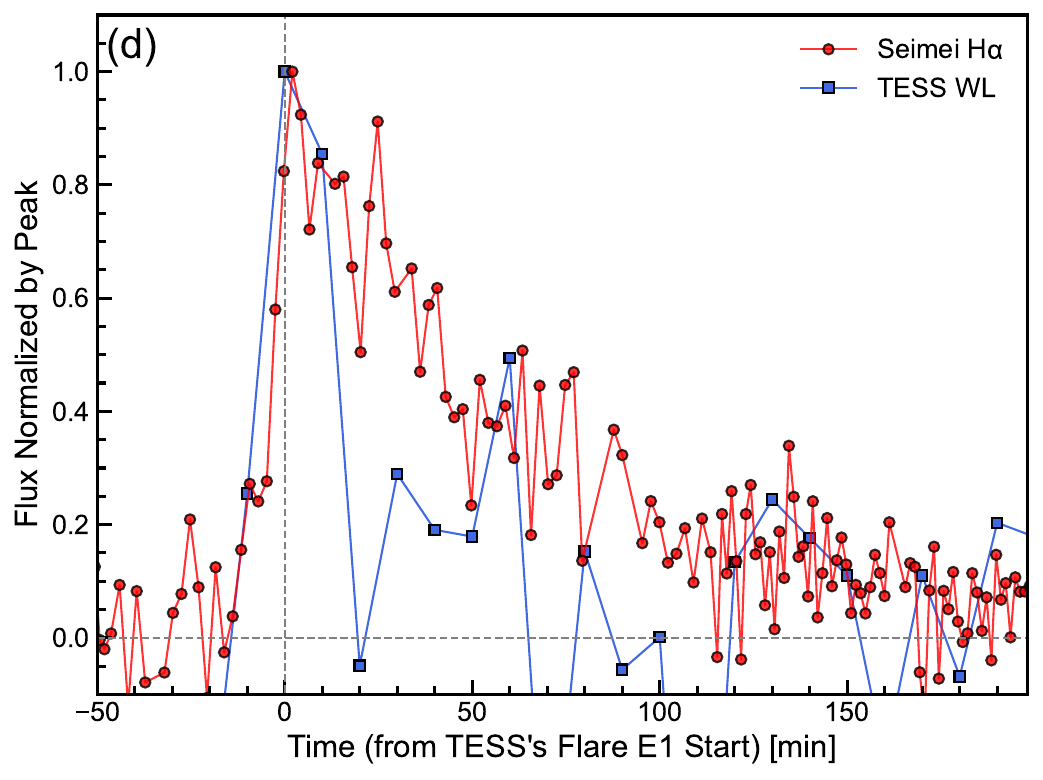}
\caption{
Light curves for the superflare observed on 2022 April 10 (E1). (a) The red line is pre-flare subtracted H$\alpha$ equivalent width ($\Delta$EW = $-$ (EW $-$ EW$_{\rm preflare}$)) with the grayed error bars. Note that positive $\Delta$EW values correspond to emissions. The H$\alpha$ EW is calculated within a range of 6562.8$\pm$20 {\AA}, which differs from the definition applied to flares E2 and E3 (i.e., 6562.8$\pm$10 {\AA}). (b) Background-subtracted \textit{TESS} WL (= ($F$ $-$ $F_{\rm bkg}$)/$F_{\rm av}$ ). (c) Soft X-ray count rate (0.5 -- 3 keV). In panel (c), each ISS orbit on this day is labeled as block 1--6. (d) H$\alpha$ $\Delta$EW and WL, each normalized by their respective peak values. In each panel, the vertical dashed line marks the \textit{TESS} flare onset time, and the horizontal dashed line depicts the basal level for each wavelength (H$\alpha$: median value of --80$\sim$--20 min, soft X-ray: mean value of block 3).
}
\label{fig:2}
\end{figure}

\begin{figure}
\epsscale{0.5}
\plotone{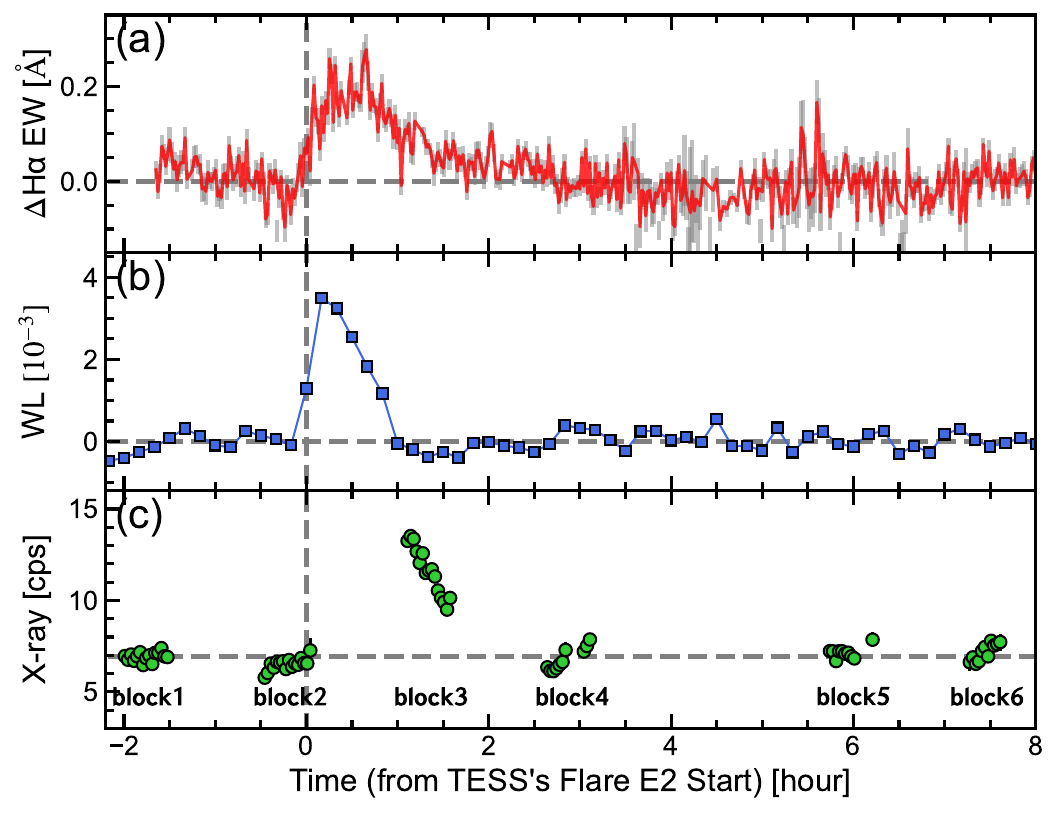}
\plotone{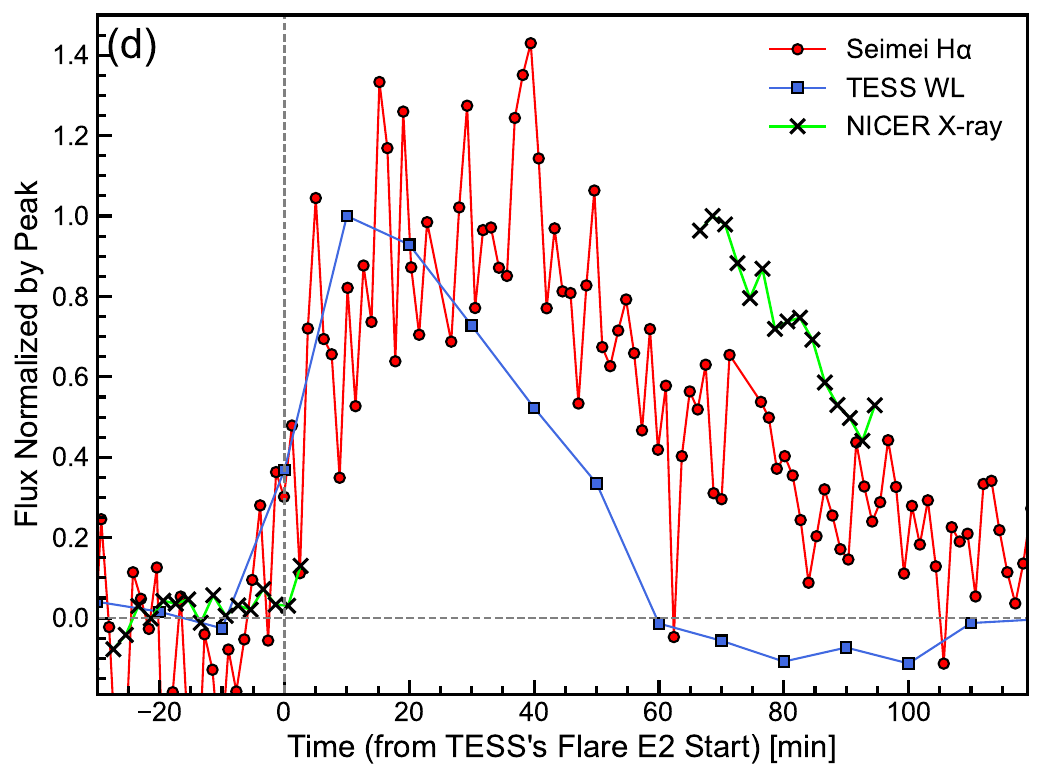}
\caption{The same as Figure \ref{fig:2}, but for the superflare observed on 2022 April 16 (E2). 
In panel (a), the H$\alpha$ EW is calculated within a range of 6562.8$\pm$10 {\AA}. Panel (d) displays H$\alpha$ $\Delta$EW, WL, and soft X-ray, each normalized by their respective peak values. The basal level of H$\alpha$ $\Delta$EW is a median value of --45$\sim$--15 min, while the basal level of the soft X-ray is the mean value of block 2.}
\label{fig:3}
\end{figure}

\begin{figure}
\epsscale{0.5}
\plotone{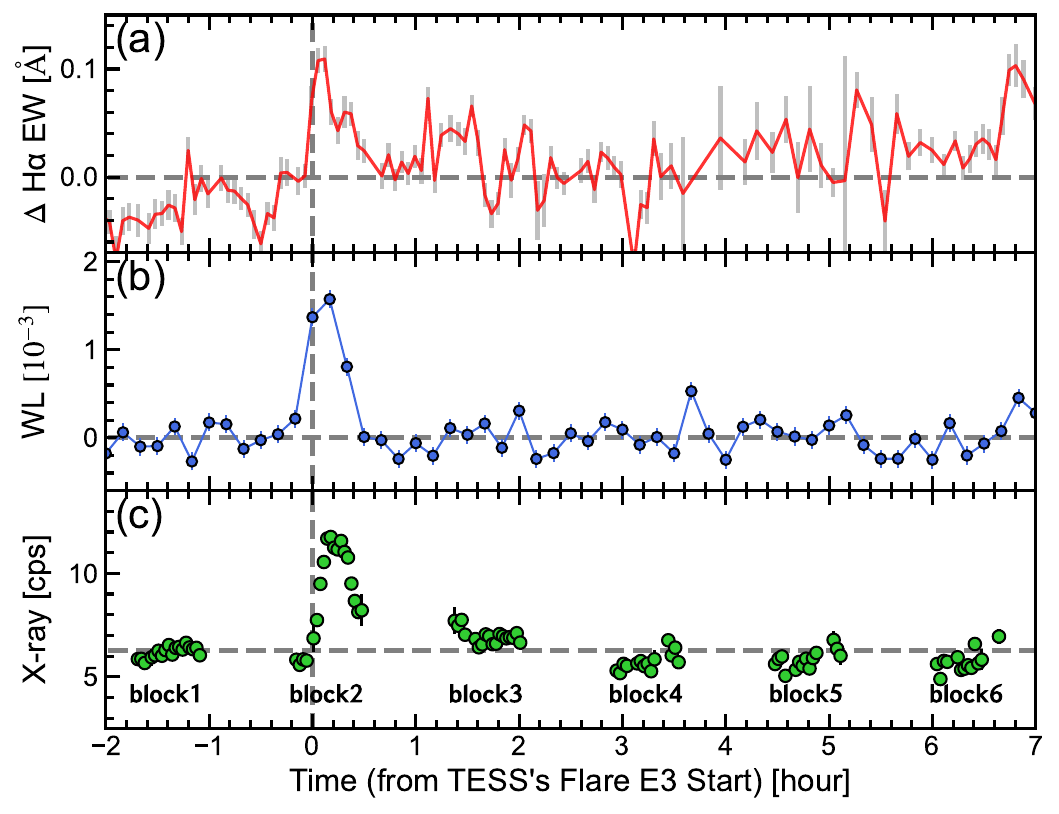}
\plotone{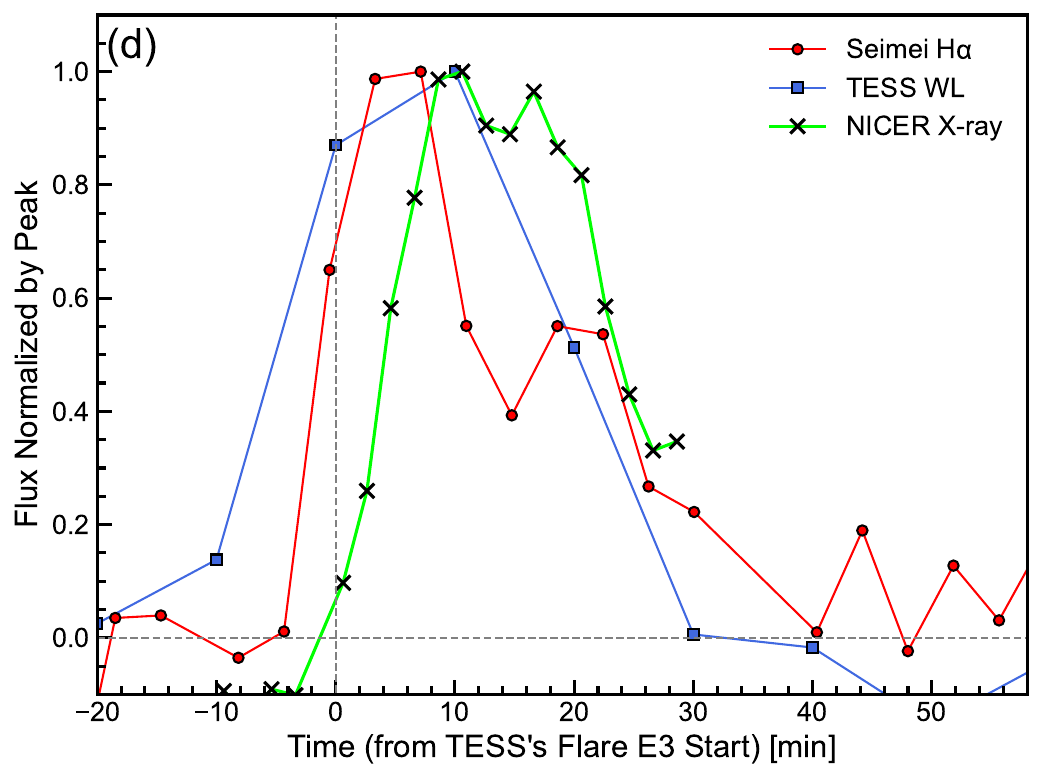}
\caption{The same as Figures \ref{fig:2} and \ref{fig:3}, but for the superflare observed on 2022 April 17 (E3). 
The basal level of H$\alpha$ EW is a median value of --25$\sim$--5 min, while the basal level of soft X-ray is a mean value of block 1. 
}
\label{fig:4}
\end{figure}

Figure \ref{fig:1} overviews the observed multi-wavelength light curves of EK Dra.
As shown in the figure, we have labeled the superflares, targeted in this paper, as ``E1" to ``E3" (Table \ref{tab:flare-basic}).
We are targeting the events where there is an increase in H$\alpha$ or X-ray that occurs simultaneously with the flares from \text{TESS}.
Figures \ref{fig:2} to \ref{fig:4} show the enlarged light curves of flares E1 to E3. 
Flare E1 has shown the largest H$\alpha$ emission increase ($\Delta$EW of $\sim$0.7 {\AA}) ever reported for solar-type stars, and an increase has been confirmed in both H$\alpha$ and white light (WL). Simultaneous observation data in X-rays do not exist for flare E1 due to the observation gap of the ISS, but data before and after the flare do exist also in X-ray. 
Flare E2 shows a simultaneous increase in H$\alpha$ and WL, and an increase during its decay phase has been detected in X-ray. 
Furthermore, for Flare E3, an increase has been detected in H$\alpha$, WL, and X-rays altogether.
In the following, we will describe the \textit{TESS} data analyses and results in Section \ref{sec:3:tess}, H$\alpha$ spectral analyses and results in Section \ref{sec:3:ha}, and X-ray light curve and spectral analyses and results in Section \ref{sec:3:x}.

\subsection{TESS White-light Flare Analysis}\label{sec:3:tess}

We quantified the white-light flare (WLF) emission in the \textit{TESS}-band, following the procedures as taken in \cite{2022NatAs...6..241N,2022ApJ...926L...5N}. 
The detection of flares first requires the removal of background rotational brightness variations. 
We removed it by using the Fast Fourier Transformation \textsf{numpy.fft.fft} and a low-pass filter with the cut-off frequency at 3.0 d$^{-1}$. 
The detrended light curves are shown in panels (a) of Figures \ref{fig:2}, \ref{fig:3}, and \ref{fig:4}.
This study did not perform an automatic flare detection, but simply identified the presence of significant WLFs in the \textit{TESS} light curve, specifically those occurring simultaneously with H$\alpha$ and X-ray flares.
In this context, a ``significant" flare refers to an event where more than two consecutive data points exceed the photometric error (approximately 100 ppm) as computed by \textsf{eleanor}. 
As noted in Section \ref{sec:3:lc}, we confirmed the presence of significant flare increases corresponding to each of the H$\alpha$ and X-ray flares E1, E2, and E3, as in the vertical lines in panels (a) of Figures \ref{fig:2}, \ref{fig:3}, and \ref{fig:4}, respectively.
Throughout this paper, we define the first flare point in TESS light curve as the flare standard time (BJD-2459680.0329, 2459685.9980, 2459687.0119 for flares E1, E2, and E3, respectively).

The calculation of bolometric WL flare energy follows the conventional technique proposed by \cite{2013ApJS..209....5S}. 
We assumed a 10,000 K blackbody radiation spectrum for the flares and computed the bolometric radiation energy. 
As a result, we derived the WLF energy ($E_{\rm WL,bol}$) of flare E1 to be $1.5_{\pm 0.1}\times 10^{33}$ erg, flare E2 to be $1.22_{\pm 0.02}\times 10^{34}$ erg, and flare E3 to be $3.4_{\pm 0.1}\times 10^{33}$ erg; hence, all of these are classified as ``superflares". 
\cite{2022ApJ...926L...5N} reported the largest energy scale of superflares on EK Dra is $\sim5\times10^{34}$ erg, so these are relatively smaller than the upper limit. 
Also, we derived the full-width at half maximum (FWHM) duration ($t_{\rm WL}$) and e-folding decay time of WLFs ($\tau_{\rm WL}$). 
The e-folding decay times ($\tau_{\rm WL}$) were obtained by fitting the decay phase of WLFs with an exponential function and \textsf{scipy.optimize.curve\_fit}. 
These values are presented in Table \ref{tab:flare-basic}.

To better constrain the white light flare energy, we need to discuss the assumption of 10,000 K blackbody temperature.
Most of solar flare studies that employ spatially-resolved data reported the emission temperature of 5,000--7,000 K \citep[e.g.,][]{2013ApJ...776..123W,2014ApJ...783...98K,2016ApJ...816...88K,2017ApJ...851...91N}, while only Sun-as-a-star observations suggest the hotter temperature of the blackbody spectrum $\sim$9,000 K \citep{2010NatPh...6..690K,2011A&A...530A..84K}. 
On the other hand, a historically well-known stellar superflare on an M-dwarf AD Leonis shows the broad-band stellar spectra with the effective temperature of 9,000--10,000 K (e.g., \citealt{1992ApJS...78..565H}), and thus many stellar flare studies use this temperature as a template.
Recent statistical survey shows that the flare temperatures vary depending on flare energies for M-dwarf flares \citep{2020ApJ...902..115H}. 
More recently, \textit{COROT}'s two-band photometric observations constrained the temperatures of G-dwarfs superflares as 3,570--24,300 K \citep{2022AJ....164..223R}. 
Hence, no unified model for radiation temperature of solar and stellar WLFs is currently available. 
\cite{2017ApJ...851...91N} concluded that the uncertainty in the emission temperature does not largely affect the estimation of the WLF energy because the decrease in the temperature leads to the decrease in surface luminosity ($\propto T^{4}$) and the increase in emission area ($A\propto T^{-4}$), both of which are cancelled by each other when using the Equation (1) of \cite{2013ApJS..209....5S} (i.e., $E=\sigma T^{4}A$). 
Therefore, our assumption may provide an uncertainty of the flare energy by a factor of few. 
We need to perform multi-band observations of superflares in U, B, V and near-UV bands to derive better constraints on the flare energy \citep{2023ApJ...944....5B}.

\subsection{H$\alpha$ Flare Analysis}\label{sec:3:ha}

\subsubsection{Overview of H$\alpha$ Flare Spectra}\label{sec:3:ha-1}

\begin{figure}
\epsscale{0.55}
\plotone{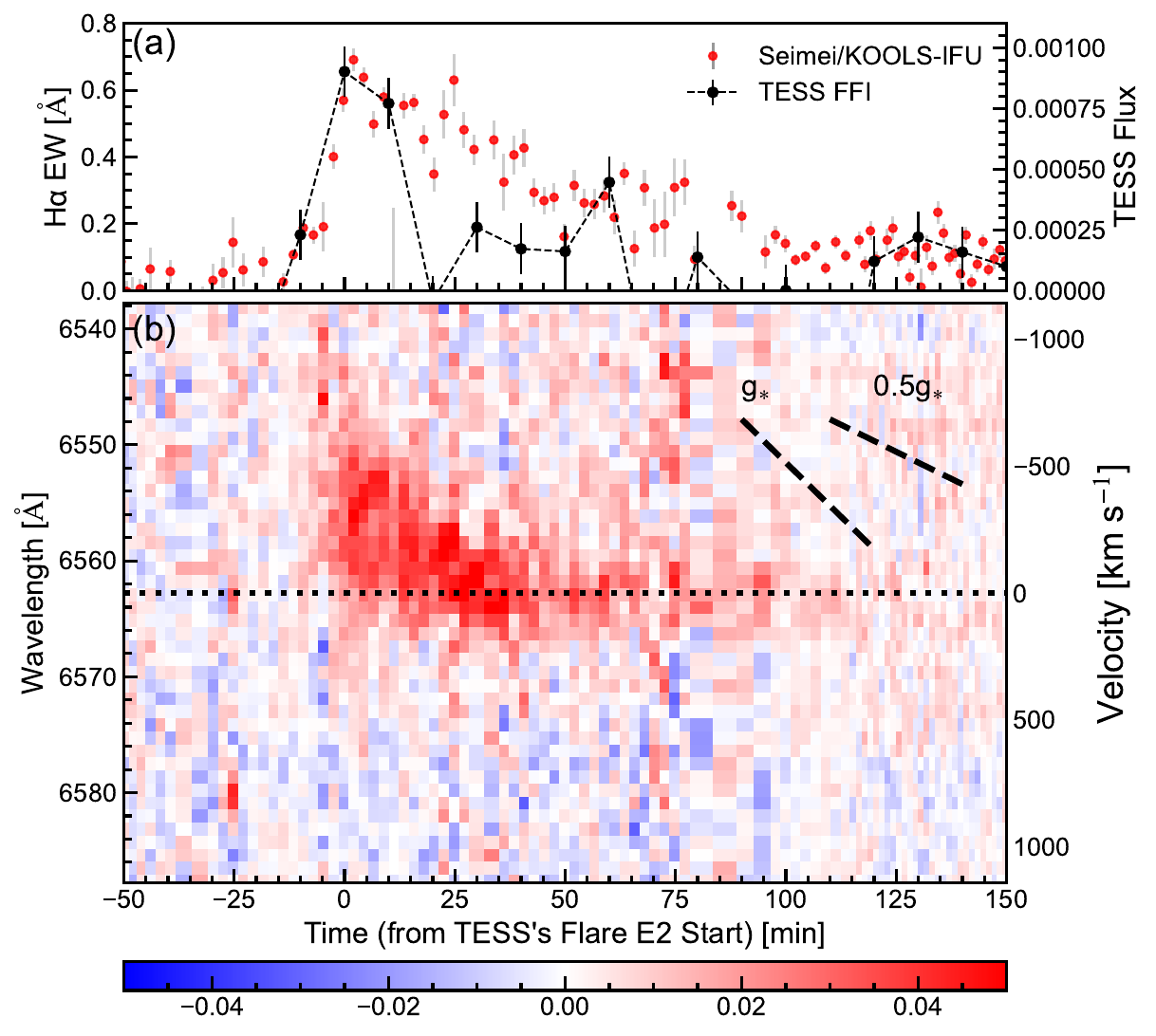}
\epsscale{0.45}
\plotone{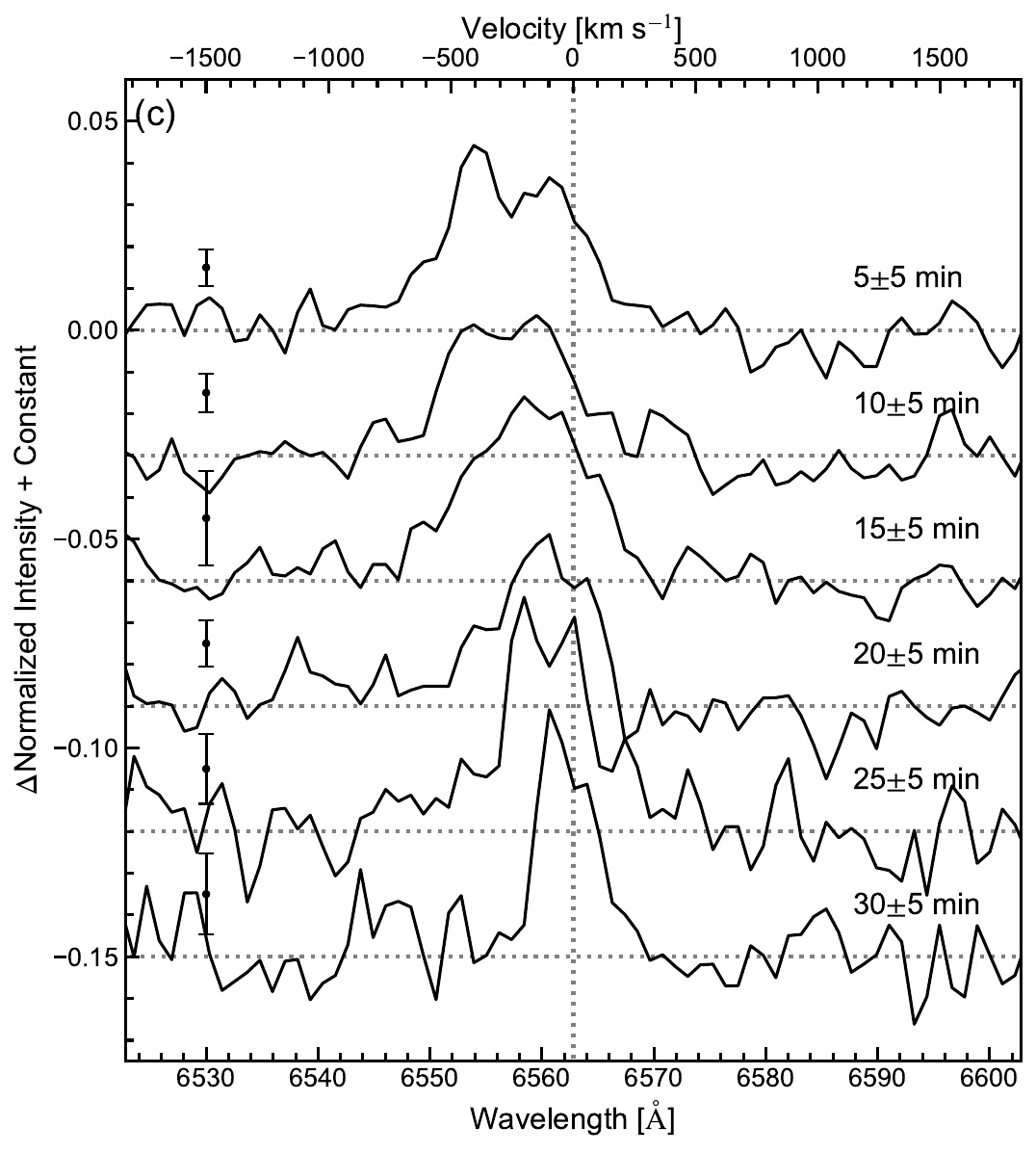}
\caption{Time evolution of pre-flare subtracted H$\alpha$ spectra of the superflare observed on 2022 April 10 (E1). 
(a) Reference light curves in H$\alpha$ $\Delta$EW (scaled on the left vertical axis) and white light  (scaled on the right vertical axis).
(b) A dynamic spectrum of the pre-flare subtracted H$\alpha$ spectrum, plotted on the time-wavelength plane. The red (blue) color represents emission (absorption) signals, with the upper part indicating shorter wavelengths (i.e., blueshift), following the unified format used in our series of papers \citep{2022NatAs...6..241N,2022ApJ...926L...5N}. 
The color bar shows the non-dimensional intensity which is normalize by the continuum level.
The dashed lines indicate the stellar surface gravity ($g_{\ast}$) and half of the surface gravity ($0.5\cdot g_{\ast}$).
(c) Time-binned, pre-flare subtracted H$\alpha$ spectra during the superflare. Detailed spectra can be seen in Figure \ref{fig:8}. The 1-$\sigma$ errors in continuum level is plotted as error bars for reference.
}
\label{fig:5}
\end{figure}

\begin{figure}
\epsscale{0.55}
\plotone{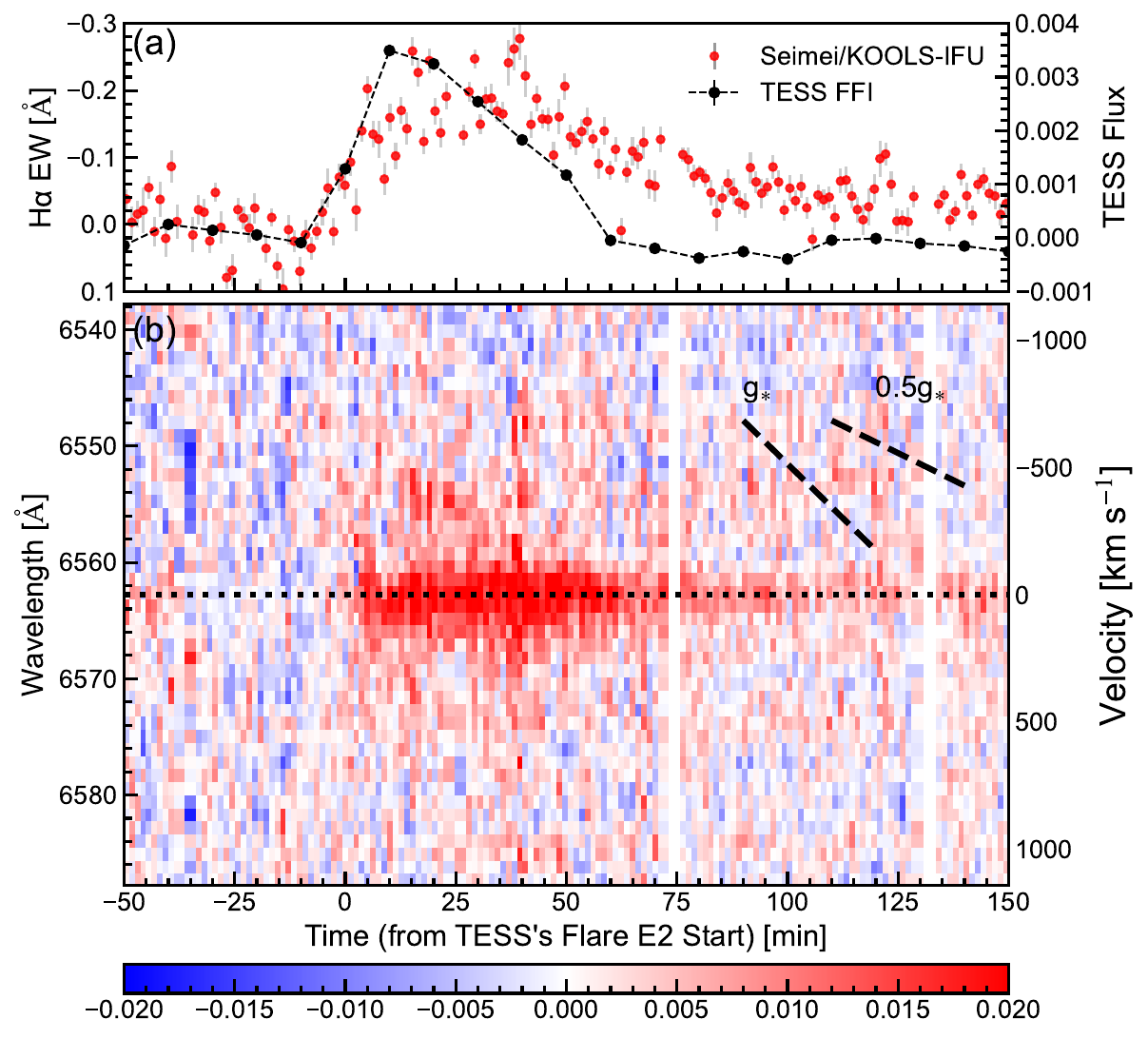}
\epsscale{0.45}
\plotone{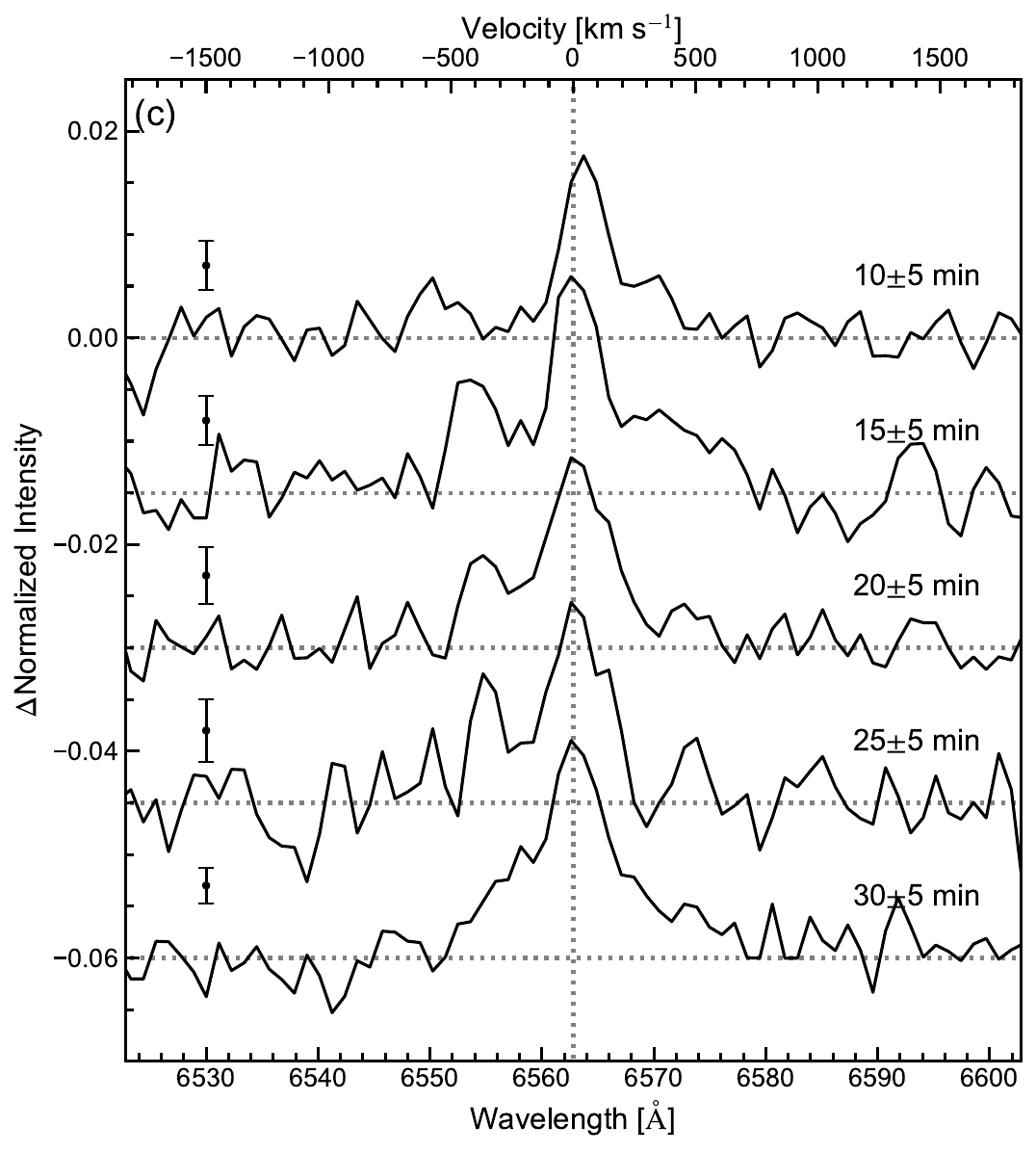}
\caption{The same as Figure \ref{fig:5}, but for the superflare on 2022 April 16 (E2).}
\label{fig:6}
\end{figure}

\begin{figure}
\epsscale{0.55}
\plotone{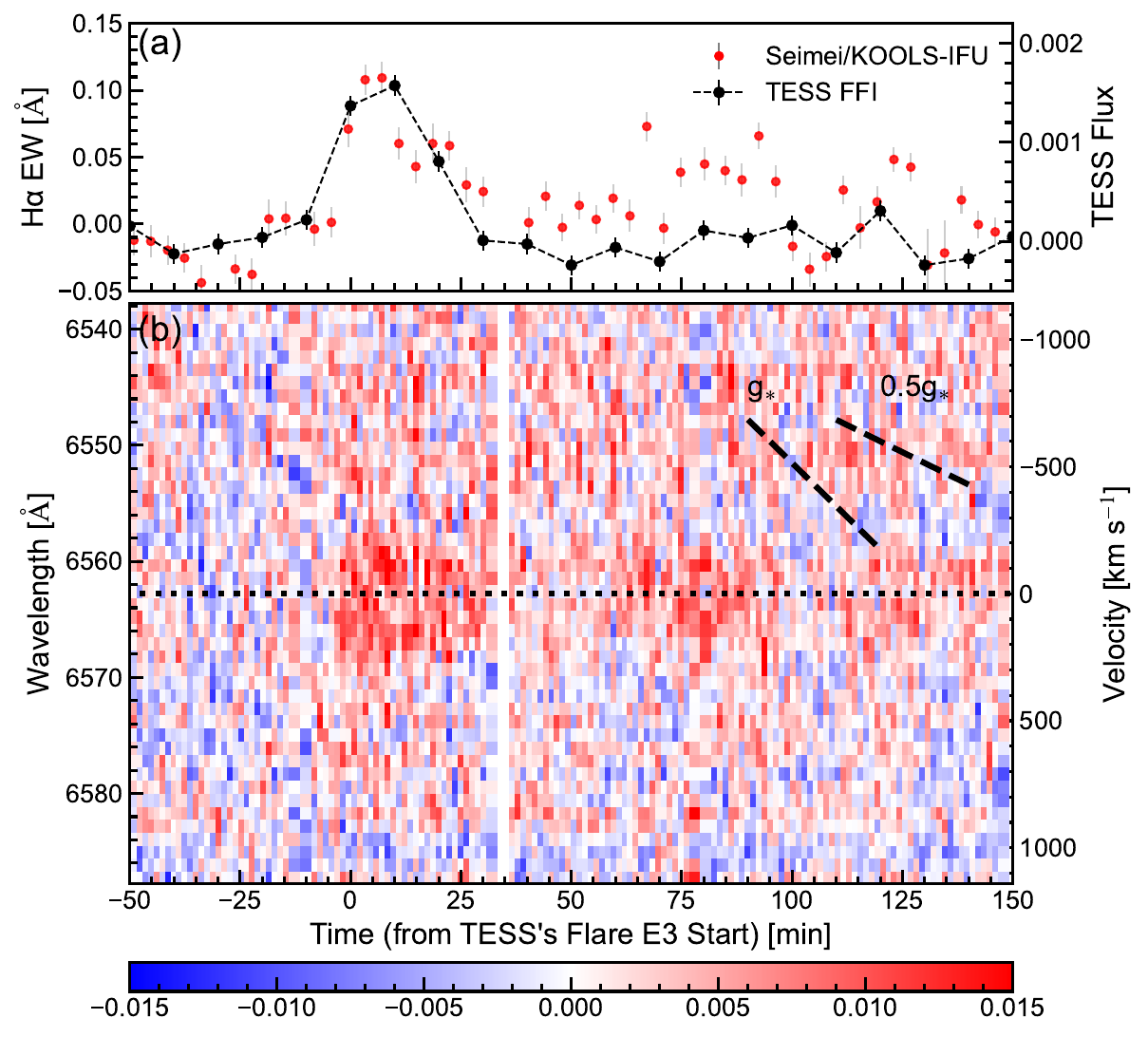}
\epsscale{0.45}
\plotone{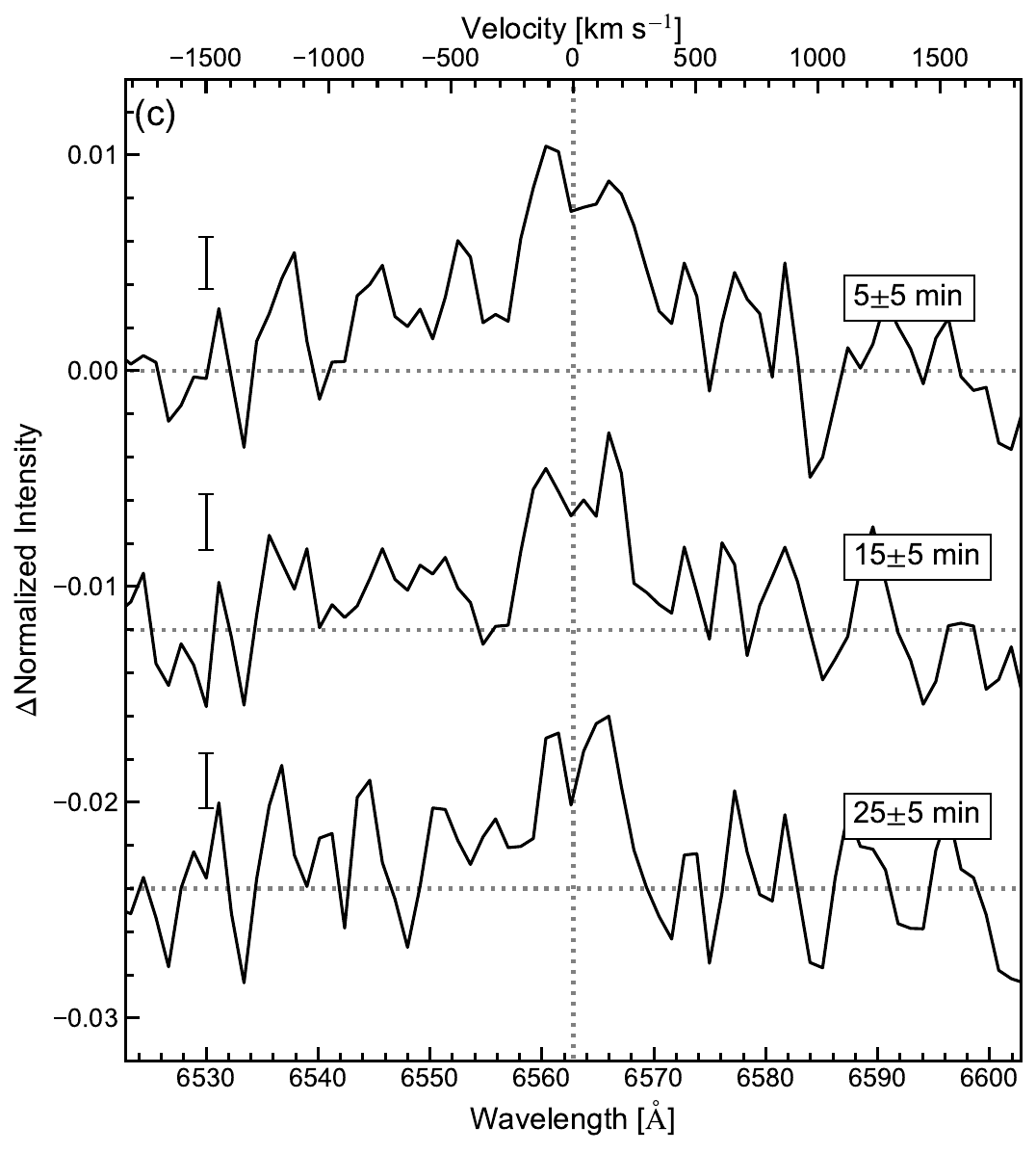}
\caption{The same as Figure \ref{fig:6}, but for the superflare on 2022 April 10 (E3).}
\label{fig:7}
\end{figure}

Figure \ref{fig:5}, \ref{fig:6}, and \ref{fig:7} shows pre-flare subtracted H$\alpha$ spectra of flare E1, E2, and E3. Each flare show different flare spectra as follows:
\begin{itemize}
\item[(E1)] The flare E1 exhibits significant blueshifted emission profiles during the flare. During the impulsive phase of WLF, it does not demonstrate strong emission in the H$\alpha$ line center and most of the radiation comes from the blueshifted component (see, Section \ref{sec:3:ha-3}). Following the noticeable blueshift profiles, the blueshift wavelength gradually reverts to the H$\alpha$ central wavelength, disappearing within approximately 34 min (see, Section \ref{sec:3:ha-2}).
\item[(E2)] The superflare E2 primarily manifests emission in the H$\alpha$ line center. However, the period spanning roughly 10 to 15 minutes from the TESS flare onset displays significant blueshifted emission profiles that exceed the 3$\sigma$ noise level of the far-wing continuum.
\item[(E3)] The superflare E3 exhibits the weakest H$\alpha$ EW enhancement and does not display any significant H$\alpha$ line asymmetry.
\end{itemize}

\subsubsection{Spectral Fitting}\label{sec:3:ha-2}

\begin{figure}
\epsscale{0.33}
\plotone{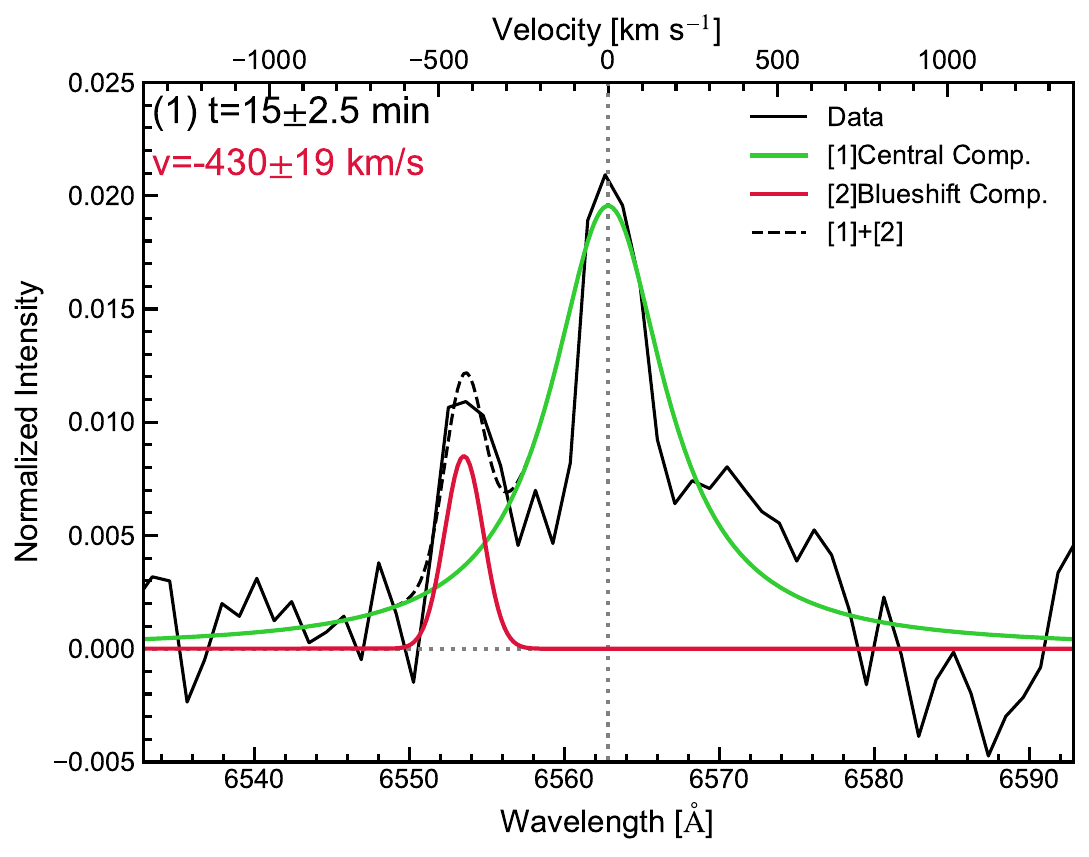}
\plotone{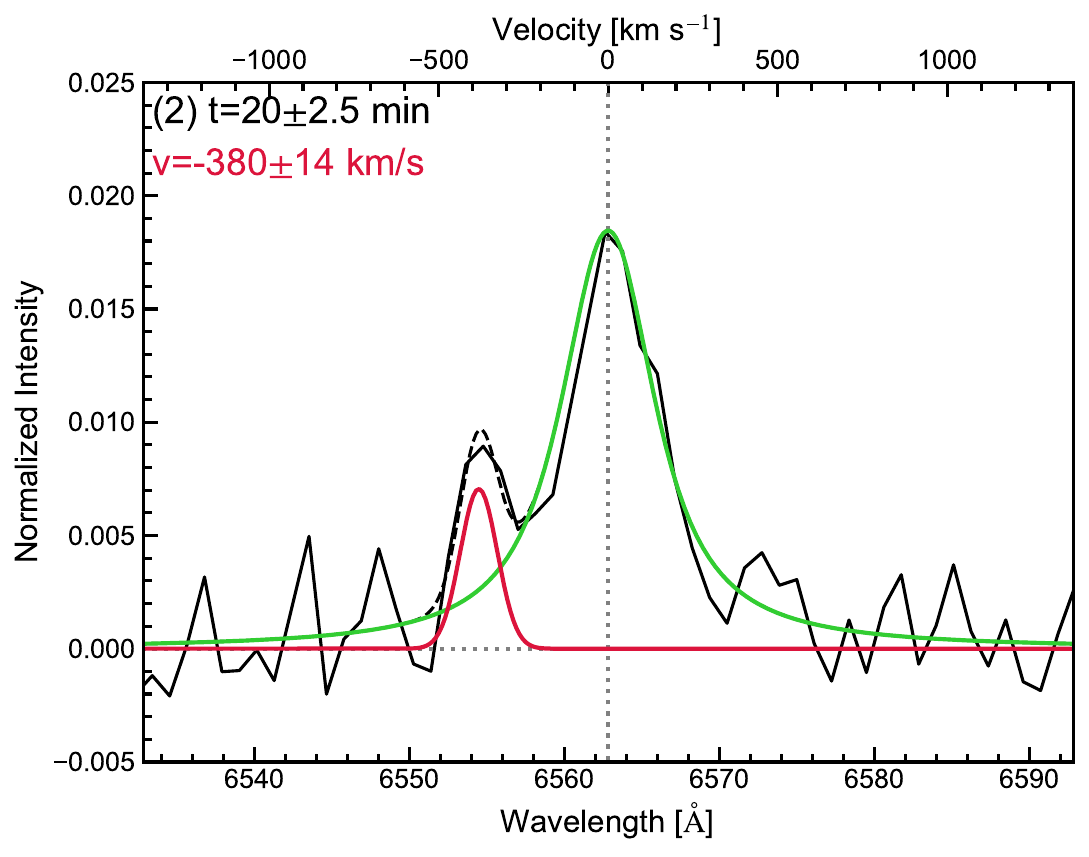}
\plotone{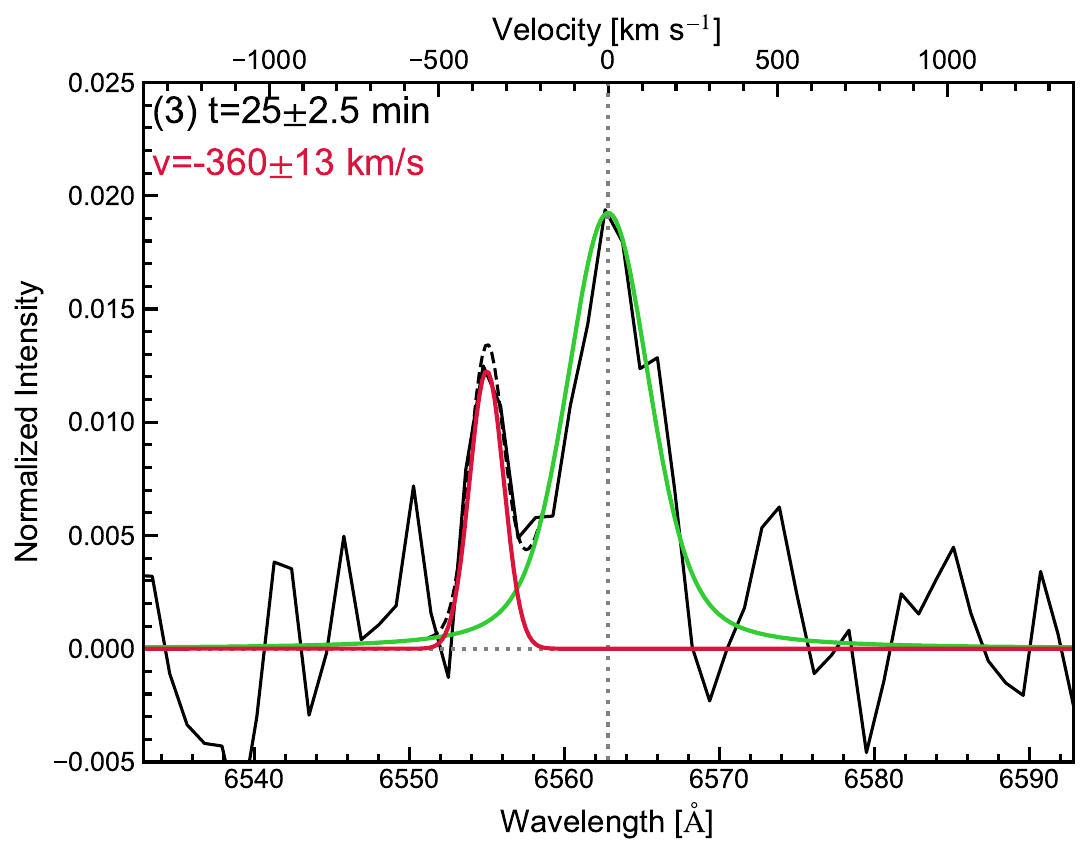}
\caption{Temporal evolution of pre-flare-subtracted spectra during superflare on 2022 April 16 (E2). 
To improve the signal-to-noise ratio for fitting, the data has been binned into 5-minute intervals. The magenta and green lines represent the results of fitting with a two-component Gaussian, and the dashed lines correspond to the sum of the two Gaussians. Please see Figure \ref{fig:6} for the other spectra of flare E2 without significant blueshifted components.}
\label{fig:9}
\end{figure}

In regard to flares E1 and E2, which demonstrated explicit blueshift phenomena, we performed a spectral fit using Gaussian functions to derive the velocity, velocity dispersion, and EW of the blueshifted component. 
In this process, we conducted two types of fits based on whether or not to isolate and fit the central flare component. 
The method (1): the first method simply fits the emission component with a single Gaussian function (referred to as ``one component" fit). 
The method (2): the second method first fitted the central component by using the spectra only the longer wavelength of H$\alpha$ line center ($>$6562.8 {\AA}) with a Gaussian function to obtain the central H$\alpha$ component. 
Following that, we subtracted this fit function from the original spectrum and then fitted the residual spectrum again with a Gaussian function, which we defined as the blueshifted component (referred to as ``two components" fit). 
In both methods, we evaluated the errors on the y-axis based on the data scattering of the far-wing continuum level and performed the fitting using \textsf{scipy.optimize.least\_squares}. 
The method (2) has also been employed in studies such as \cite{2020PASJ..tmp..253M} and \cite{2023ApJ...945...61N}. 
We use a Gaussian function which involves fewer parameters than a Voigt function because of the low signal-to-noise ratio (S/N) of our data. Figures \ref{fig:8} and \ref{fig:9} show the fitting results for the flare E1 and E2, respectively.
The following is the summary of the obtained results of fitting:

\begin{figure}
\epsscale{1.}
\plotone{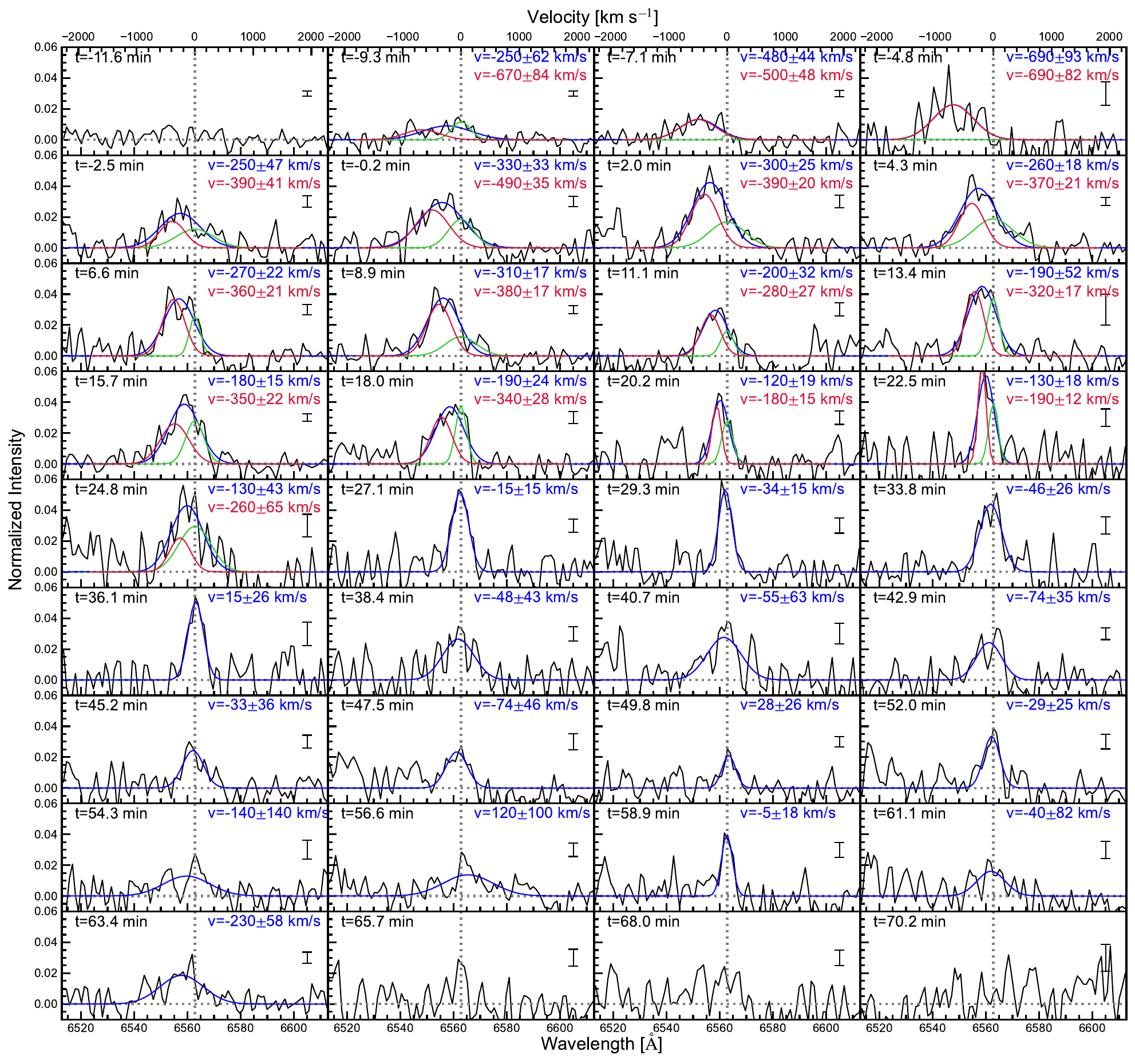}
\caption{Temporal evolution of pre-flare-subtracted spectra during superflare on 2022 April 10 (E1). 
Observations were carried out with an exposure time of 2 (or 1) minute(s) and a readout time of 17 seconds. The 1$\sigma$ continuum-level scattering in each spectrum is represented by error bars. The blue line corresponds to the results of fitting with a single-component Gaussian, while the red and green lines denote the results of fitting with a two-component Gaussian. The fitting process was conducted using \textsf{scipy.optimize.least\_squares}, taking into account the error bars.
}
\label{fig:8}
\end{figure}

\begin{itemize}
\item[(E1)] For flare E1, we utilized both methods 1 and 2. 
This is because the profile seems to be totally blueshifted in its initial phase.
Figure \ref{fig:8} displays the fitting results up to $t$=63 min, where method 1 fits successfully. Regarding method 2, we show the fitting results up to $t$=25 min, where the significant blueshifted component still remained after removing the central component.
The maximum blueshifted velocity $V_{max}$, defined as $|max\{V(t)\}|$, was 690$_{\pm 82}$ km s$^{-1}$ at t = --4.8 min for both methods. However, the period from t = --9.3 min to --4.8 min, where relatively high velocities are present, does not yet have a high signal compared to noise. 
We therefore sometimes describe the maximum velocity as $V_{max}$ = 330$_{\pm 35}$--690$_{\pm 92}$ km s$^{-1}$ (one component) and 490$_{\pm 33}$--690$_{\pm 93}$ km s$^{-1}$ (two component) to include a conservative estimate (where the conservative value is the maximum velocity after t $>$ --2.5 min).
\item[(E2)] For superflare E2, considering that the blueshifted component is clearly separated from the central component, we only applied method 2. The maximum velocity $V_{max}$ is 430$_{\pm 19}$ km s$^{-1}$
\end{itemize}
These values are summarized in Tables \ref{tab:prominence-1} and \ref{tab:prominence-2}.

\begin{deluxetable*}{lcccccccccc}
\label{tab:prominence-1}
\tablecaption{Properties of the prominence and filament eruptions on EK Dra.}
\tablewidth{0pt}
\tablehead{
\colhead{} & \colhead{H$\alpha$ Asymmetry} & $V_{\rm max}$ & $V_{\rm typical}$ & $V_{\rm disp,typ}$ & --d$V$/d$t$ & $L_{\rm H\alpha, blue}$ & $M_{\rm p}$ & $E_{\rm kin}$ &  Ref.  \\
\colhead{} & \colhead{} & \colhead{[km s$^{-1}$]}  & \colhead{[km s$^{-1}$]} &  \colhead{[km s$^{-1}$]} & \colhead{[km s$^{-2}$]} & \colhead{[$10^{28}$ erg s$^{-1}$]} &  \colhead{[10$^{18}$ g]}  &  \colhead{[10$^{32}$ erg]} &   
}
\startdata
2022 Apr 10 (E1) & Blue emission & 330-690 & 300 & 300 & 0.34$_{\pm 0.15}$ & 16.6$_{\pm 1.7}$ & $130^{+290}_{-90}$ & $580^{+1280}_{-400}$ & (1)  \\
 & (two comp.)$^\S$ & (490-690) & (390) & (230) & (0.23$_{\pm 0.14}$) & (9.8$_{\pm 1.3}$) & (--) & (--) &  \\
2022 Apr 16 (E2) & Blue emission & 430 & 380  & 55 & 0.12$_{\pm 0.20}$ & 0.47$_{\pm 0.16}$   & $1.7^{+3.7}_{-1.1}$ & $12^{+27}_{-8}$ & (1) \\
2022 Apr 17 (E3) & No & -- & --  & -- & -- & -- &  -- & --   & (1) \\
\hline
2020 Apr 05 (E4) & Blue absorption & 510 & 260 & 220 & 0.34$_{\pm 0.04}$ & -- & 1.1$^{+4.2}_{-0.9}$ & 3.5$^{+14.0}_{-3.0}$ & (2) \\
2020 Mar 14 (E5) & No (Red?) & -- & -- & -- & -- & -- & --   & -- & (3) \\
\enddata
\tablecomments{
$V_{\rm max}$ is the maximum blueshifted velocity ($max(-V)$).
$V_{\rm typical}$ is the typical blueshifted velocity around when the EW takes a maximum value.
$V_{\rm typical}$ is the typical velocity dispersion ($\sigma$) in the fitted Gaussian function.
--d$V$/d$t$ is the deceleration of the blueshifted component. EK Dra's surface gravity is 0.30$_{\pm 0.05}$ km s$^{-2}$.
$L_{\rm H\alpha, blue}$ is the luminosity of the blueshifted component.
$M_{\rm p}$ is the mass of the erupted prominences/filament.
$E_{\rm kin}$ is the kinetic energy of the prominence/filament eruptions.
$^\S$The values in parentheses ``()"  are the values of the two component fitting of the flare E1.
References: (1) This study, (2) \cite{2022NatAs...6..241N}, (3) \cite{2022ApJ...926L...5N}.
}
\end{deluxetable*}

\begin{deluxetable*}{lcccccccccccc}
\label{tab:prominence-2}
\tablecaption{Properties of the prominence and filament eruptions (continued from Table \ref{tab:prominence-1}) and the possible X-ray dimming.}
\tablewidth{0pt}
\tablehead{
\colhead{} & \colhead{$t_{\rm WLF}$} & \colhead{$t_{\rm blue}$} & \colhead{ Timing$_{\rm Blue Asym.}$ } & \colhead{X-ray dimming} & \colhead{ Timing$_{\rm X.dim.}$ } & \colhead{$t_{\rm X.dim.}$} & \colhead{amp$_{\rm X.dim.}$} & \colhead{$\Delta EM_{\rm X.dim.}$} & \colhead{$L_{\rm X.dim.}$}  \\
\colhead{} & \colhead{[min]} & \colhead{[min]} & \colhead{} & \colhead{} & \colhead{} &  \colhead{[min]} & \colhead{[\%]} & \colhead{[10$^{51}$ cm$^{-3}$]} & \colhead{[10$^{10}$ cm]}
}
\startdata
(E1) & 20 & 34$_{\pm 2}$ & $\sim$start of WLF & Yes?(block5,6) & 133 min after WLF & $>$91 & 5.8-8.5 &  (--)  & (--)  \\
 & (--) & (--) & (--) & (bkg subtracted) & (--) & (--) & 8.8-14.7 &  7.2--12.7  & 0.90--5.0  \\
(E2) & 40 & 10$_{\pm 5}$ & $\sim$peak of WLF &  No & -- &   --  & -- &   --  & -- \\
(E3) & 30 & -- & -- & No  & -- &  --  & --  &   --  & -- \\
\hline
(E4) & 6.0 & 58-86 & $\sim$end of WLF & (No obs.) &  -- & --  & -- &   --  & -- \\
(E5) & 130 & -- & -- & (No obs.) & -- &  --  & -- &   --  & -- \\
\enddata
\tablecomments{$t_{\rm WLF}$ is the FWHM duration of WLFs as a reference. 
$t_{\rm blue}$ is the duration of blueshifted components. For E1, the duration from t = $-$7.1 min to t = 25.0 min. For E2, the duration from t = 15$\pm$2.5 min to t = 25$\pm$2.5 min. For E3, the lower limit is the duration where the velocity $<$ 0 km s$^{-1}$ and the upper limit is the duration where the velocity is $<$ --100 km s$^{-1}$.
``Timing$_{\rm Blue Asym.}$" is the timing when the blue shift components appear relative to WLFs.
``Timing$_{\rm X.dim.}$" is the timing when the possible X-ray dimming appears relative to WLFs.
$t_{\rm X.dim.}$ is the duration of the possible X-ray dimming.
${\rm amp}_{\rm X.dim.}$ is the X-ray dimming amplitude for the X-ray count rates (0.5--3 keV). 
The value in parentheses ``()"  is the dimming amplitude for the background-subtracted X-ray count rates.
$\Delta EM_{\rm X.dim.}$ is the change of emission measure due to the possible X-ray dimming.
$L_{\rm X.dim.}$ is the length scale of the possible X-ray dimming region.
}
\end{deluxetable*}

\begin{figure}
\epsscale{0.33}
\plotone{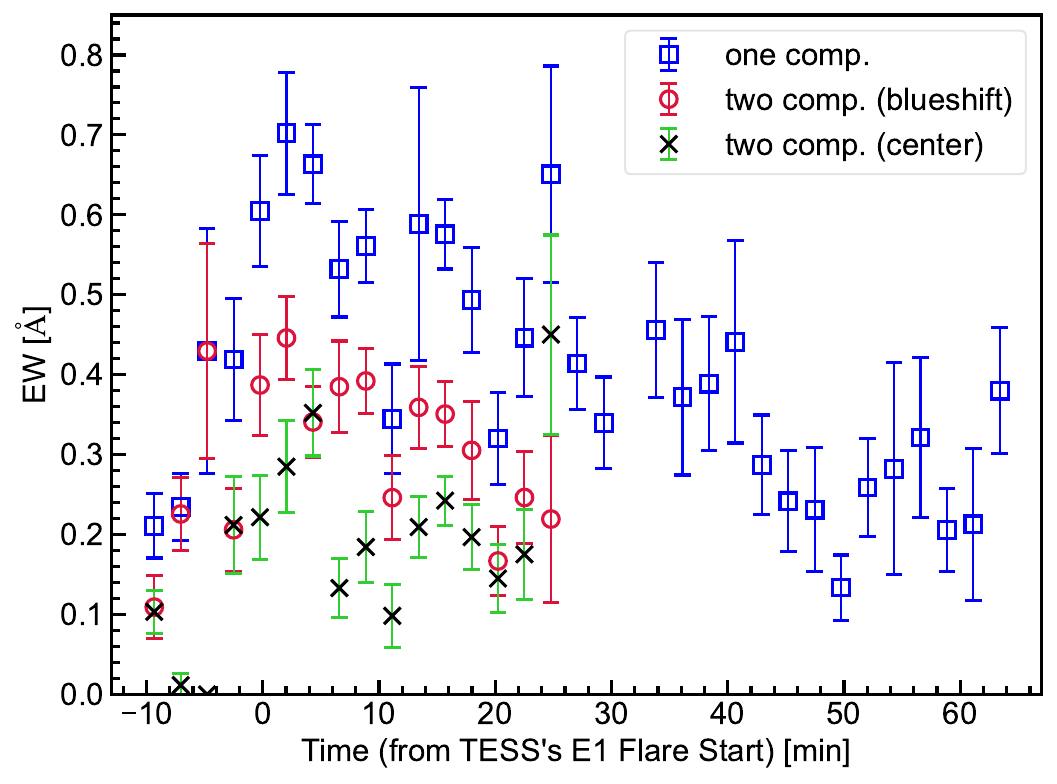}
\plotone{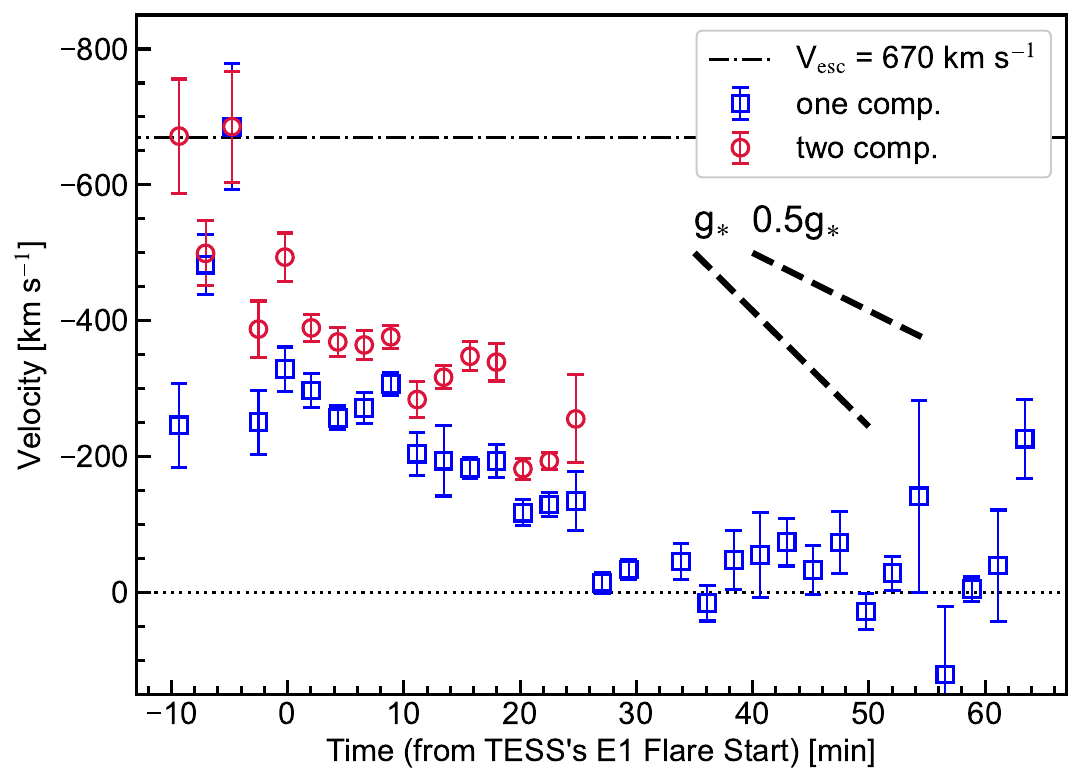}
\plotone{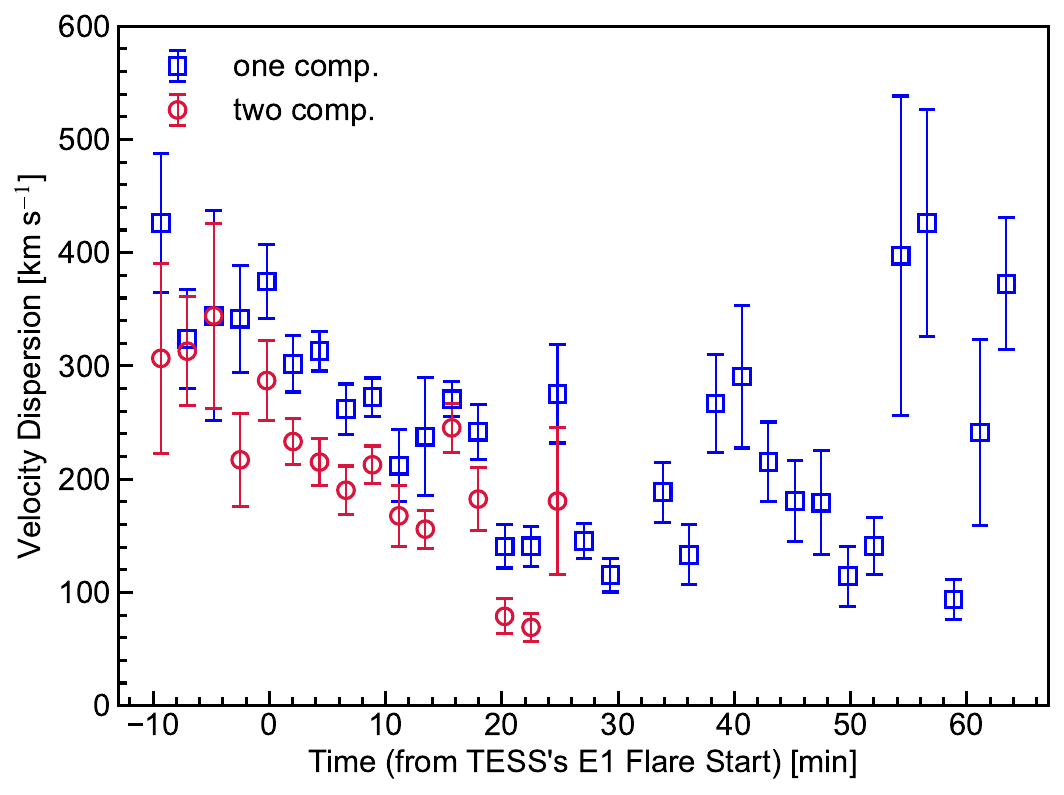}
\plotone{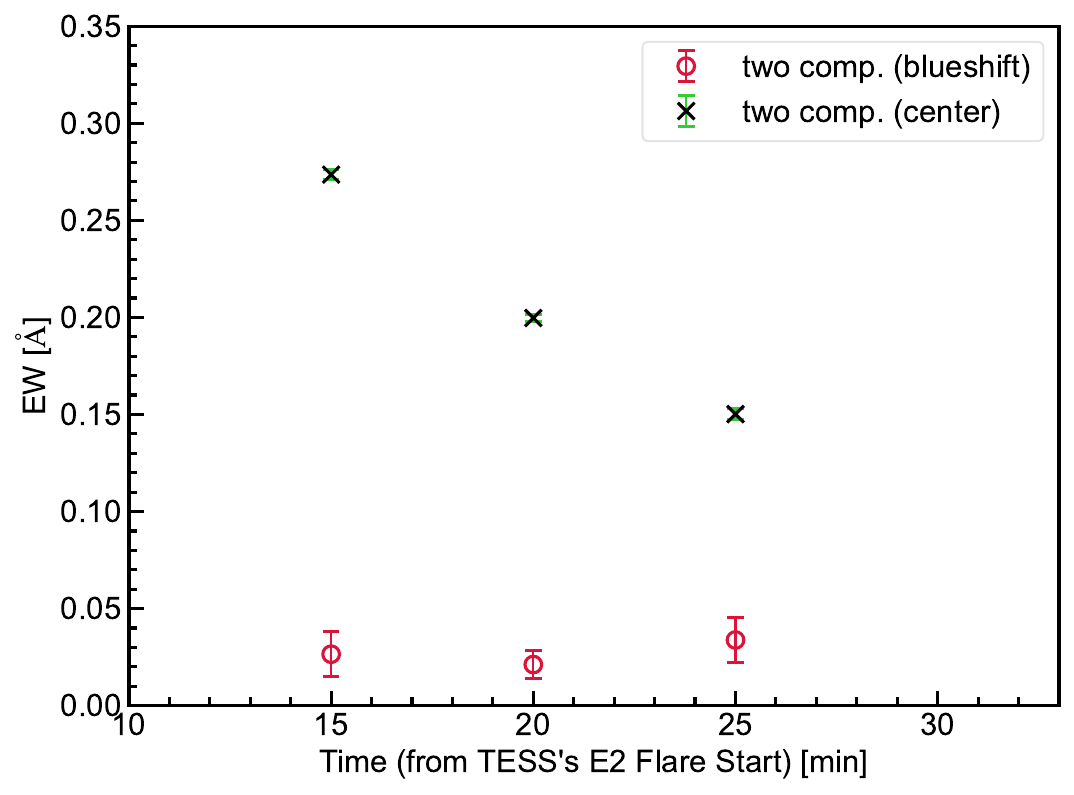}
\plotone{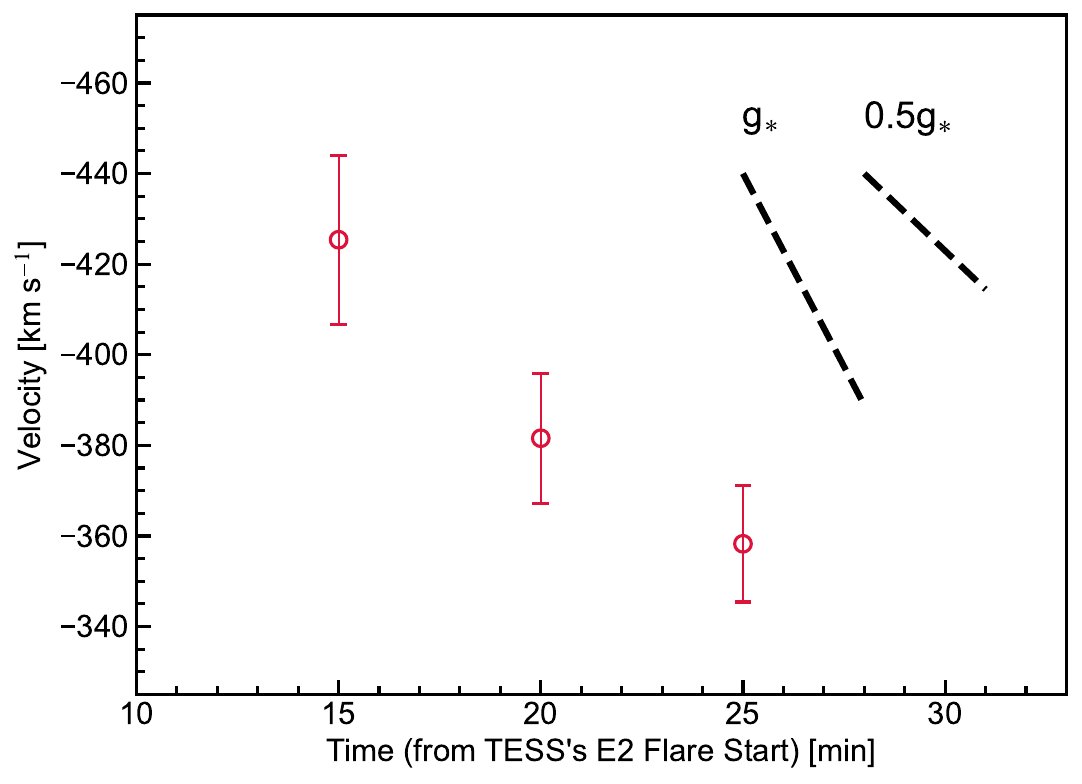}
\plotone{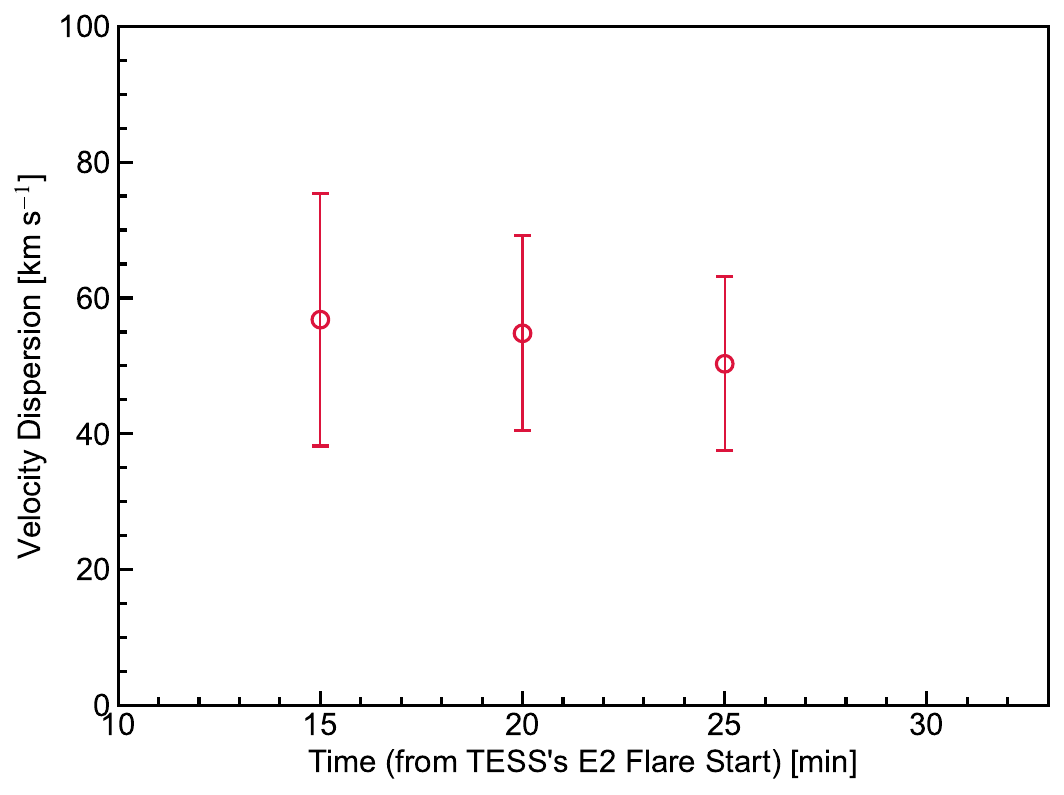}
\caption{Temporal evolution of the blueshifted emission line components for (Upper) flare E1 on 2022 April 10 and (Lower) superlare E2 on 2022 April 16. 
(Left) Temporal variation in EW (note that positive values mean emissions), (Middle) Temporal variation in velocity, (Right) Temporal variation in velocity dispersion (Gaussian's standard deviation $\sigma$). The blue squares represent the result of fitting with a one-component Gaussian, while the red and green lines represent the blueshifted component and the central component, respectively, from the two-component Gaussian fit. 
}
\label{fig:10}
\end{figure}

Figure \ref{fig:10} shows the temporal evolution of the EW, velocity, and velocity dispersion of the fitted parameters for the superflares E1 and E2. 
For the flare E1, both results for one component and two component fitting are plotted.
The flare E1 shows clear temporal decrease in the blueshifted EW, velocity, and velocity dispersion for both fitting results.
The superflare E2 shows a weak decreasing trend of the blueshifted velocity, but no clear evolution for blueshifted EW and velocity dispersion.

\begin{figure}
\epsscale{0.5}
\plotone{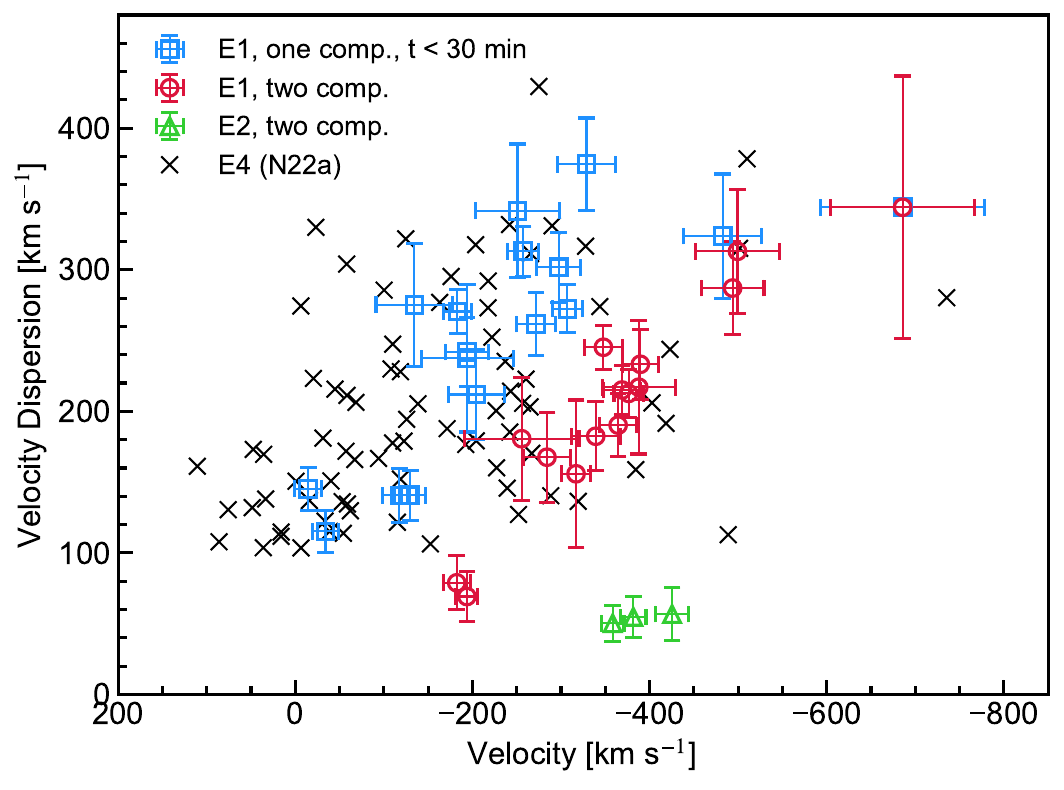}
\caption{Relationship between the velocity and velocity dispersion of the blueshifted component. The data for E1 and E2 are from Figure \ref{fig:10}. The filament eruption (blueshifted absorptions) on EK Dra on 2020 April 5 (labeled as ``E4") was taken from \cite{2022NatAs...6..241N} (labeled as ``N22a"). 
}
\label{fig:11}
\end{figure}

Figure \ref{fig:11} shows the relation between velocity and velocity dispersion for blueshifted velocity of the flare E1 and E2. 
As a reference, the filament eruption event on EK Dra reported by \cite{2022NatAs...6..241N} (``E4" events) is added in Figure \ref{fig:11}.
The superflare E2 and E4 show clear positive correlation between these two parameters.

Based on the temporal evolution of the blueshifted velocity, some basic values of blueshifted components are estimated and summarized in Tables \ref{tab:prominence-1} and \ref{tab:prominence-2}. 
We estimated the typical blueshifted velocity $V_{\rm typical}$ and velocity dispersion $V_{\rm disp,typ}$ as follows: 
For flare E1, the values are estimated when the EWs of the blueshifted component are most prominent, while for flare E2, the values are estimated as a median value. 
The deceleration of the blueshifted components $-$d$V$/d$t$ (= $-$A/$\tau$) are obtained by fitting the time evolution of velocity with an exponential function = $A e^{-(t-t_0)/\tau}$. 
Also, the duration and timing of the apperance of the blueshifted components are summarized in Table \ref{tab:prominence-2}.

\subsubsection{Estimation of Energy and Timescale}\label{sec:3:ha-3}

The H$\alpha$ radiated energy is estimated from the light curve of $\Delta$EW whose emission is larger than the one fifths of the peak values.
It is calculated by multiplying the enhanced H$\alpha$ $\Delta$EW by the continuum flux and integrating in time.
The continuum flux of EK Dra around H$\alpha$ is derived as 1.57 W m$^{-2}$ nm$^{-1}$ at 1 AU \citep{2022NatAs...6..241N} from flux-calibrated EK Dra's spectrum and the stellar distance given by Gaia Data Release 2 \citep{2018A&A...616A...2L}.
As a result, the H$\alpha$ flare radiation energy were estimated as 4.9$_{\pm 0.2}\times 10^{32}$ erg, 1.8$_{\pm 0.1}\times 10^{32}$ erg, and 2.9$_{\pm 0.3}\times 10^{31}$ erg for flare E1, E2, and E3, respectively (Table \ref{tab:flare-basic}).

Note that the above radiation energies of the flares E1 and E2 include blueshifted components.
For the flare E1, the radiation energy of the possible central component is estimated as 9.1$\times 10^{31}$ erg, which is only 19 \% of the total radiation energy, meaning that most of the radiations come from blueshifted component.
Unlike the flare E1, for the flare E2, the radiation energy of the possible central component is estimated as 1.7$\times 10^{31}$ erg, which is 97 \% of the total radiation energy, meaning that most of the radiations come from the central component.

The duration of the H$\alpha$ flares are estimated as the one at FWHM of the smoothed flaring light curves of H$\alpha$ $\Delta$EW, summarized in Table \ref{tab:flare-basic}.
Note that the most of the radiation of the flare E1 comes from blueshifted component, so the duration of the flaring emission from footpoints or flare loops would be different from the obtained here.


\subsection{X-ray Flare Analysis}\label{sec:3:x}

\subsubsection{Spectral Fitting for X-ray Spectra}

\begin{figure}
\epsscale{0.33}
\plotone{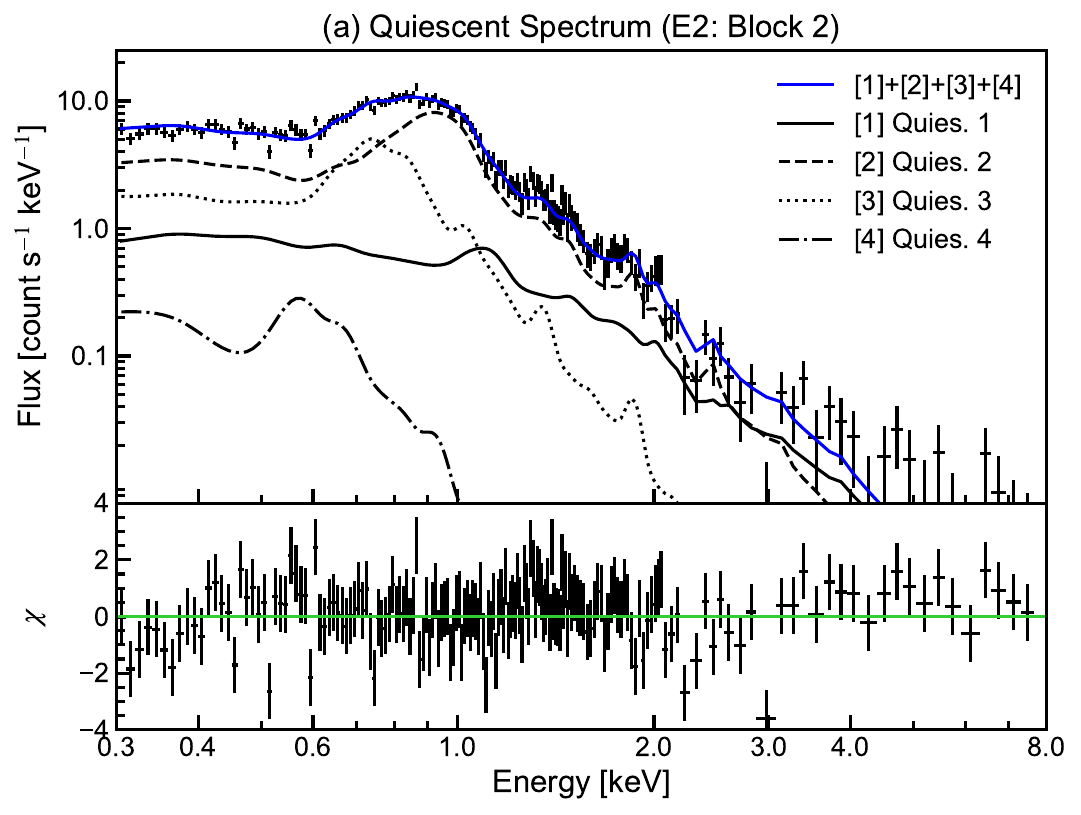}
\plotone{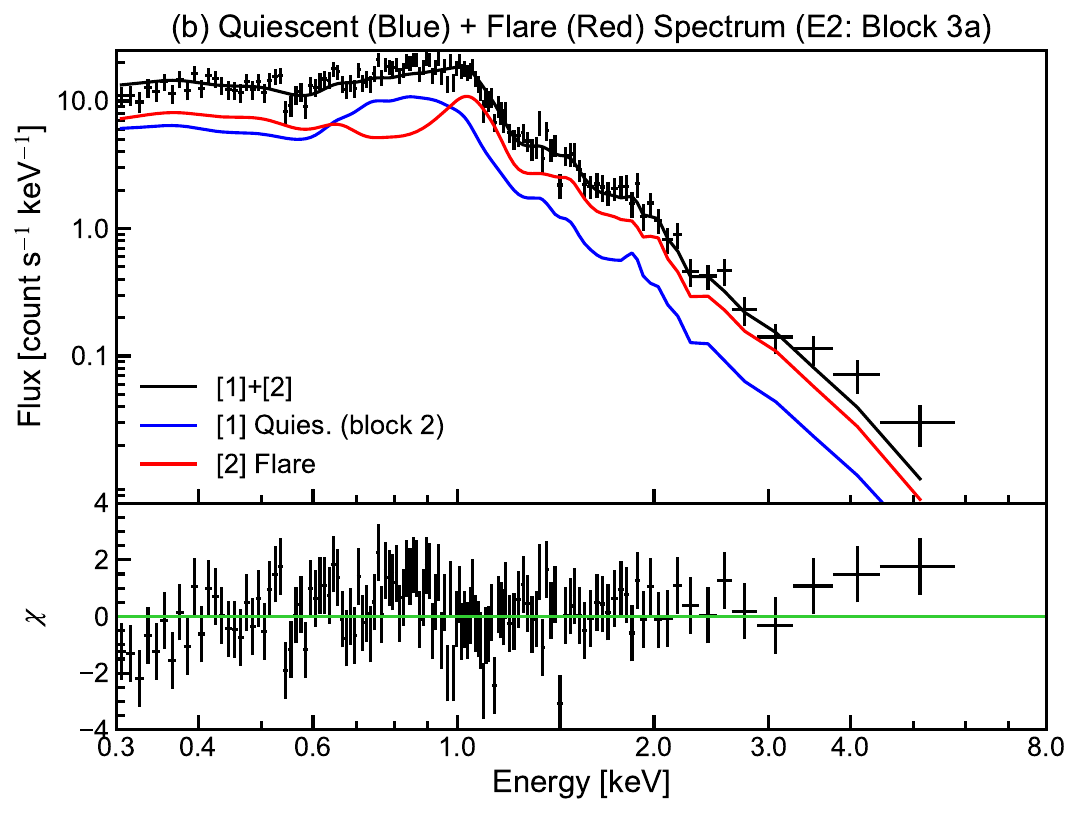}
\plotone{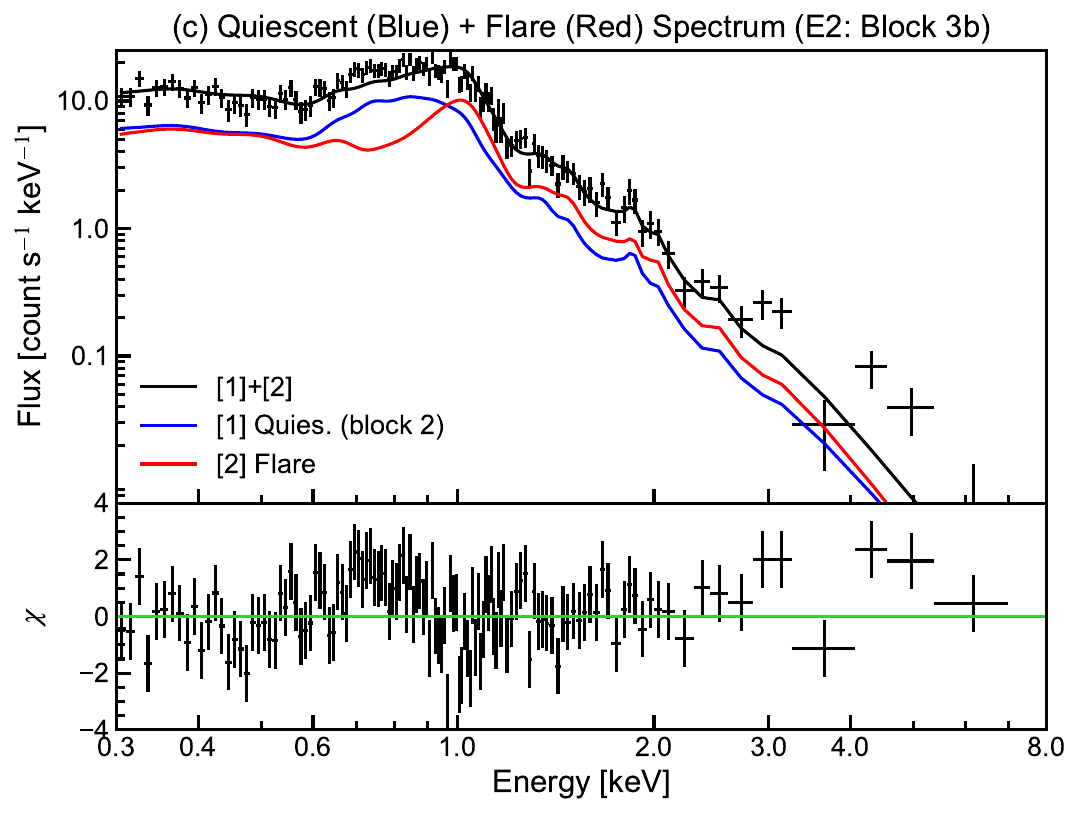}
\plotone{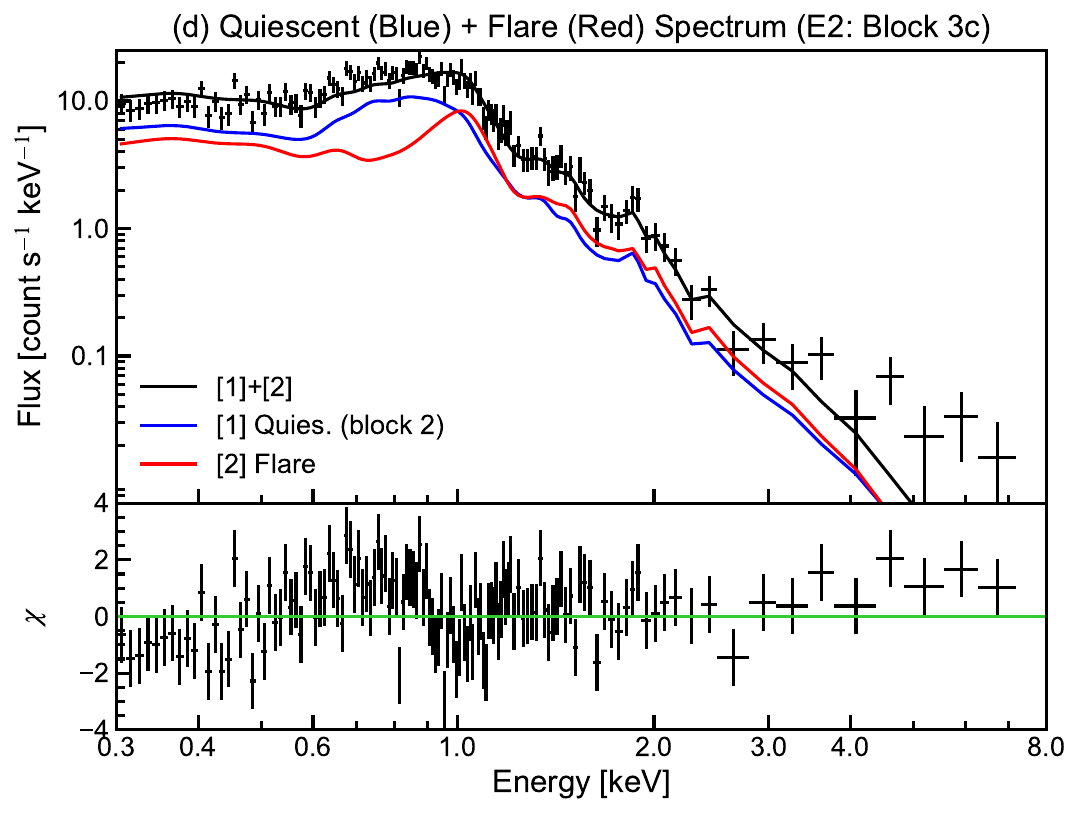}
\plotone{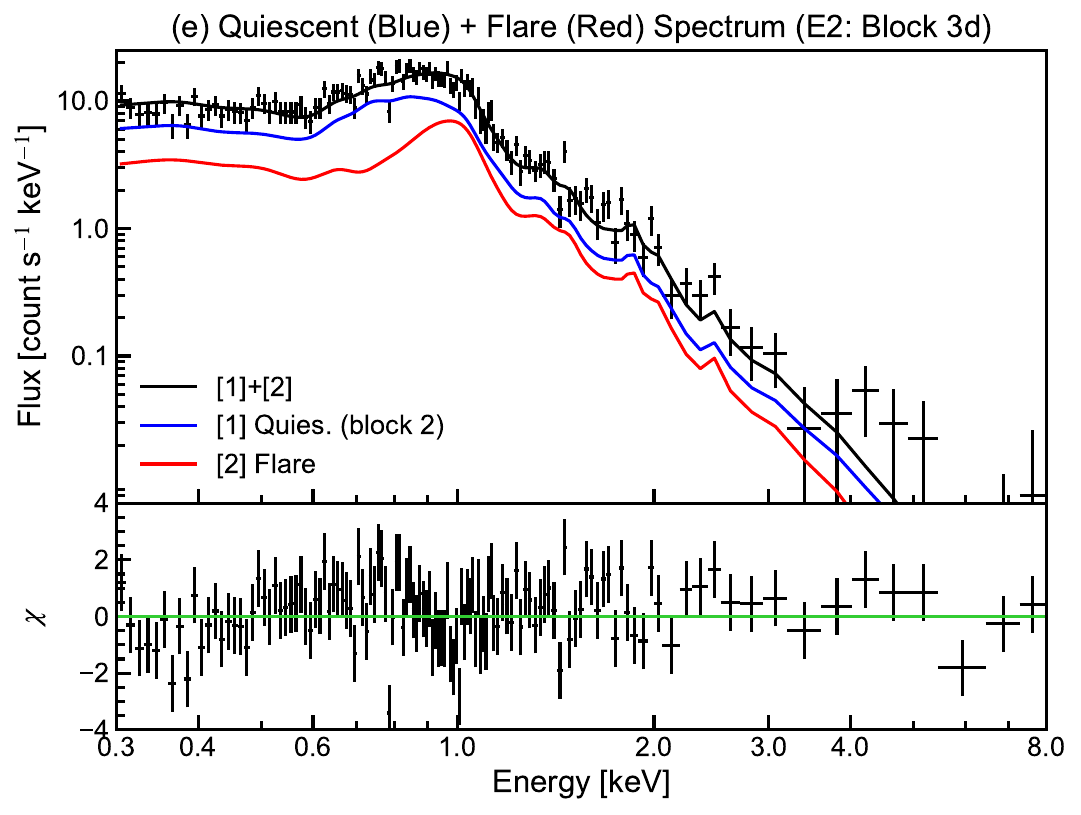}
\plotone{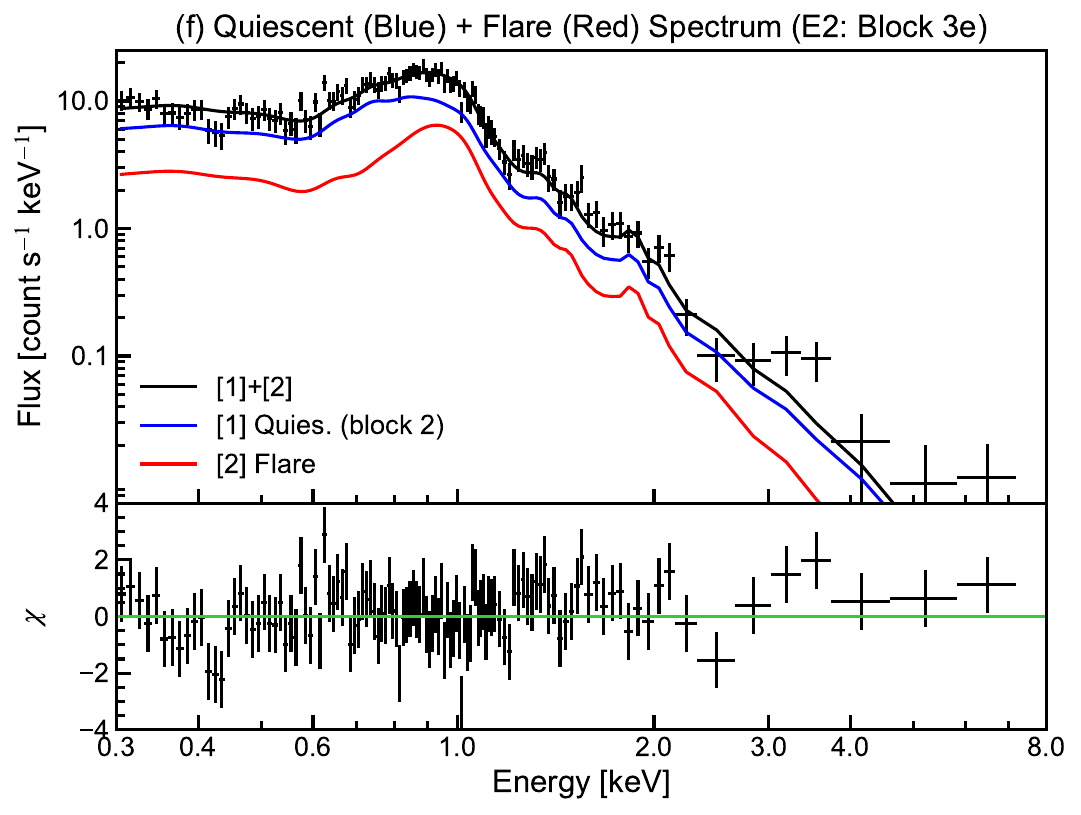}
\caption{NICER X-ray spectrum of the superflare on 2022 April 16 (E2) and the corresponding fitting results (applied to the range of 0.3--8.0 keV).
(a) Fitting results for the pre-flare spectrum of block 2 of E2, fitted with four \textsf{apec} components (each represented by a black line). The blue line represents the total model spectrum, which is used as the pre-flare spectrum in panels (b) through (f).
(b-f) Observed and model spectra when flare block 3 is divided into blocks 3a through 3e.
Refer to Figure \ref{fig:14} for the division times and Tables \ref{tab:2} and \ref{tab:3} for the fitting parameters.}
\label{fig:12}
\end{figure}

\begin{figure}
\epsscale{0.33}
\plotone{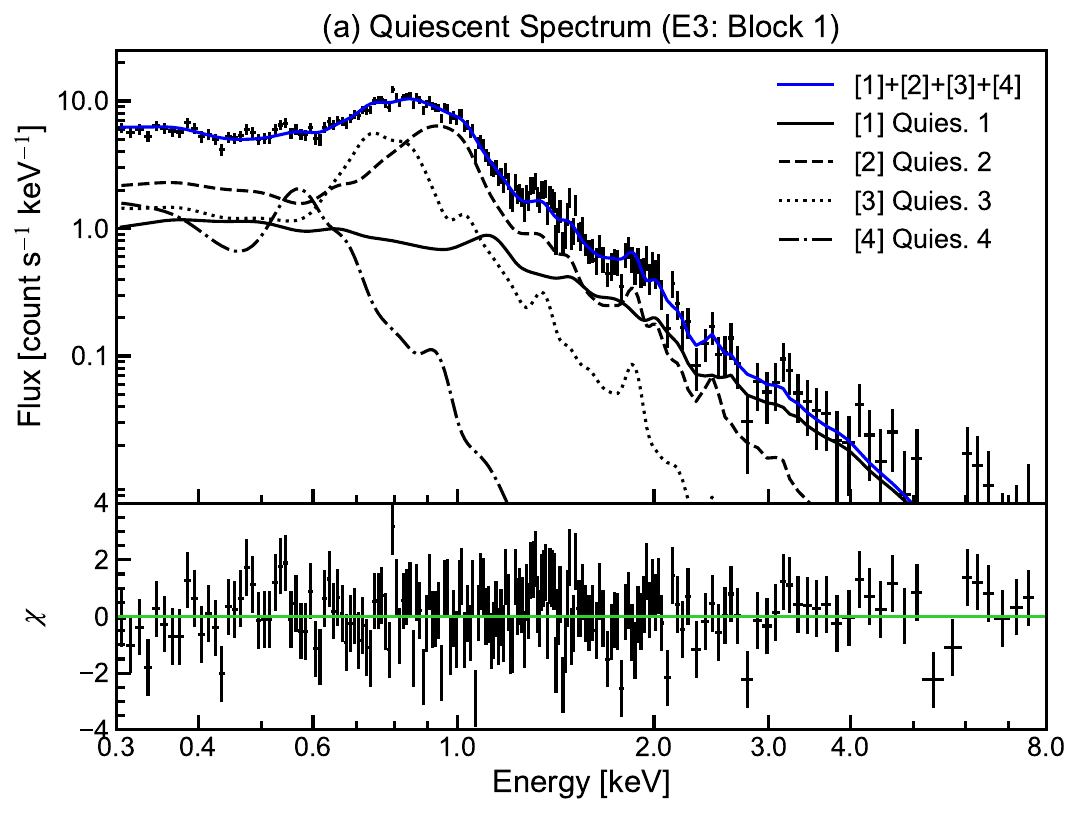}
\plotone{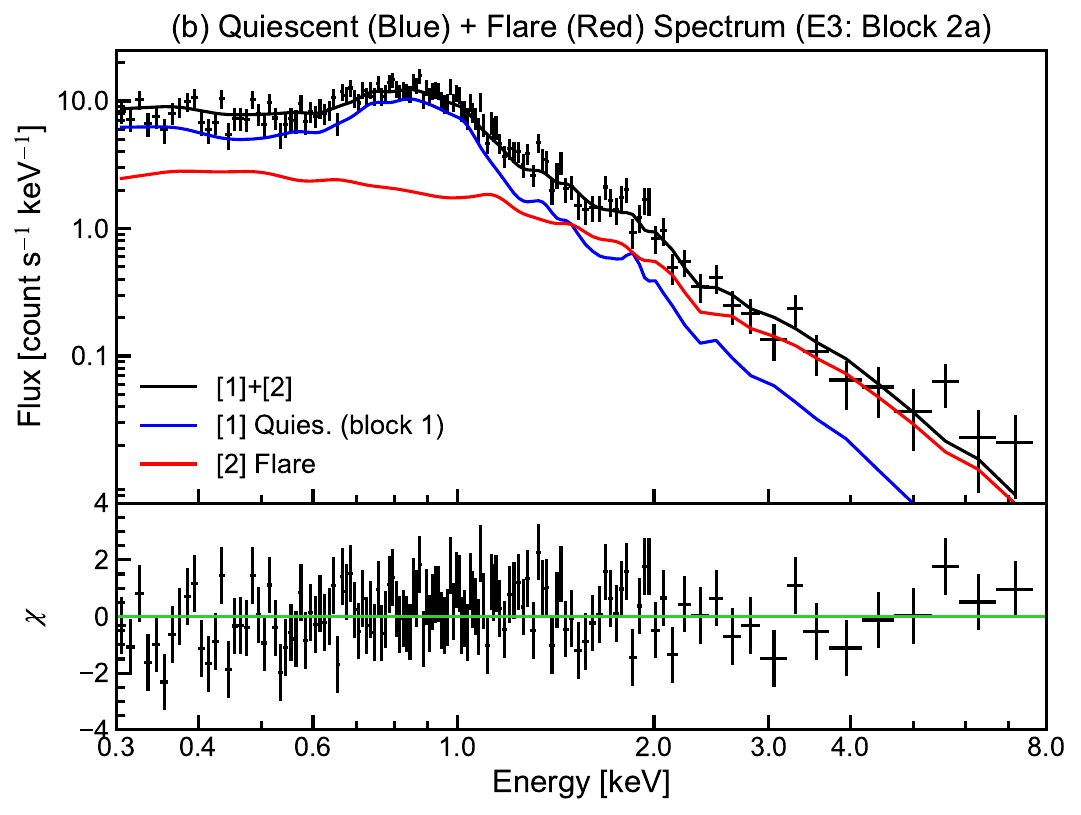}
\plotone{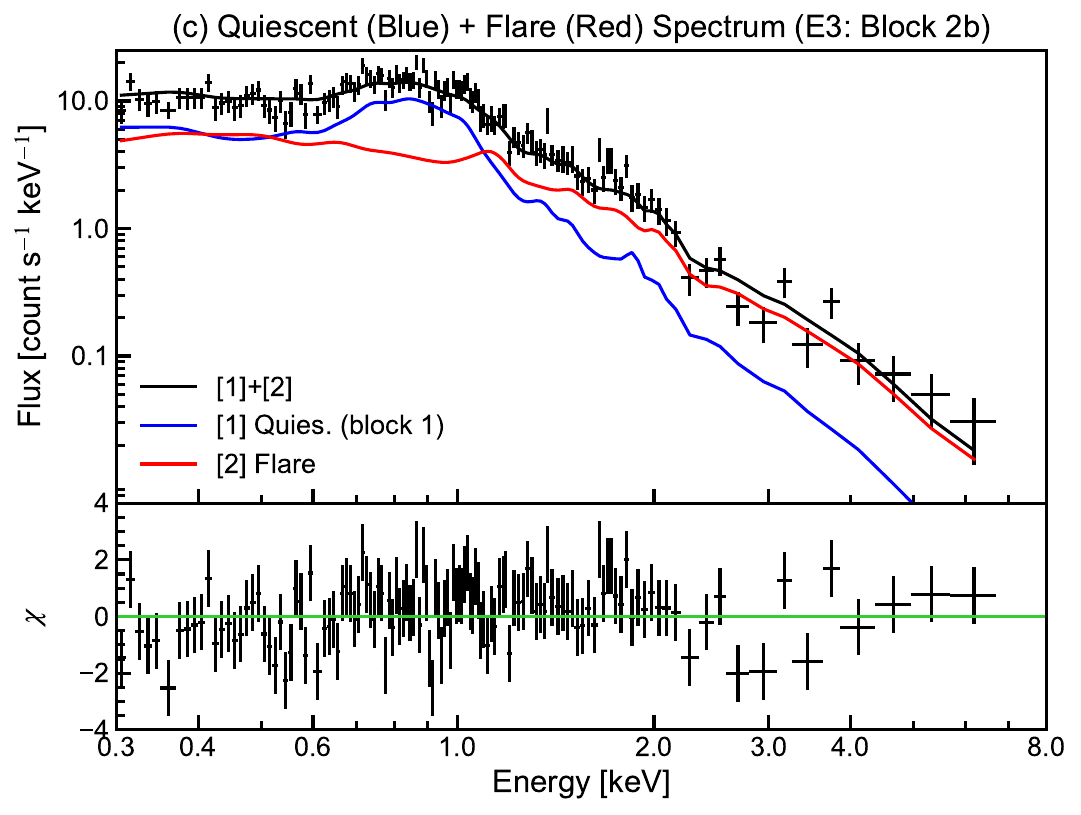}
\plotone{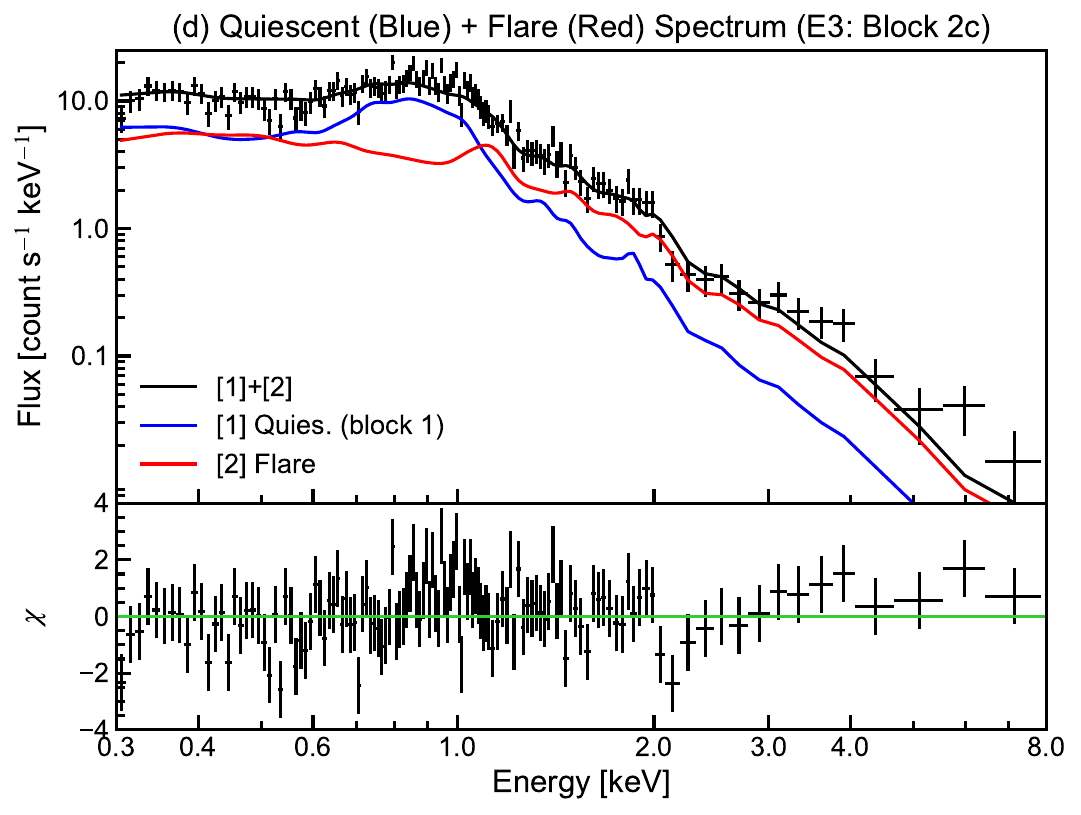}
\plotone{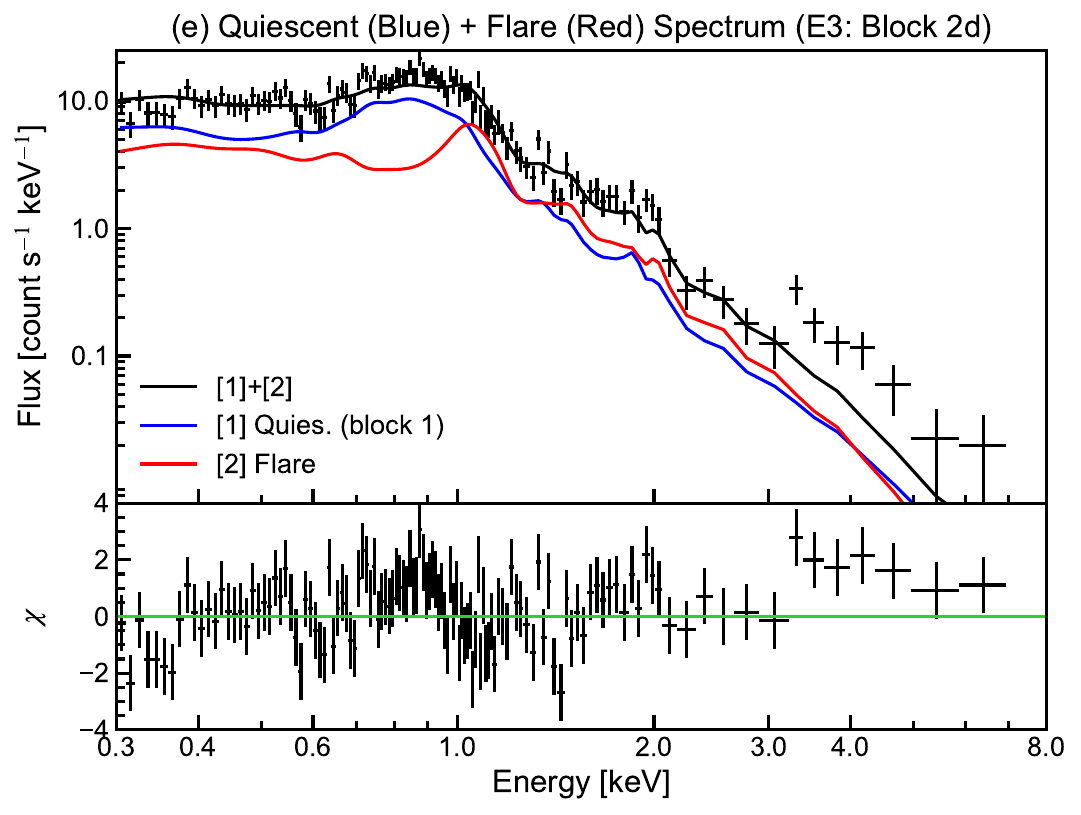}
\plotone{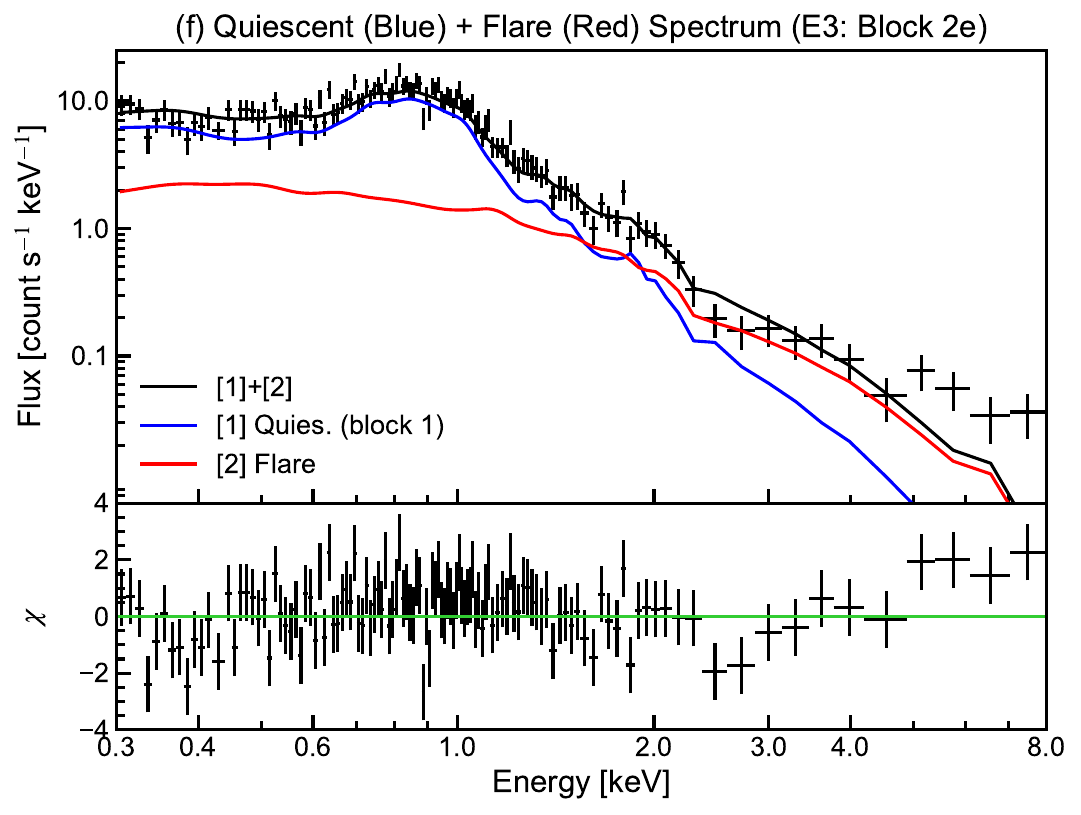}
\caption{The same as Figure \ref{fig:12}, but for the superflare on 2022 April 17 (E3).}
\label{fig:13}
\end{figure}

For the flares E2 and E3, an increase in intensity was detected in NICER's X-ray data, simultaneously with TESS WL and H$\alpha$ flares. 
We performed a spectral analysis on the increased intensity during the flares using an emission spectrum from collisionally-ionized diffuse gas calculated from the AtomDB atomic database (\textsf{apec})\footnote{\url{https://heasarc.gsfc.nasa.gov/xanadu/xspec/manual/node134.html}} and The Tuebingen-Boulder ISM absorption model (\textsf{TBAbs})\footnote{\url{https://heasarc.gsfc.nasa.gov/xanadu/xspec/manual/node268.html}} in \textsf{xspec}. 
First, we defined the data from the ISS orbit prior to the flares as ``pre-flare orbits" (``block 2" for flare E2 in Figure \ref{fig:3}(c), ``block 3" for flare E3 in Figure \ref{fig:4}(c)) and carried out a spectral fit of the pre-flare spectra. 
The objective is to remove the pre-flare component from the flare component, thus we utilized as many as four \textsf{apec} models and \textsf{TBAbs}.
Three \textsf{apec} models did not adequately reproduce the overall spectral shape.
Only the ISM hydrogen column density $N_{\rm H}$ is fixed as 3$\times10^{18}$ cm$^{-2}$ based on the previous study \citep{1995A&A...301..201G} since it is too small to consistently determine for each frame. 
\cite{2022AJ....164..106L} reported a similar $N_{\rm H}$ value of $1.7\times 10^{18}$ cm$^{-2}$.
In each case, the $N_{\rm H}$ value is so small that the factor 2 difference does not significantly affect the fitting results.
The fitted spectra are presented in Figures \ref{fig:12}(a) and \ref{fig:13}(a), while the parameters are summarized in Table \ref{tab:2}. The error bars are obtained as a result of the \textsf{xspec} fitting.

\begin{deluxetable*}{ccccccccc}
\label{tab:2}
\tabletypesize{\footnotesize}
\tablecaption{Best-fit values of the pre-flare NICER X-ray spectra for E2 and E3.}
\tablewidth{0pt}
\tablehead{
\colhead{Component} & \multicolumn{4}{c}{Flare E2 (block 2)} & \multicolumn{4}{c}{Flare E3 (block 1)} \\
 & T [keV] & (T [$10^{7}$ K]) & EM [10$^{52}$ cm$^{-3}$] & Abund. & T [keV] &  (T [$10^{7}$ K])  & EM [10$^{52}$ cm$^{-3}$]  & Abund.
}
\startdata
1 & 2.06$_{\pm 0.53}$ & (2.39$_{\pm 0.62}$) & 1.58$_{\pm 0.52}$ & 0.28$_{\pm 0.04}$ & 2.77$_{\pm 0.62}$ & (3.21$_{\pm 0.71}$) & 2.12$_{\pm 0.38}$ & 0.38$_{\pm 0.06}$ \\
2 & 0.94$_{\pm 0.02}$ & (1.09$_{\pm 0.03}$) & 5.28$_{\pm 0.74}$ & --  & 0.96$_{\pm 0.04}$ & (1.11$_{\pm 0.05}$) & 3.30$_{\pm 0.42}$&  -- \\
3 & 0.40$_{\pm 0.08}$ & (0.47$_{\pm 0.09}$) & 2.63$_{\pm 0.77}$ & --  & 0.50$_{\pm 0.09}$ & (0.58$_{\pm 0.10}$) & 1.94$_{\pm 0.33}$&  -- \\
4 & 0.19$_{\pm 0.38}$ & (0.22$_{\pm 0.44}$) & 0.18$_{\pm 0.62}$ & --  & 0.17$_{\pm 0.03}$ & (0.20$_{\pm 0.04}$) & 0.99$_{\pm 0.15}$&  -- \\
\hline
$\chi^2$(dof) & \multicolumn{4}{c}{184(168)} & \multicolumn{4}{c}{190(177)}  \\
reduced $\chi^2$ & \multicolumn{4}{c}{1.10} & \multicolumn{4}{c}{1.07} \\
\enddata
\tablecomments{The hydrogen column density $N_{\rm H}$ in \textsf{TBAbs} is fixed as $3\times10^{18}$ cm$^{-2}$.
}
\end{deluxetable*}

\begin{figure}
\gridline{
\fig{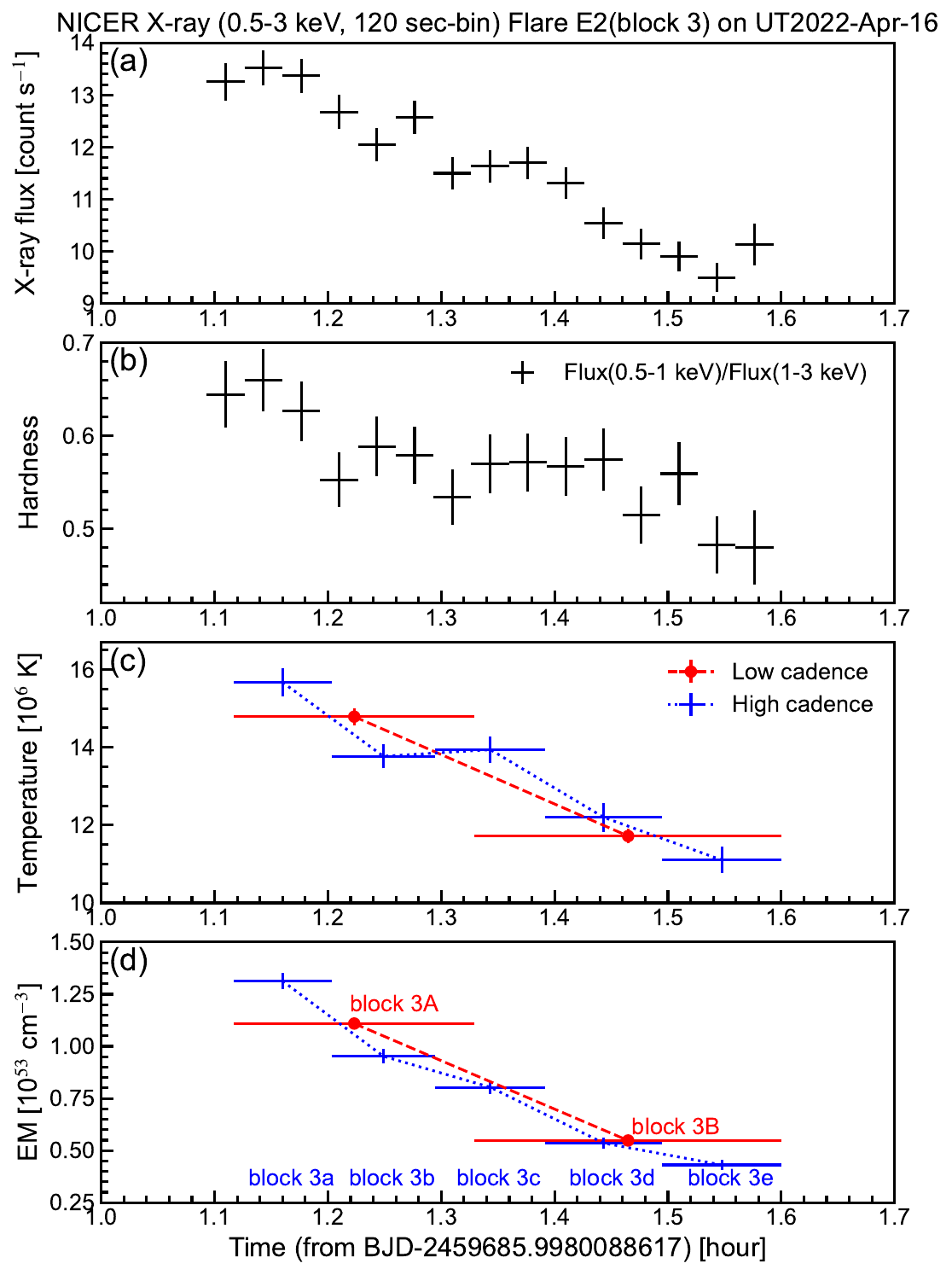}{0.45\textwidth}{\vspace{0mm} (A)}
\fig{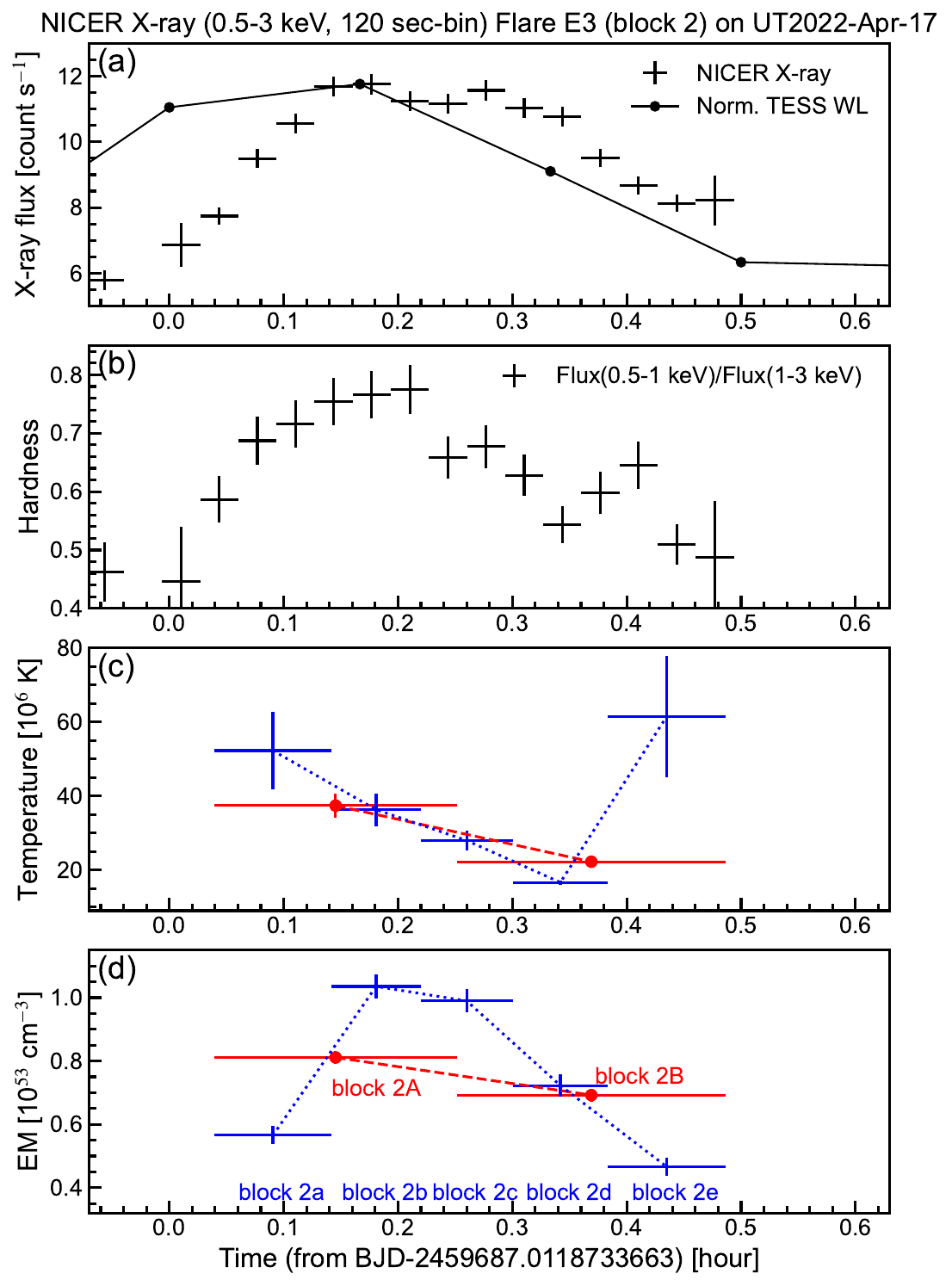}{0.45\textwidth}{\vspace{0mm} (B)}
}
\caption{Temporal evolution of the X-ray spectral fitting results for superflares E2 (A) and E3 (B).
(a) Light curves of X-ray count rates with a time resolution of 2 minutes in the 0.5-3 keV range.
(b) Hardness ratio, which is obtained by dividing the flux in the 0.5-1 keV range by the flux in the 1-3 keV range.
(c) Temporal change in temperature, with different colors indicating different time resolutions.
(d) Temporal change in the Emission Measure (EM).
}
\label{fig:14}
\end{figure}

We fixed the parameters of the obtained pre-flare spectrum and added a new \textsf{apec} model to perform a spectral fit of the flare component.
The flare duration was divided into either two or five segments (named ``block"). 
Initially, 1-sec-cadence light curve was created in 0.5-3 keV. Based on this, time bins were generated to divide the total count value of the orbit during the flare by half or into fifths. 
For flare E2 (or E3), the two segmented bins are labeled as blocks 2A -- 2B (or blocks 3A -- 3B for flare E3) and the five segmented bins are labeled as blocks 2a -- 2e (or blocks 3a -- 3e for flare E3). 
See Figure \ref{fig:14} for more details of the time bins.
A spectral fit was conducted for all these bins. 
We estimated temperature ($T$) and emission measure ($EM$) as free parameters.
The coronal abundance was fixed as the pre-flare values since it was not determined consistently when it was set as a free parameter.
The results of the spectral fits are summarized in Figure \ref{fig:12} and \ref{fig:13}, while the fitting parameters are compiled in Table \ref{tab:3}.
The time-resolved temperature ($T$) and emission measure ($EM$) are plotted in Figure \ref{fig:14}. 
Since the procedure ``\textsf{nicerarf}" does not work for the barycentric corrected event data, the above spectral fit was performed for non barycentric corrected event data. After the spectral analysis, we corrected the times for a barycentric correction.

\begin{deluxetable*}{cccccccccc}
\label{tab:3}
\tabletypesize{\footnotesize}
\tablecaption{Results of NICER X-ray spectral fit for flares E2 and E3.}
\tablewidth{0pt}
\tablehead{
\colhead{Flare E2's block} & \colhead{3} & \colhead{3A} & \colhead{3B} & \colhead{3a} &\colhead{3b} & \colhead{3c} & \colhead{3d} & \colhead{3e} & \colhead{--} }
\startdata
T [keV] & 1.14$_{\pm 0.01}$ & 1.27$_{\pm 0.02}$ & 1.01$_{\pm 0.02}$ & 1.35$_{\pm 0.03}$ & 1.19$_{\pm 0.03}$ & 1.20$_{\pm 0.03}$ & 1.05$_{\pm 0.03}$ & 0.96$_{\pm 0.03}$ & --  \\
(T [10$^{7}$ K]) & 1.32$_{\pm 0.01}$ & 1.48$_{\pm 0.02}$ & 1.17$_{\pm 0.02}$ & 1.57$_{\pm 0.04}$ & 1.38$_{\pm 0.03}$ & 1.39$_{\pm 0.03}$ & 1.22$_{\pm 0.04}$ & 1.11$_{\pm 0.03}$ & --  \\
EM [10$^{52}$ cm$^{-3}$] & 8.02$_{\pm 0.14}$ & 11.09$_{\pm 0.24}$ & 5.48$_{\pm 0.16}$ & 13.13$_{\pm 0.40}$ & 9.52$_{\pm 0.34}$ & 8.03$_{\pm 0.32}$ & 5.36$_{\pm 0.27}$ & 4.31$_{\pm 0.23}$ & --  \\
$\chi^{2}$(dof) & 414(220) & 250(168) & 232(177) & 141(126) & 166(125) & 185(129) & 152(127) & 112(118) & --  \\
reduced $\chi^{2}$ & 1.88 & 1.49 & 1.31 & 1.12 & 1.33 & 1.44 & 1.19 & 0.95 & --  \\
\hline
\hline
\colhead{Flare E3's block} & \colhead{2} & \colhead{2A} & \colhead{2B} & \colhead{2a} &\colhead{2b} & \colhead{2c} & \colhead{2d} & \colhead{2e} & \colhead{3} \\
\hline
T [keV] & 2.67$_{\pm 0.16}$ & 3.22$_{\pm 0.28}$ & 1.92$_{\pm 0.10}$ & 4.50$_{\pm 0.90}$ & 3.13$_{\pm 0.38}$ & 2.41$_{\pm 0.24}$ & 1.44$_{\pm 0.05}$ & 5.30$_{\pm 1.42}$ & 1.00$_{\pm 0.06}$ \\
(T [10$^{7}$ K]) & 3.09$_{\pm 0.18}$ & 3.74$_{\pm 0.33}$ & 2.22$_{\pm 0.11}$ & 5.23$_{\pm 1.04}$ & 3.63$_{\pm 0.44}$ & 2.80$_{\pm 0.27}$ & 1.66$_{\pm 0.06}$ & 6.15$_{\pm 1.64}$ & 1.16$_{\pm 0.07}$ \\
EM [10$^{52}$ cm$^{-3}$] & 6.09$_{\pm 0.12}$ & 8.11$_{\pm 0.22}$ & 6.92$_{\pm 0.20}$ & 5.66$_{\pm 0.29}$ & 10.35$_{\pm 0.38}$ & 9.90$_{\pm 0.36}$ & 7.23$_{\pm 0.35}$ & 4.66$_{\pm 0.28}$ & 0.70$_{\pm 0.07}$ \\
$\chi^{2}$(dof) & 388(243) & 246(184) & 278(177) & 125(128) & 151(127) & 145(128) & 177(123) & 138(120) & 279(203) \\
reduced $\chi^{2}$ & 1.60 & 1.34 & 1.57 & 0.97 & 1.19 & 1.13 & 1.44 & 1.15 & 1.37 \\
\enddata
\tablecomments{The hydrogen column density $N_{\rm H}$ in \textsf{TBAbs} is fixed as $3\times10^{18}$ cm$^{-2}$. The abundance in \textsf{apec} is fixed as a pre-flare value of each flare in Table \ref{tab:2} (0.28 for flare E2 and 0.38 for flare E3).
}
\end{deluxetable*}

The flare radiation fluxes and energies are estimated in the energy band of 0.5--3 keV (an observed band), 0.1--100 keV (referred to as ``bolometric X-ray" in this study; \citealt{2022PASJ...74..477K}), 0.5--7.9 keV \citep{2019AA...622A.210G}, 0.2--12 keV \citep{2021ApJ...912...81K,2022AA...667L...9S}, and 1.55--12.4 keV (=1--8{\AA}, \textit{GOES} band, \citealt{2022AA...667L...9S}) by using \textsf{flux} command in \textsf{xspec}.
We calculated the flare energies in these different bands to compare with previous studies. 
Table \ref{tab:4} shows the peak X-ray flux and total X-ray energy in each band.

\begin{deluxetable*}{ccccccccc}
\label{tab:4}
\tabletypesize{\footnotesize}
\tablecaption{Summary of NICER X-ray flare parameters.}
\tablewidth{0pt}
\tablehead{
\colhead{Flare} & \colhead{$EM_{\rm peak}$} & \colhead{$T_{\rm EM,peak}$} & \colhead{$kT_{\rm slope}$} &\colhead{$\tau_{\rm rise}$} & \colhead{$\tau_{\rm decay}$} & \colhead{$\xi_{\rm EM,T}$} & \colhead{$L_{\rm X}$} & \colhead{$E_{\rm X}$}  \\
\colhead{} & \colhead{[10$^{52}$ cm$^{-3}$]} & \colhead{[10$^{7}$ K]} & \colhead{[keV ksec$^{-1}$]} &\colhead{[ksec]} & \colhead{[ksec]} & \colhead{} & \colhead{[10$^{29}$ erg s$^{-1}$]} & \colhead{[10$^{32}$ erg]}   \\
 &  &   &   &  \multicolumn{2}{c}{(0.5--3 keV)} & & \multicolumn{2}{c}{(0.5--3 keV)} 
}
\startdata
E2 & 13.1$_{\pm 0.4}$ & 1.57$_{\pm 0.04}$ & -0.26$_{\pm 0.91}$ & ($<$3.96) & 1.95$_{\pm 0.19}$ & 0.603 & 6.3$_{-0.4}^{+0.2}$ & 22.9$_{-1.0}^{+0.5}$\\
(ext.) &  &  &  &  &  &  & (26.6$_{-1.7}^{+1.0}$) & (61.9$_{-2.6}^{+1.4}$)\\
E3 & 10.4$_{\pm 0.4}$ & 3.63$_{\pm 0.44}$ & -3.32$_{\pm 1.49}$ & 0.7$_{\pm0.2}$ & 0.63$_{\pm 0.10}$ & 4.28 & 4.9$_{-0.2}^{+0.3}$ & 6.9$_{-0.1}^{+0.3}$\\
\hline
\hline
\colhead{Flare} & \colhead{$L_{\rm X}$} & \colhead{$E_{\rm X}$} &\colhead{$L_{\rm X}$} & \colhead{$E_{\rm X}$} & \colhead{$L_{\rm X}$} & \colhead{$E_{\rm X}$} &\colhead{$L_{\rm X}$} & \colhead{$E_{\rm X}$}   \\
\colhead{} & \colhead{[10$^{29}$ erg s$^{-1}$]} & \colhead{[10$^{32}$ erg]} &\colhead{[10$^{29}$ erg s$^{-1}$]} & \colhead{[10$^{32}$ erg]} & \colhead{[10$^{29}$ erg s$^{-1}$]} & \colhead{[10$^{32}$ erg]} &\colhead{[10$^{29}$ erg s$^{-1}$]} & \colhead{[10$^{32}$ erg]}   \\
& \multicolumn{2}{c}{(0.1--100 keV)} & \multicolumn{2}{c}{(0.5--7.9 keV)} & \multicolumn{2}{c}{(0.2--12 keV)} & \multicolumn{2}{c}{(1.55--12.4 keV/GOES)} \\
\hline
E2 & 8.2$_{-0.5}^{+0.4}$ & 29.3$_{-1.2}^{+0.7}$& 5.9$_{-0.3}^{+0.2}$ & 21.0$_{-0.7}^{+0.5}$&7.4$_{-0.5}^{+0.3}$ & 26.1$_{-1.1}^{+0.7}$&1.9$_{-0.1}^{+0.2}$ & 5.3$_{-0.3}^{+0.4}$\\
(ext.) & (34.5$_{-2.1}^{+1.5}$) & (79.1$_{-3.4}^{+2.0}$)& (24.8$_{-1.2}^{+1.0}$) & (56.6$_{-2.0}^{+1.4}$)&(31.1$_{-2.1}^{+1.4}$) & (70.6$_{-2.9}^{+2.0}$)&(8.0$_{-0.6}^{+0.8}$) & (14.3$_{-0.8}^{+1.1}$)\\
E3 & 8.1$_{-0.6}^{+0.7}$ & 11.4$_{-0.5}^{+0.6}$& 6.2$_{-0.4}^{+0.5}$ & 8.6$_{-0.3}^{+0.4}$&7.6$_{-0.6}^{+0.7}$ & 10.5$_{-0.5}^{+0.5}$&4.0$_{-0.5}^{+0.6}$ & 5.4$_{-0.5}^{+0.4}$\\
\enddata
\tablecomments{
``ext." means the extrapolated value by the light curve fits.
$EM_{\rm peak}$ is the peak emission measure in the observational window.
$T_{\rm EM,peak}$ is the temperature when the emission measure (EM) is maximum value  in the observational window.
$kT_{\rm slope}$ is the slope of the temperature ($kT$) evolution.
$\tau_{\rm rise}$ is the rise time of 0.5-3 keV light curve from its onset to peak.
$\tau_{\rm decay}$ is the e-fodling decay time of 0.5-3 keV light curve.
$\xi_{\rm EM,T}$ is the power-law index of the relation between temperature ($T$) and emission measure ($EM$) in decay phase in the formula of $T\propto (EM^{0.5}) ^{\xi_{\rm EM,T}}$.
$L_{\rm X}$ is the X-ray luminosity and $E_{\rm X}$ is the X-ray energy. The energy and flux values are not corrected for light curve extrapolations in this table.
}
\end{deluxetable*}

\subsubsection{Light Curve Fitting}

\begin{figure}
\plottwo{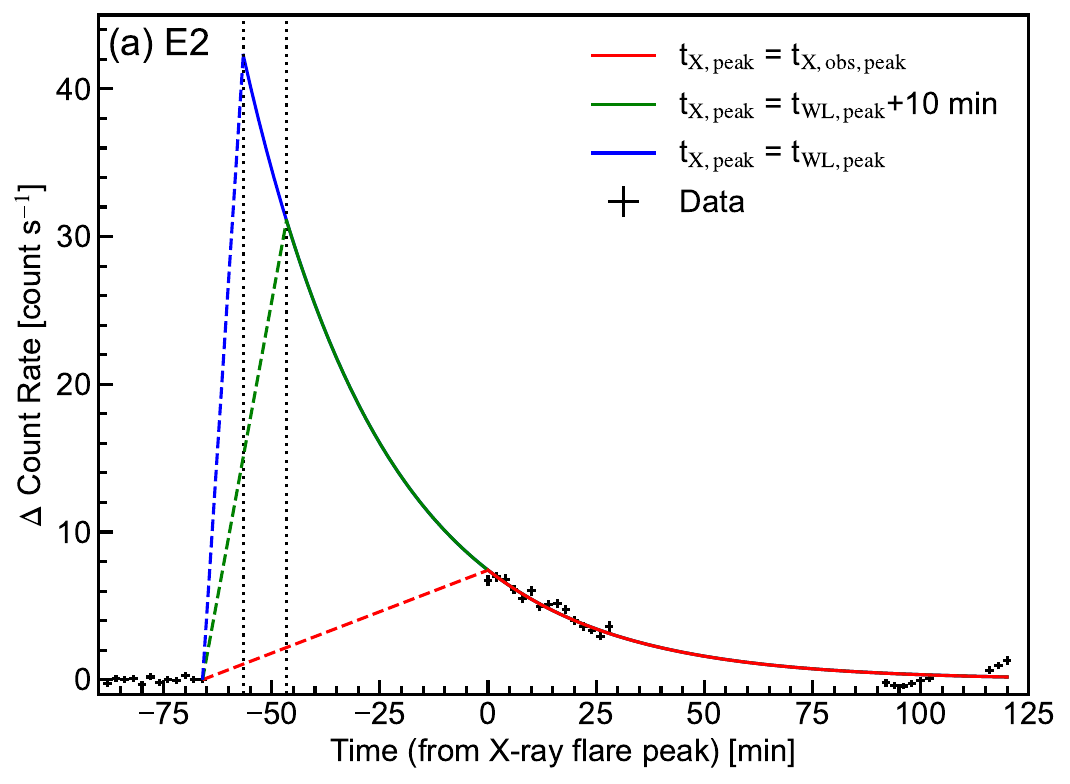}{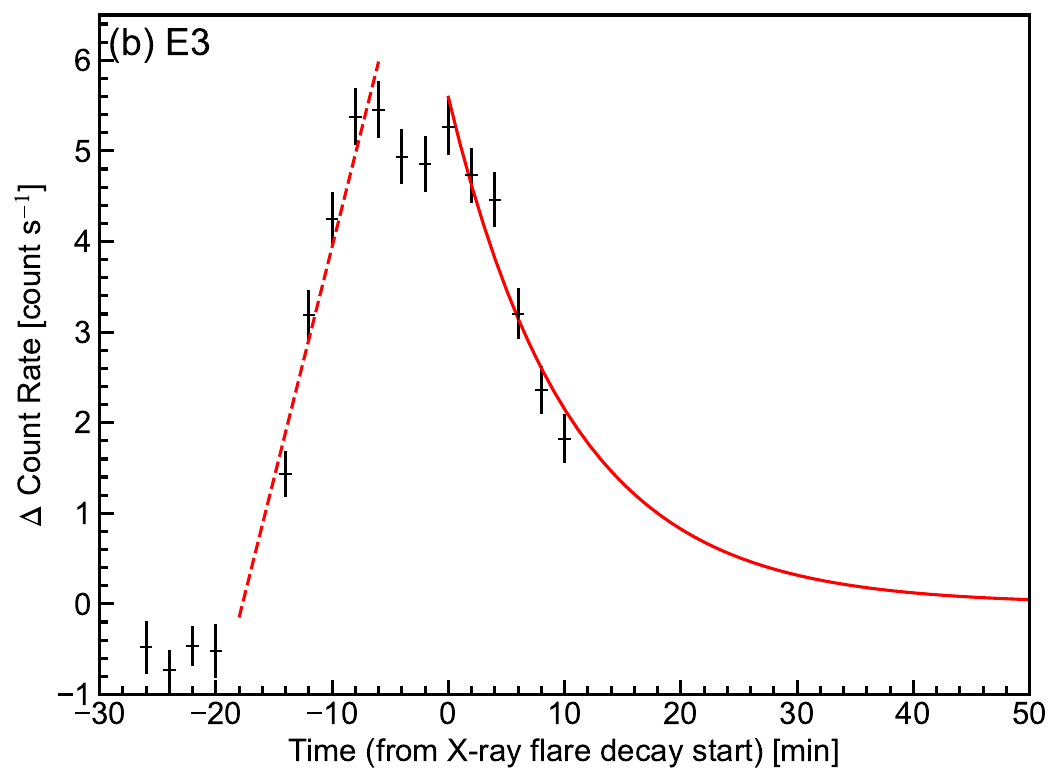}
\caption{
Fitting results of the X-ray light curves for superflares E2 (a) and E3 (b). The decay phase is fitted with an exponential function (colored solid lines), and the rise phase with a linear function (colored dashed lines), using \textsf{curve\_fit}. (left) For flare E2, two patterns of light curve extrapolation are presented: assuming the peak time is the same as that of the white light (blue), and assuming the peak time is delayed by 10 minutes compared to the white light. The X-ray peak of flare E2 could possibly be higher by a factor of 4.2 (green) and 5.7 (blue) than the observed peak. The X-ray energy of flare E2 could possibly be higher by a factor of 2.7 (green) and 3.2 (blue) than the observed energy.
}
\label{fig:16}
\end{figure}

\textit{NICER}'s orbit did not enough cover the total flare evolution, so we have extrapolated the light curve by fitting the decay phase with exponential function and fitting the rise phase with linear function (see Figure \ref{fig:16}).
The rise time and decay time are summarized in Table \ref{tab:4}.
For flare E2, the flare rise phase was not observed, so the rise timescale is just an upper limit.
By assuming that the flare peak time is 0 min or 10 min later than the peak time of the \textit{TESS}'s WLF, we extrapolated the possible peak X-ray flux.
As a result of the extrapolation, the X-ray peak of flare E2 is possibly higher by a factor of 4.2--5.7 than the observed peak. 
The X-ray energy of flare E2 is possibly higher by a factor of 2.7--3.2 than the energy calculated from the observed light curve.
In this study, we use the possible X-ray flare peak value of flare E2 as the flux when the flare peak time is 10 min later than the peak time of the \textit{TESS}'s WLF.
For flare E3, we did not perform any extrapolation for the flux and energy because most of the flare evolution was covered by the one \textit{NICER}'s orbit. 

\section{Time Evolution, Energetics, and Length Scales of Superflares}\label{sec:4}

\subsection{Time Evolution of Optical and X-ray Flares}\label{sec:4-1}

\begin{figure}
\epsscale{0.5}
\plotone{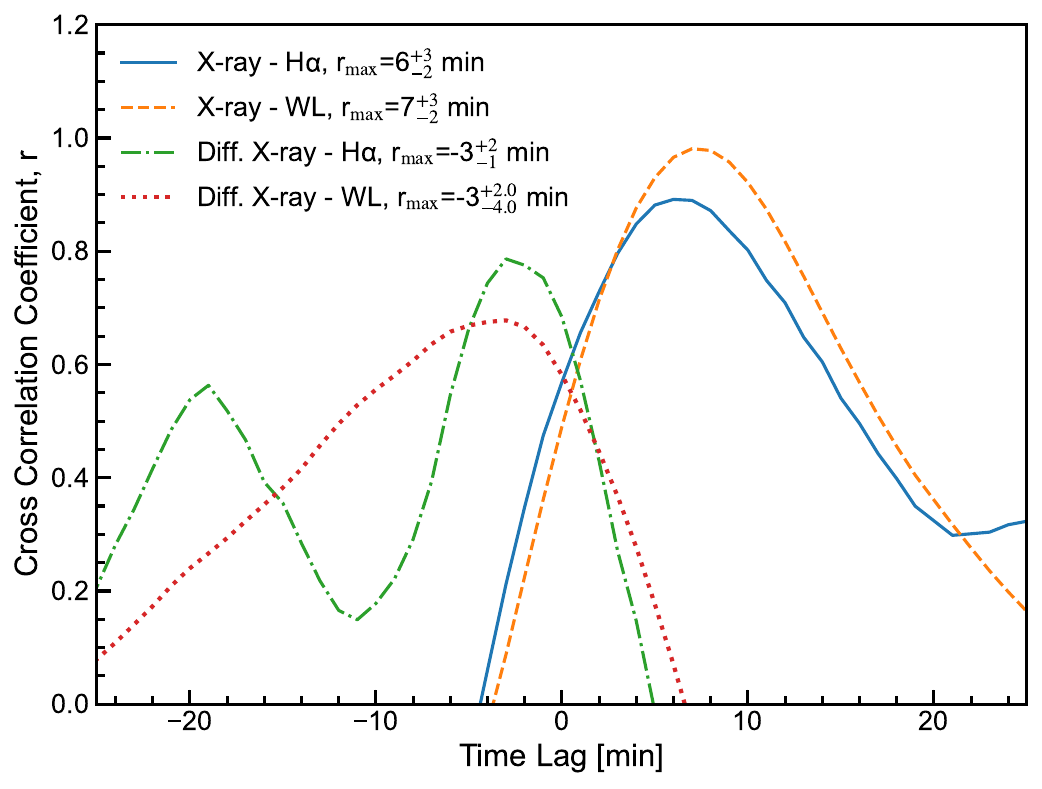}
\caption{The cross-correlation coefficients for different lag times among light curves for the superflare on 2022 April 17 (E3). The time lags of maximum correlation coefficients, r$_{\rm max}$, is defined as the most likely value. The 90\% percentile gives the lag time uncertainty. The time delay of X-ray against H$\alpha$ is  6$_{ -2 }^{+3 }$ min, the time delay of X-ray against WL is  7$_{ -2 }^{+3 }$ min, the time delay of differential X-ray against H$\alpha$ is  --3$_{ -2 }^{ +1 }$ min, and the time delay of differential X-ray against WL is  --3$_{ -4 }^{+2 }$ min. See \cite{2023ApJ...951...33T} for the method. }
\label{figs:16}
\end{figure}

Here we discuss the time evolution of flaring emissions in multi-wavelength, particularly focusing on flare E3.
Figure \ref{fig:4} reveals that the rise of the soft X-rays appears to be clearly delayed in comparison to the white light and H$\alpha$ emissions.
We used the method employed by \cite{2005A&A...431..679M} (later, \citealt{2023ApJ...951...33T}) to estimate the most likely delay time of soft X-ray flare $L_{\rm X}(t)$ and time-derivative soft X-ray d$L_{\rm X}$(t)/d$t$ ($>$0).
Figure \ref{figs:16} shows the cross correlation coefficient as a function of time lags. 
The time lags of maximum of the Pearson's correlation coefficients, r$_{\rm max}$, is defined as the most likely value. 
The time delay of X-ray against H$\alpha$ is 6$_{ -2 }^{+3 }$ min, and the time delay of X-ray against WL is 7$_{ -2 }^{+3 }$ min.
The error bar is obtained the same way as \cite{2023ApJ...951...33T}.
Also, we calculated the time-derivative soft X-ray light curve, as a good proxy of the non-thermal radiation (d$L_{\rm X}$/d$t$, \citealt{2002A&A...382.1070V}).
The time delay of time-derivative X-ray against H$\alpha$ is  --3$_{ -2 }^{ +1 }$ min, and the time delay of differential X-ray against WL is  --3$_{ -4 }^{+2 }$ min.
The time-derivative soft X-ray light curve (d$L_{\rm X}$/d$t$) shows an increase almost simultaneous with the WL and H$\alpha$, with an overall delay of a few minutes for the entire light curve, but peaking at nearly the same time. 

The close correspondence between the impulsive phase of WL and H$\alpha$ with the time-derivative X-ray light curve may be an indication of the Neupert effect \citep{1968ApJ...153L..59N,2023ApJ...951...33T}.
The theoretical Neupert effect is usually expressed as a delay of soft X-rays (thermal emission) against hard X-rays (non-thermal radiation) (cf. \citealt{2002A&A...382.1070V}), which is the evidence of the heating by non-thermal electrons followed by the chromospheric evaporation increasing the coronal density of the flaring loops \citep{2011LRSP....8....6S}.
Since the WL emissions (and some of the H$\alpha$ emissions) are known to be a good proxy of hard X-ray emission in the Sun \citep{2011ApJ...739...96K,2017ApJ...851...91N},
the time delay of soft X-ray against WL (and possibly H$\alpha$) for flare E3 is interpreted as the ``empirical" Neupert effect \citep[e.g.,][]{2023ApJ...951...33T}.
This interpretation suggests that flare E3 occurs through the same physical mechanism as solar flares \citep[the same conclusion has been obtained for M-dwarf flares;][]{1995ApJ...453..464H,2023ApJ...951...33T}.
\cite{2022A&A...666A.198P} found a clear signature of the ``empirical" Neupert effect in 200--300 nm near-UV band and soft X-ray relation for superflares on a young solar-type star DS Tuc A (age of 40 Myr, spectral type of G6).
\cite{2022A&A...666A.198P} and this study have the same conclusion about the flare heating mechanism on young solar-type stars.

Furthermore, H$\alpha$ and white light potentially appear to lag slightly behind the time-derivative soft X-rays, or seem to have a longer decay.
This suggests that in addition to the heating caused by non-thermal electrons corresponding to hard X-rays, heating due to thermal conduction and/or radiative backwarming also contribute to H$\alpha$ and WL emissions in decay \citep{1989SoPh..124..303M,2007PASJ...59S.807I,2017ApJ...851...91N,2020PASJ...72...68N,2022ApJ...926L...5N}. 
These multi-wavelength light curve variations should be compared with the future numerical calculations.

\subsection{Flare Radiative Energy Partition}\label{sec:4-2}

\begin{figure}
\gridline{
\fig{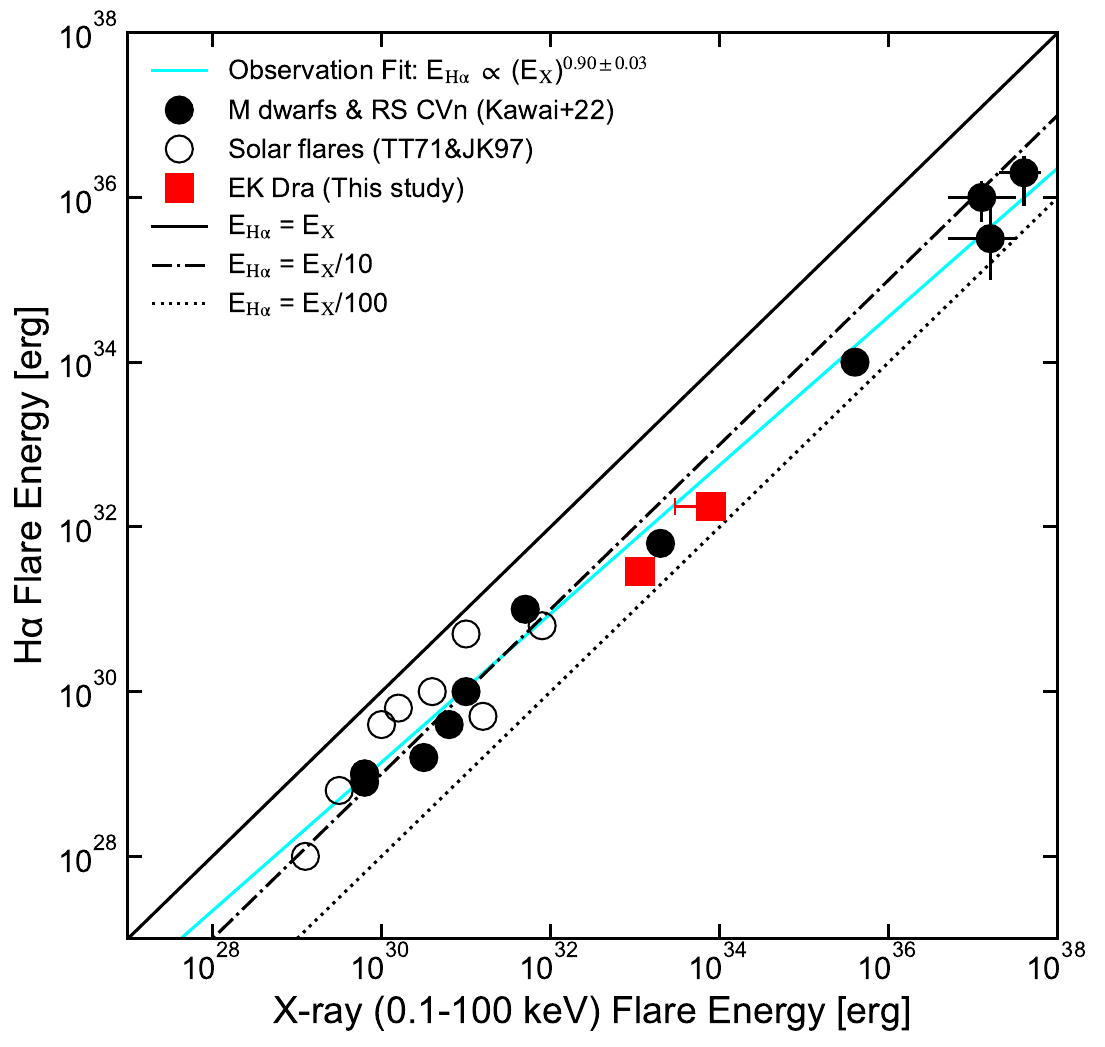}{0.5\textwidth}{\vspace{0mm} (a)}
\fig{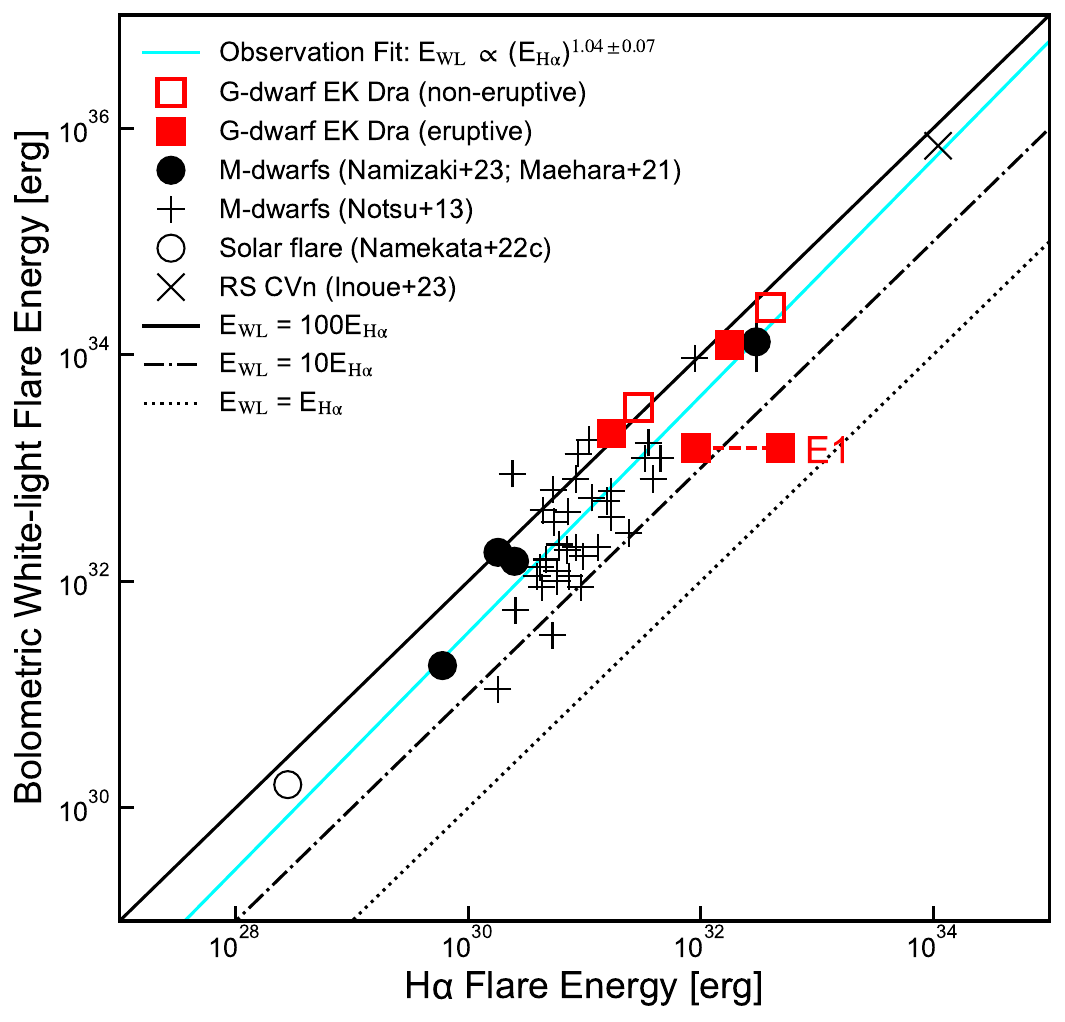}{0.5\textwidth}{\vspace{0mm} (b)}
}
\caption{Relations between H$\alpha$, X-ray, and white-light energies for solar and stellar flares. 
(left) The relation between H$\alpha$ and Bolometric X-ray energies (0.1--100 keV = 0.124--124 {\AA}). Data of superflares on the G-dwarf EK Dra detected in this study are represented by red squares. 
Other solar and stellar data are taken from \cite{2022PASJ...74..477K}. 
The function fitted to the observational data with a power-law is shown by the cyan line. 
(right) The relation between bolometric white-light and H$\alpha$ energies. Data of superflares on the G-dwarf EK Dra, which are associated with filament/prominence eruptions, are represented by filled red squares, whereas those without eruptions are represented by open red squares (this study, \citealt{2022NatAs...6..241N}, \citealt{2022ApJ...926L...5N}). 
For E1, the total energy including the blueshift component is plotted as the upper limit, with the radiative energy of the central component from the two-component fit being plotted as the lower limit.
Data of M-dwarf flares are taken from \cite{2020PASJ..tmp..253M}, \cite{2023ApJ...945...61N}, and \citep{Notsu2023}. Solar flare data are taken from \cite{2022ApJ...933..209N}. The function fitted to the observational data, excluding flare E1, with a power-law is shown by the cyan line.}
\label{figs:17}
\end{figure}


\begin{figure}
\epsscale{0.5}
\plotone{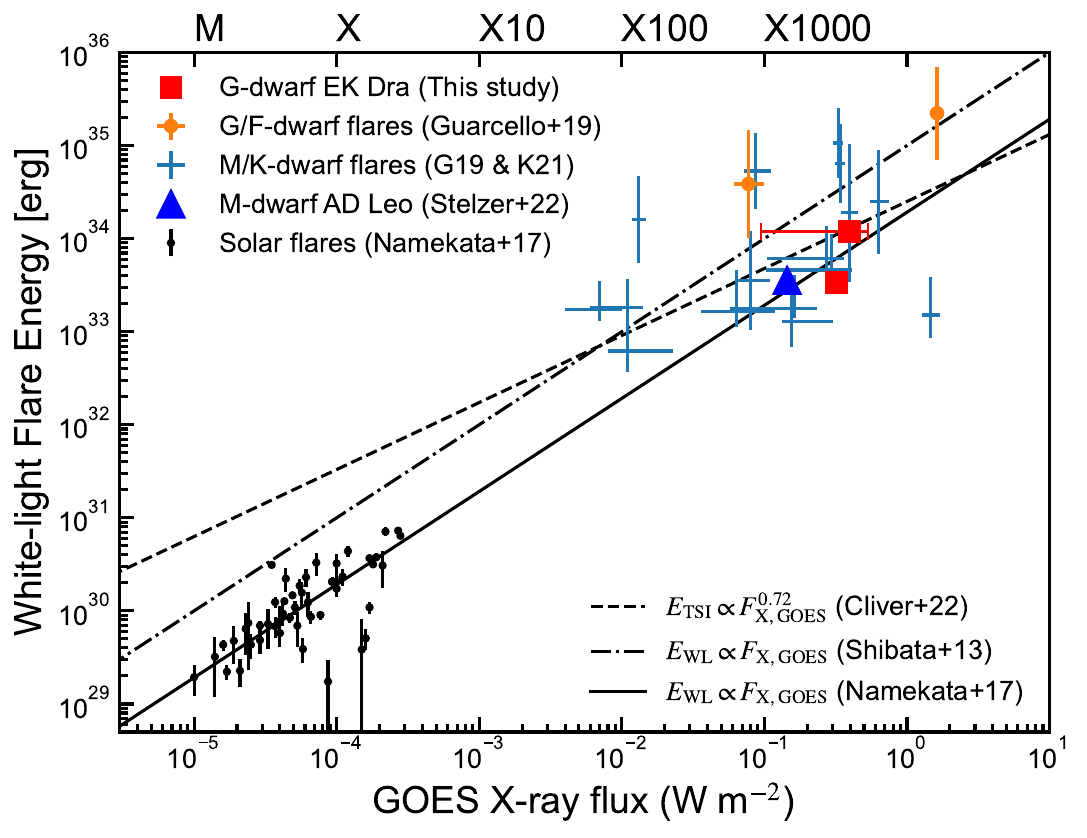}
\caption{Relation between the bolometric white-light flare energy and the peak GOES-band soft X-ray flux (1--8 {\AA} = 1.55--12.4 keV) for solar and stellar flares. Data for the superflares (E2, E3) from the G-dwarf EK Dra derived in this paper are plotted as red squares. Other data for superflares from G/F-dwarfs (orange circles) are taken from \cite{2019AA...622A.210G}. Data for superflares for M/K-dwarfs are taken from \cite{2019AA...622A.210G}, \cite{2021ApJ...912...81K}, and \cite{2022AA...667L...9S}. Note that data from \cite{2019AA...622A.210G} have been converted to GOES band flux and bolometric WLF flare energy. The solar observational scaling laws from \cite{2022LRSP...19....2C} (dashed line) and \cite{2017ApJ...851...91N} (solid line), as well as the empirical law from \cite{2013PASJ...65...49S} (dash dotted line), are also plotted.}
\label{figs:18}
\end{figure}

Figure \ref{figs:17}(a) shows the relation between H$\alpha$ and the bolometric soft X-ray flare radiation energy for solar and stellar flares, while Figure \ref{figs:17}(b) shows that between the bolometric WL and H$\alpha$ flare radiation energy
.
We found that the data of the superflares on EK Dra are almost consistent with a trend from solar flares and stellar flares on M-dwarf and RS CVn stars.
This suggests that the multi-wavelength emission mechanisms of the flares of EK Dra we detected are equivalent to those occurring in flares on other types of stars.

In Figure \ref{figs:17}(b), the H$\alpha$ energy of flare E1 is plotted (labeled as ``E1") with the upper limit as the total emission energy including the blueshifted spectrum, and the lower limit as the emission energy assuming the presence of a central component. 
The upper limit (and even possibly the lower limit) is clearly 1-1.5 orders of magnitude larger in H$\alpha$ emission energy relative to WL emission energy than the other EK Dra flares. 
This can be attributed to the fact that the emission source of the blueshifted H$\alpha$ line in flare E1 is different from the other EK Dra flares (not the flare loop or footpoints), suggesting a radiation contribution from stellar prominence eruption, as discussed in Section \ref{sec:5:disc}.

Incorporating the data from EK Dra (except for flare E1), we derived empirical rules regarding the relationship between H$\alpha$ and X-ray, as well as WL and H$\alpha$. 
By linking these, we deduced an empirical scaling relation for energy distributions at the bolometric X-ray (0.1-100 keV) and bolometric WL wavelengths, which are releasing a large amount of energy of flares.
\begin{eqnarray}
\frac{E_{\rm H\alpha}}{10^{33} \, \rm erg} &=& 10^{-1.15 \pm 0.08}\cdot \left( \frac{E_{\rm X,bol}}{10^{33} \, \rm erg} \right)^{0.90 \pm 0.03}, \label{eq:ha-x}\\
\frac{E_{\rm WL,bol}}{10^{33} \, \rm erg} &=& 10^{1.68 \pm 0.15}\cdot \left(\frac{E_{\rm H\alpha}}{10^{33} \, \rm erg} \right)^{1.04 \pm 0.07}, \label{eq:wl-ha}
\end{eqnarray}
Equations \ref{eq:ha-x} and \ref{eq:wl-ha} are plotted in cyan lines in Figure \ref{figs:17}. From Equations \ref{eq:ha-x} and \ref{eq:wl-ha}, an empirical relation between the X-ray ($E_{\rm X,bol}$) and WL ($E_{\rm WL,bol}$) flare energy can be deduced as 
\begin{eqnarray}
\frac{E_{\rm WL,bol}}{10^{33} \, \rm erg} = 10^{0.48 \pm 0.19}\cdot\left( \frac{E_{\rm X,bol}}{10^{33} \, \rm erg} \right)^{0.94 \pm 0.07}. \label{eq:WL-x}
\end{eqnarray}
While this relationship appears to be almost linear, there may be a tendency for X-ray emission energy to slightly dominate as the flare scale increases. This relationship would be useful for statistically investigating the high-energy radiation environment around host stars by connecting to the large statistics of WLFs established by \textit{TESS} and \textit{Kepler} \citep{2012Natur.485..478M,2016ApJ...829...23D,2019ApJ...876...58N,2019PASP..131i4502F,2020arXiv201102117O,2020ApJ...890...46T,2021ApJS..253...35T,2021A&A...652A.107V}.

Furthermore, assuming that the emission energy at different wavelengths can be described by a linear relationship, a corresponding relationship is derived below for reference: 
\begin{eqnarray}
E_{\rm H\alpha} = 10^{-1.06 \pm 0.09} \cdot E_{\rm X,bol}, \label{eq:ha-x-ratio}\\
E_{\rm WL,bol} = 10^{1.59 \pm 0.06} \cdot E_{\rm H\alpha}, \label{eq:wl-ha-ratio}
\end{eqnarray}
From Equations \ref{eq:ha-x-ratio} and \ref{eq:wl-ha-ratio}, another empirical relation can be deduced as 
\begin{eqnarray}
E_{\rm WL,bol} = 10^{0.54 \pm 0.11} \cdot E_{\rm X,bol} \label{eq:WL-x-ratio}
\end{eqnarray}

Note that the scaling factor in Equation \ref{eq:WL-x-ratio} is so different from that in Equation \ref{eq:WL-x} because Equation \ref{eq:WL-x-ratio} is a linear relation while Equation \ref{eq:WL-x} is a non-linear one. In the above, we collected and compared X-ray data in the 0.1--100 keV energy range.

Many other studies derive emission energy in different energy bands, causing inconsistency in comparison (see, Section \ref{sec:3:x}). 
In this paper, to compare with other previous studies, we recalculated the flare energy of EK Dra in all energy bands shown in previous studies. 
The results are summarized in Table \ref{tab:difenergy-balance}. 
As a result, it was found that the white-light-to-X-ray energy ratios ($E_{\rm WL}/E_{\rm X}$) were comparable to those of previous studies in all energy bands. 
However, in the data compared with solar flares in the GOES band, it was found that solar flares show relatively larger $E_{\rm WL}/E_{\rm X}$ ratios ($\sim$70--100) than superflares on EK Dra ($\sim$6.2 and 7.0--22.6) and an M-dwarf AD Leo ($\sim$13). 
This may suggest, as mentioned above, that in large-scale stellar superflares ($10^{33-34}$ erg), X-ray emission energy may become stronger than small-scale solar flares ($<10^{32}$ erg). 
In particular, the GOES band is located at a higher energy compared to other energy bands compared in Table \ref{tab:difenergy-balance}, and therefore is sensitive to the flare's temperature increase around 10$^7$ K. 
Since it is widely known that the flare temperature increases as the energy scale increases \citep{1998ApJ...503..894T,2016PASJ...68...90T,1999ApJ...526L..49S,2002ApJ...577..422S,2021ApJ...920..154G}, the energy in the GOES band may tend to increase more than proportionally.

\begin{deluxetable*}{ccccccc}
\label{tab:difenergy-balance}
\tabletypesize{\footnotesize}
\tablecaption{Summary of X-ray and bolometric white-light energy of stellar flares in this study and other studies.}
\tablewidth{0pt}
\tablehead{
\colhead{X-ray band}& \colhead{$E_{\rm WL,bol}/E_{\rm X}$}  & \colhead{$E_{\rm WL,bol}/E_{\rm X}$} & \colhead{$E_{\rm WL,bol}/E_{\rm X}$} & \colhead{$E_{\rm WL,bol}$ [erg]} & Sp. Type & \colhead{References} \\
 & \colhead{(E2=1.2$\times 10^{34}$ erg)} &  \colhead{(E3=3.4$\times 10^{33}$ erg)} & \colhead{(Previous study)} &  \colhead{(Previous study)} &  \colhead{(Previous study)} &  
}
\startdata
\hline
0.5--7.9 keV & 1.8-5.7  & 3.9 & 0.9-10$^{(\ast)}$ & 0.72-46$\times$10$^{33}$$^{(\ast)}$ & M-F stars  & (1)  \\
0.3--3.0 keV & 1.6-5.2  & 4.9 & $>$1-40$^{(\S)}$ & 0.32-16$\times$10$^{31}$$^{(\S)}$ & EV Lac (M dwarf) & (2) \\
0.2--12 keV & 1.4-4.6  & 3.2 & 0.1,$\sim$1.5,$\sim$3-4 & 0.62--6.2$\times$10$^{33}$  & M/K dwarfs & (3) \\
0.2--12 keV & 1.4-4.6  & 3.2 & 4.4 & 5.6$\times$10$^{33}$ & AD Leo (M dwarf) & (4) \\
1.55--12.4 keV/GOES & 7.0-23  & 6.2 & 13 & 5.6$\times$10$^{33}$ & AD Leo (M dwarf) & (4) \\
1.55--12.4 keV/GOES & 7.0-23  & 6.2 & 70--100 & 10$^{29-32}$ & The Sun (G2V) & (5), (6) \\
\enddata
\tablecomments{
$E_{\rm WL,bol}$ is the bolometric white-light flare energy under the assumption of 10,000 K blackbody radiation. $E_{\rm X}$ is the X-ray flare energy in the energy band of the first culmn. 
``Sp. Type" is the spectral type of the star.
$^{\ast}$The Kepler-band flare energy is reported. 
$^\S$The TESS-band flare energy is reported. The authors also mentioned that the $E_{\rm WL,bol}/E_{\rm X}$ values are lower limits because all of the X-ray flare evolution were not covered by NICER.
References: (1) \cite{2019AA...622A.210G}, (2) \cite{2021ApJ...922...31P}, (3) \cite{2021ApJ...912...81K}, (4) \cite{2022AA...667L...9S}, (5) \cite{2012ApJ...759...71E}, (6) \cite{2015ApJ...809...79O}.
}
\end{deluxetable*}

Lastly, Figure \ref{figs:18} shows the relation between the \textit{GOES} X-ray peak flux of the flare ($F_{\rm GOES}$) and the energy of the white light flare ($E_{\rm WLF,Bol}$). 
Conventionally, some studies have used an empirical relationship $E_{\rm WLF,Bol} = 10^{35} \cdot F_{\rm GOES}$ \citep[][]{2013PASJ...65...49S} to compare the energy scale of solar and stellar flares because most solar flares are evaluated by $F_{\rm GOES}$ while most stellar flares are by $E_{\rm WLF,Bol}$. 
Later, \cite{2017ApJ...851...91N} confirmed observationally that this power-law index becomes linear ($\sim 1$) for solar flare observations, but its absolute value is slightly different from \cite{2013PASJ...65...49S}'s law, and the relationship was $E_{\rm WLF,Bol} = 10^{34.3} \cdot F_{\rm GOES}$. 
\cite{2017ApJ...851...91N} suggest that this linear relationship can be explained by well-known Neupert effect, by assuming that WLF is a good proxy of hard X-ray radiation. 
On the other hand, a non-linear scaling relation for total solar irradiance (TSI) and \textit{GOES} X-ray peak flux of solar flares has been reviewed by \cite{2022LRSP...19....2C}, and the relationship $E_{\rm TSI} = 0.32\times10^{32} \cdot (F_{\rm GOES}/(10^{-4} \rm W m ^{-2}))^{0.72}$ has been derived. 
While all of these relationships are scaling laws describing solar flares, they differ slightly, and a consistent view has not yet been established \citep[cf.][]{2022LRSP...19....2C}.
Then, which scaling law can describe superflares on stars?
Figure \ref{figs:18} shows the comparison between the scaling laws and stellar superflares on G/M dwarfs, including EK Dra's superflares with red color.
As a result, we found that all the scaling laws intersect without significant contradiction with the stellar data. 
In detail, it can be said for now that the two superflares of EK Dra prefer the scaling laws of \cite{2017ApJ...851...91N} and \cite{2022LRSP...19....2C}. 
In the future, it is expected that multi-wavelength observations of superflares with $10^{37-39}$ ergs in RS CVn type binary stars, for example, could lead to an understanding of the extension capability of these scaling laws.

\subsection{Length Scale of Flare Loop} \label{sec:flare-length}

The length scales of the coronal flare loops are estimated by using (1)  \cite{1999ApJ...526L..49S,2002ApJ...577..422S}'s model (see also \citealt{2017PASJ...69....7N}), 
(2) \cite{2007AA...471..271R}'s model, and (3) \cite{2017ApJ...851...91N}'s method. See Table \ref{tab:length-scale} for the result summary.

\begin{deluxetable*}{lccccccccc}
\label{tab:length-scale}
\tablecaption{Length scale and area of the superflares, prominence/filament eruptions, and starspots on EK Dra in this study and past observations.}
\tablewidth{0pt}
\tablehead{
\colhead{} & \multicolumn{3}{c}{Flare Loop} & \multicolumn{2}{c}{Prominence} & \colhead{BVAmp} & \multicolumn{2}{c}{Spot} & \colhead{Rot. Phase}  \\
\colhead{} & \multicolumn{3}{c}{[10$^{10}$ cm]} & \colhead{[10$^{10}$ cm]} & \colhead{[10$^{21}$ cm$^2$]} & \colhead{[$F_{\rm av}$]} & \colhead{[10$^{10}$ cm]} & \colhead{[10$^{21}$ cm$^2$]} & \\
\colhead{(Methods)} & \colhead{(S\&Y99)$^\S$} & \colhead{(R07)$^{\ss}$} & \colhead{(N17)$^{\%}$} \\
\colhead{(Input)} & \colhead{(X-ray)$^\S$} & \colhead{(X-ray)} & \colhead{(WLF)$^{\%}$} 
}
\startdata
2022 Apr 10 (E1) & -- & -- & 0.73  & 8.3-300 & 18--63  & 0.024 & 2.1 & 0.42 & Decline-Loc.min \\
2022 Apr 16 (E2) & (10.08) & -- & 1.16  & 1.4-51 & 0.53--1.9  & 0.033 & 2.4 & 0.57 & Rise \\
2022 Apr 17 (E3) & 2.06 & 0.68 & 0.67  & -- & --  & 0.032 & 2.4 & 0.56 & Decline \\
\hline
2020 Apr 05 (E4)$^\dagger$ & -- & -- & 0.48  & 1.4-125 & 0.2--16  & 0.025 & 2.1 & 0.43 & Loc.max \\
2020 Mar 14 (E5)$^{\ast}$ & -- & -- & 1.50  & -- & -- & 0.021 & 1.9 & 0.36 & Decline \\
\enddata
\tablecomments{
``BVAmp" is the brightness variation amplitude relative to average flux in \textit{TESS} band due to the stellar rotation. ``Rot. Phase" is the timing when the superflares occur relative to stellar rotational phase. ``Loc.min" and ``Loc.max" is the local minimum and local maximum of the light curve, respectively, and ``Rise" and  ``Decline" is the rising and decline phase of the light curve, respectively.
As a reference, stellar radius is 0.94 times solar radius $\sim$6.55$\times 10^{10}$ cm, and stellar disk area is $\sim$13.5$\times 10^{21}$ cm$^{2}$.
$^\S$The loop length is estimated with the method by \cite{1999ApJ...526L..49S,2002ApJ...577..422S} under the assumption of coronal electron density $n_e=10^{11}$ cm$^{-3}$.
$^{\%}$The loop length is estimated with the method by \cite{2017ApJ...851...91N} under the assumption of coronal density $n=10^{2} n_{\odot}\sim 10^{11}$ cm$^{-3}$.
$^{\ss}$The estimation from rise timescale of \cite{2007AA...471..271R}.
$^\dagger$The data is taken from \cite{2022NatAs...6..241N}.
$^{\ast}$The data is taken from \cite{2022ApJ...926L...5N}.
}
\end{deluxetable*}

\cite{1999ApJ...526L..49S,2002ApJ...577..422S}'s model is based on the two dimensional magnetohydrodynamic (MHD) simulations of flares performed in \cite{1998ApJ...494L.113Y}.
They assume the balance between heating by magnetic reconnection and cooling by thermal conduction in the flare loop and derived the scaling relation to estimate the flare loop length in the following equation
\begin{eqnarray}
L_{\rm loop} &=& 10^{9} \left( \frac{EM_{\rm peak}}{10^{48} \rm cm^{-3}} \right)^{3/5} \left( \frac{n_{\rm e}}{10^9 \rm cm^{-3}} \right)^{-2/5} \left( \frac{T_{\rm EM,peak}}{10^7 \rm K} \right)^{-8/5} \rm cm, 
\end{eqnarray}
where $EM_{\rm peak}$ is the peak EM, $T_{\rm EM,peak}$ is the temperature at the EM peak.
This reliable method is later validated by solar flare observations \citep{2017PASJ...69....7N}.
\cite{1995A&A...301..201G} reported that the coronal density of EK Dra is $>4\times 10^{10}$ cm$^{-3}$ , and \cite{2004A&A...427..667N} also reported the consistent value \citep[e.g.,][]{2004A&ARv..12...71G}. Therefore, the assumption of coronal density of $\sim 10^{11}$ cm$^{-3}$ would be a reasonable value for EK Dra's coronal active region.
Under the assumption of the pre-flare coronal density in active region $n_{\rm e}$ = 10$^{11}$ cm$^{-3}$ , the loop length of flare E2 and flare E3 are derived to be 1.0$\times10^{11}$ cm and 2.1$\times10^{10}$ cm, respectively (see, Table \ref{tab:length-scale}).
The advantage of this approach lies in the fact that information on temporal variation is not required as long as the flare peak is captured. However, as in the case of flare E2 where the flare peak is not captured, it should be noted that accurate values may not be derived (therefore, the value in Table \ref{tab:length-scale} is in bracket).

\cite{2007AA...471..271R} applied their one-dimensional hydrodynamic flare model to the rise phase of the flare, suggesting an
equation to estimate the flare loop (half-)length:
\begin{eqnarray}
L_{\rm loop} &=& 950 \sqrt{T_{\rm peak}} \tau_{\rm rise} \psi \; \rm cm, 
\end{eqnarray}
where $T_{\rm peak}$ is the maximum temperature and $\psi = T_{\rm peak}/T_{\rm EM,peak}$.
This gives the estimate of the loop length of the flare E3 as 0.68$\times10^{10}$ cm. 
Note that \cite{1997AA...325..782R} initially derived the scaling relation to derive the flare loop length from the decay phase of the X-ray flare, but it requires the instrument-calibrated parameters, and no NICER's parameters are available for now. But future development may enable us to estimate the loop length from the decay phase, we derived the required parameters in Table \ref{fig:4}.

\cite{2017ApJ...851...91N} derived the scaling relation to derive the flare loop length from the WLF's energy and duration.
They used the relationship that the WLFs timescale is proportional to reconnection timescale ($\propto$Alfven timescale) and flare energy $E_{\rm flare}$ is determined by the magnetic energy. The derived relation is in the formula of
\begin{eqnarray}
L_{\rm loop} &\sim& 1.64\times 10^{9} \left( \frac{\tau_{\rm WL}}{100 \rm s} \right)^{2/5} \left( \frac{E_{\rm WLF}}{10^{30} \rm erg} \right)^{1/5} \left( \frac{n}{n_{\odot}} \right)^{-1/5} \rm cm,
\end{eqnarray}
where the coefficient was calibrated by solar coronal loop observation, $\tau_{\rm WL}$ is the e-folding decay time of WLFs, $E_{\rm WLF}$ is the bolometric WLF energy, and $n$ is the pre-flare coronal number density.
Note that the dependence on $n$ is not explicitly included in some previous studies \citep[e.g.,][]{2022ApJ...926L...5N}, but we included it because we need to consider it to discuss the results consistently with other methods (cf., \citealt{1999ApJ...526L..49S,2002ApJ...577..422S}'s model assumed the pre-flare coronal density).
The normalized factor of $n$ is assumed as solar value $n_\odot$, which is usually considered as $\sim10^9$ cm$^{-3}$ . Under the assumption of $n \sim 100n_{\odot} \sim 10^{11}$ cm$^{-3}$ \citep{1995A&A...301..201G}, the loop length of flare E1, E2, and flare E3 are derived to be 0.73$\times10^{10}$ cm, 1.16$\times10^{10}$ cm, and 0.67$\times10^{10}$ cm, respectively (see, Table \ref{tab:length-scale}).

In summary, the flare loop length estimated from X-ray and WL flares are listed in Table \ref{tab:length-scale}.
The estimated loop length is roughly $\sim 0.5-2\times10^{10}$ cm, which corresponds to $\sim 8-30$ \% of stellar radius.
We found that the loop lengths derived from the X-ray rise time by  \cite{2007AA...471..271R}'s methods are consistent with those derived from WLFs energy-duration relationship proposed by \cite{2017ApJ...851...91N}. 
The method proposed by \cite{1999ApJ...526L..49S,2002ApJ...577..422S} also yields a close value that are approximately twice as large as those from other methods for flare E3 (excluding flare E2 where the flare peak was not captured). 
Thus, these consistency supports the prediction capability of these different methods which are widely used but not often simultaneously applied.
Particularly, in this campaign, as all flares have been observed by \textit{TESS}, the estimations from WLFs by \cite{2017ApJ...851...91N} appear to be very useful to compare between flares

The length of the flare loops derived here will be compared with the length scale of prominences and starspots in the following Section \ref{sec:picture}. 
This will enable new discussions on the spatial scale ratios of different aspects of eruptive events. 
Moreover, the length of the flare loops is important for constraining the initial height of prominence eruptions, and is expected to be useful in future comparisons with numerical calculations in our paper II (see Section \ref{sec:future}).

\section{Prominence Eruptions on Young Solar-type Star}\label{sec:5}

\subsection{Discovery of Blueshifted ``Emission" Spectra as Evidence of Stellar Prominence Eruptions}\label{sec:5:disc}

As revealed in Section \ref{sec:3}, among the three H$\alpha$ superflares we detected (labeled E1, E2, E3), 
two superflares (E1, E2) exhibited a significant blueshifted H$\alpha$ ``emission" profile.
This is the first discovery of the blueshifted H$\alpha$ emission profile on solar-type star (G dwarf), while similar emission profile has been frequently reported on M dwarfs.
The profile exhibits a more prominent blueshifted component compared to those observed in M-type stars: In flare E1, the entire spectrum is blueshifted in the initial phase, and in flare E2, a bump-like structure is observed at a wavelength different from the H$\alpha$ center. This suggests that something different from the usual phenomena seen at the H$\alpha$ center (i.e. flare ribbons) is occurring and moving in the LOS direction.
As mentioned in the introduction, \cite{2022NatAs...6..241N} detected a blueshifted ``absorption" component moving at a maximum velocity of approximately --510 km s$^{-1}$ after a superflare (labeled as flare ``E4" in Table \ref{tab:prominence-1} and \ref{tab:prominence-2}), and comparison with solar observations confirmed that this was due to the filament eruption occurring inside the stellar disk. 
Our new discovery in this study proposed a picture similar to the Sun, where on a solar-type star EK Dra, the H$\alpha$ line appears as an absorption component when cool plasma are present within the stellar disk (called filament) and as an emission component when they extend beyond the disk (called prominence). Given this context, the blueshifted emission component detected this time could likely be the first evidence of prominence eruption on a solar-type star (i.e. a G dwarf).

The temporal variation of the velocity strongly supports that the observed blueshifted emission component are stellar prominence eruptions.
From the temporal evolution of the shifted velocity, it is possible to derive the initial deceleration rate (discussed in more detail in Section \ref{sec:5-2}). 
For flare E1, the deceleration rate is calculated to be 0.34$\pm$0.15 km s$^{-2}$ for the one component fit (or 0.23$\pm$0.14 km s$^{-2}$ for the two component fit). 
This corresponds remarkably well with EK Dra's surface gravity of 0.30$\pm$0.05 km s$^{-2}$, strengthening the picture that a prominence-like cool plasma is the ejected and decelerated by the surface gravity of the star.
Interestingly, this value is very similar to those of the filament eruption events E4 of 0.34$\pm0.04$  km s$^{-2}$ \citep{2022NatAs...6..241N}, indicating the similar phenomenon is occurring except for their radiation signature.
As for flare E2, the deceleration rate comes out as 0.12$\pm$0.20 km s$^{-2}$. Although the error is large due to the time resolution of the spectral fit and the weakness of the signal, the order of magnitude matches the surface gravity.

What are other possible causes for the observed blueshift emission profile? 
First, the active regions/quiescent prominences moving in LOS due to stellar rotation \citep[cf.][]{2020MNRAS.491.4076J} can potentially contribute to the blueshift, but given that the stellar rotational velocity ($v$sin$i$) is 16.4$\pm$0.1 km s$^{-1}$, a possibility of the sudden appearance of active regions/quiescent prominences due to stellar rotation can be rejected. 
For instance, the co-rotation radius at a velocity of 500 km s$^{-1}$ is 30 stellar radius. Given the sudden occurrence of blueshift associated with the WLF originating from the stellar surface, events at 30 stellar radii are likely negligible.
Second, the presence of a cool plasma above the chromospheric evaporation flows can cause blueshifts as discussed in flare studies from M-type stars \citep{2018PASJ...70...62H}. Very rarely in solar flares, blueshifted emission line profiles of $<100$ km s$^{-1}$ are observed in spatially resolved flare ribbon H$\alpha$ spectra \citep{2018PASJ...70..100T,1962BAICz..13...37S}.
The solar observations are interpreted by a process where high energy particles penetrate deep into the chromosphere causing evaporation and cool plasma in the upper chromosphere is accelerated before being heated \citep{2018PASJ...70..100T}. 
However, it is difficult to explain the phenomena we observed in EK Dra (especially E1) with this process. 
The occurrence of evaporation implies simultaneous chromospheric heating and condensation flows, with chromospheric radiation from denser plasma expected in the H$\alpha$ core or redshifted components. However, in flare E1, initially almost all H$\alpha$ emission components are blueshifted, and even when it decelerates and overlaps with the line core, the blueshifted component's EW is a dominant feature, which contradicts the expected time evolution. 
Therefore, it is difficult to explain the observations (especially E1) caused by evaporated flows.

Furthermore, as mentioned in Section \ref{sec:4-2}, for flare E1, the total radiation energy of the H$\alpha$ line (including the blueshifted component and line center component) is exceptionally large, around $\sim$30 \% relative to the WL energy, which is unusual considering the values of other EK Dra flares E2$\sim$E5 ($\sim$1 \%).
This suggests that the most of the H$\alpha$ radiation for flare E1 comes from a source spatially different from the WLF source, i.e. an erupted prominence.

Based on the above discussion, we concluded that we discovered ``prominence" eruptions on a solar-type star for the first time (undoubtedly for flare E1). 
Like the Sun, our study has made it clear that on solar-type stars, it is possible to determine whether the prominence/filament has erupted beyond/within the stellar visible disk based on whether it appears as an emission/absorption (see, summary in Section \ref{sec:diversity}).
In the following, we will conduct a more detailed analysis of the prominence eruptions observed in this study.

\subsection{Properties of Prominence Eruptions}\label{sec:5-2}


\subsubsection{Velocity, Velocity Dispersion, and Their Time Evolution}\label{sec:5-2-1}

The observational properties of the blueshifted component are estimated in Section \ref{sec:3:ha-1} and \ref{sec:3:ha-2}. 
Here, we review the obtained results in terms of the blushifted velocity, velocity dispersion, and their time evolution.

\begin{itemize}
\item[(E1)] This flare was the most remarkable, fastest, and long-lived prominence eruption.
The $\Delta$EW of the blueshifted component was 0.45--0.70 {\AA}.
The duration was 34$_{\pm 2}$ min, appearing at the start of the white-light flare or possibly earlier (the duration of the white light was 20 min). 
The maximum blueshifted velocity is 330$_{\pm 35}$--690$_{\pm 92}$ km s$^{-1}$  for one component fit (490$_{\pm 33}$--690$_{\pm 93}$ km s$^{-1}$ for two component).
No significant acceleration time was observed, instead, it showed an exponential decay of velocity, with an initial deceleration rate of 0.34$_{\pm0.15}$ km s$^{-2}$ (already discussed in Section \ref{sec:5:disc} in comparison wih surface gravity). 
The velocity eventually approach to zero velocity, and emission component disappeared without showing significant redshifts.
The typical velocity dispersion was 300 km s$^{-2}$, exhibiting a pattern of decay synchronous with the velocity.
\item[(E2)] This flare was the weak, slowest, short-lived, prominence eruption.
The $\Delta$EW of the blueshifted emission component was 0.034 {\AA}.
The duration was 10$_{\pm 5}$ min, appearing at the peak of the WLF. 
The maximum velocity is 430$_{\pm 19}$ km s$^{-1}$
No significant acceleration and deceleration phase was observed, although possible deceleration was estimated as 0.12$_{\pm0.20}$ km s$^{-2}$. The typical velocity dispersion was 55 km s$^{-2}$.
\end{itemize}

\begin{figure}
\epsscale{0.5}
\plotone{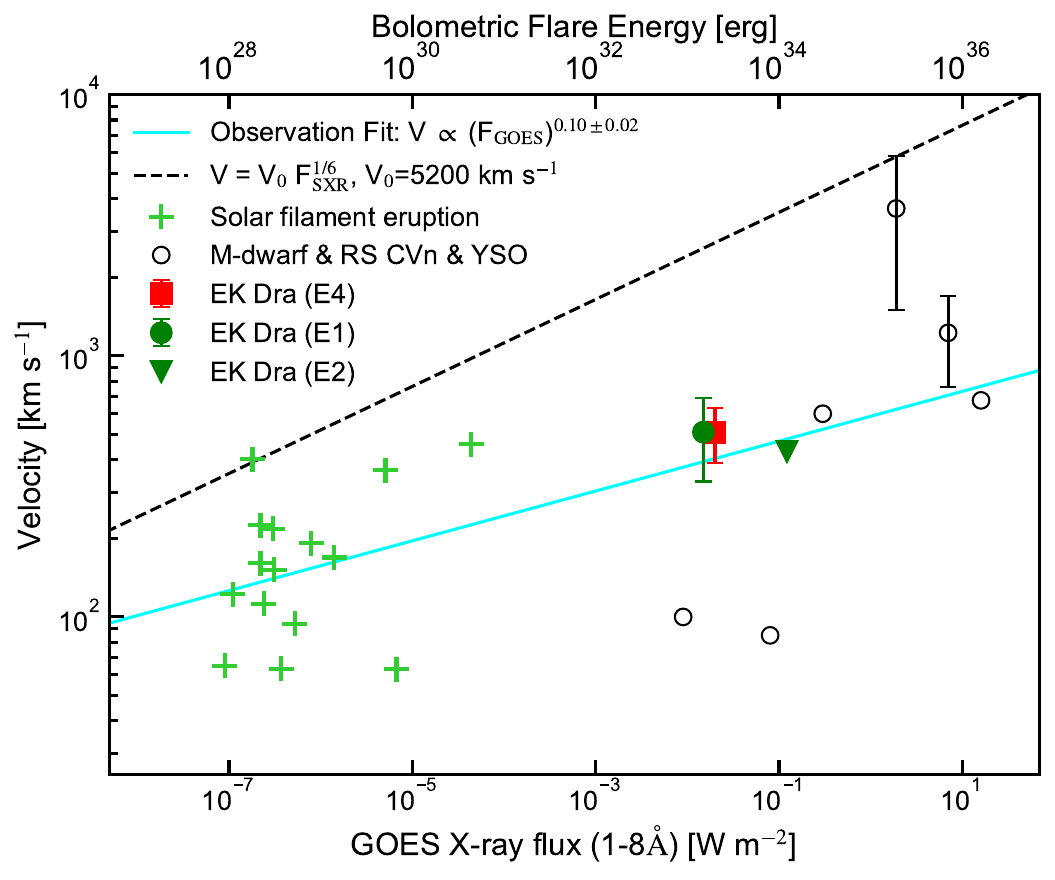}
\caption{Velocity of solar and stellar prominence/filament eruptions (blueshift events) as a function of flare GOES flux (bottom axis) or bolometric WLF energy (top axis). 
Data of filament/prominence eruptions on the G-dwarf EK Dra are derived from this study and \cite{2022NatAs...6..241N}. 
Data of solar filament eruptions are taken from \cite{2021EP&S...73...58S}. 
Data for M-dwarfs, RS CVn, and YSO are taken from \cite{2019ApJ...877..105M}, \cite{2023ApJ...948....9I}, and \cite{2020PASJ..tmp..253M}. 
The cyan solid line is a power-law fitting line for all the solar and stellar data. 
The black dashed line is a line with a power-law index of 1/6 which represents the upper edge of the solar and stellar data \cite{2016ApJ...833L...8T}.}
\label{figs:19}
\end{figure}

How can we understand the velocity in the relation of solar flares and blueshift events on other-types of stars?
Figure \ref{figs:19} shows the velocity of prominence/filament eruptions (blueshift events) as a function of flare's GOES-band flux
(bottom axis) or bolometric WLF energy (top axis) for solar and stellar flares.
The velocity of prominence/filament eruptions on EK Dra is comparable to the largest values of solar prominence eruptions while some stellar prominence eruptions from ``megaflares" (10$^{35-36}$ erg) on RS CVn-type stars/M-dwarfs seems to have a much larger velocity than solar flares \citep{1990A&A...238..249H,2023ApJ...948....9I}.
It looks that there is a weak positive trends if we include solar and stellar data as
\begin{eqnarray}
V & = & 590 \cdot (F_{\rm GOES})^{0.10\pm0.02}. 
\end{eqnarray} 
Note that this correlation may be influenced by observational bias because it is likely that only those with sufficiently high velocities are being detected for stellar flares (due to spectral resolution, flare contamination, and LOS uncertainty).
Therefore, this trend may be an apparent one and there can be a large diversity in velocity.

Given the fact that some stellar superflares exhibit prominence eruptions at much higher velocities than the Sun, it can be anticipated that the upper limits of stellar observational data is of significance.
\cite{2016ApJ...833L...8T} suggested that the upper limit of the CME speed can be theoretically described by the scaling law of $V_{\rm CME,upper} =  V_0 \cdot (F_{\rm GOES})^{1/6}$, 
based on the assumption that the flare radiation energy $E_{\rm WL,bol}\sim 10^{35} F_{\rm GOES}$ should be roughly less than CME kinetic energy $0.5\cdot M_{\rm CME}\cdot V_{\rm CME,upper}^2$.
Recently, \cite{2023ApJ...943..143K} suggested the power-law index for the relation between erupted mass and flare energy scale is the same for CMEs and prominence eruptions, i.e., $M\propto F_{\rm GOES}^{2/3}$, therefore a similar scaling law for the upper limit of the prominence velociy $V_{\rm p}$ can be also derived by following the \cite{2016ApJ...833L...8T}. 
\begin{eqnarray}
F_{\rm GOES} & \propto & \frac{1}{2}M_{\rm p} V_{\rm p, upper}^2 \propto F_{\rm GOES}^{2/3} V_{\rm p, upper}^2,  \\
V_{\rm p, upper} & = & V_0 \cdot (F_{\rm GOES})^{1/6}. \label{eq:velupper} 
\end{eqnarray} 
where $F_{\rm GOES}$ can be replaced with $10^{-35} E_{\rm WL,bol}$, $V_0 \sim$5200 km s$^{-1}$ is roughly determined by observational upper limit for solar flares/filament eruptions.
In stellar flares, there have been no observed events exceeding this empirical scaling in velocity (Equation \ref{eq:velupper}), implying that this established solar scaling law poses no contradictions in describing the upper limits of velocities for filament/prominence eruptions observed in the case of superflares.

Lastly, we would like to focus on a new aspect, the velocity dispersion of the blueshifted components. 
For flare E1, we discovered a clear decay in velocity dispersion (Figure \ref{fig:10}). 
Furthermore, it was found to have a positive correlation with the absolute value of the velocity (Figure \ref{fig:11}). 
Given that the spatial distribution of thermal and turbulent velocities of prominences are on the order of several tens of km s$^{-1}$, similar to the Sun \citep[e.g.,][]{2018PASJ...70...99S,2022NatAs...6..241N}, we cannot explain a velocity dispersion of approximately 300 km s$^{-1}$ of flare E1 and E4, even considering the velocity dispersion of the instrumental broadenings ($\sim$75 km s$^{-1}$; \citealt{2019PASJ...71..102M}). 
Actually, this broad velocity dispersion has also been detected in M-dwarf flares \citep{2016A&A...590A..11V,2022MNRAS.513.6058L}. 
Here, by assuming that prominence eruptions are a phenomenon in which loop structures expand self-similarly, we propose that this velocity dispersion may be observing components from spatially different regions. 
In other words, the slower parts could be near the foot of the loop-like prominence, and the faster parts near the top. 
Considering this self-similarly expanding loop structure, the fact that velocity dispersion has a positive correlation with the average speed is also reasonable. 
If this is true, the obtained central velocity of the prominence could be just an average one, with the velocity dispersion indicating the existence of both faster and slower plasma. 
Therefore, we suggest that the maximum speed of flare E1 could reach up to 630--990 km s$^{-1}$ (or even up to 880--1080 km s$^{-1}$), considering the velocity dispersion.

These larger velocities exceeds the escape velocity of EK Dra $\sim$670 km s$^{-1}$, and it can be suggested from the H$\alpha$ line observation alone that flare E1 (and even flare E2) could be connected to a CME.
The possibility that these are connected to CMEs will be discussed in further detail combined with other indicator in X-ray in Section \ref{sec:cancme}.

\subsubsection{Area and Length Scale}\label{sec:5-2-2}

Here, we estimate the area and length scale of the prominence.
We basically follow the method described by \cite{2023ApJ...948....9I}, but the assumption of optical depth and electron density is original based on the discussion in Section \ref{sec:5-2-4}. 
The luminosity of the blueshift component ($L_{\rm H\alpha,blue}$) at the time of maximum EW is derived as 16.6$_{\pm1.7}\times 10^{28}$ erg s$^{-1}$ for flare E1 and 0.47$_{\pm1.6}\times 10^{28}$ erg s$^{-1}$ for flare E2 as shown in Table \ref{tab:prominence-1}, based on the results of the Gaussian spectral fit. 
Assuming an optical depth $\tau=10\sim1000$ for our prominences (see, Section \ref{sec:5-2-4} for the assumption), we derive the area, length scale, mass, and kinetic energy based on the theoretical correlations between prominence plasma parameters and the emitted radiation derived by \cite{1994A&A...292..656H}.
Note that \cite{1994A&A...292..656H}'s work is based on the model of solar quiescent prominence, and the parameters in eruptive promiences on active stars could be different, although further radiative transfer modeling is beyond the scope of this paper.
The assumption of the optical depth $\tau=1\sim100$ gives the theoretical value of surface flux of the prominence $F_{\rm blue}$ as
\begin{eqnarray} \label{eq:tau-f}
F_{\rm blue}
&=&  \left\{
\begin{array}{ll}
 10^{\log_{10}(\tau) + 5} & ( \log_{10}(\tau) < -0.22 ) \\
 10^{0.55\cdot \log_{10}(\tau) + 4.9} & ( \log_{10}(\tau) > -0.22 )
\end{array} 
\, \right . \\
&=&  \left\{
\begin{array}{ll}
 2.8\times10^{5} & ( \tau = 10 ) \\
 3.5\times10^{6} & ( \tau = 1000 )
\end{array}
\, {\rm [erg\, s^{-1}\, cm^{-2}\, sr^{-1} ]}. \right.
\end{eqnarray}
Using these values, the projected surface area of the prominence $A_{\rm p}$ can be expressed as
\begin{eqnarray}
A_{\rm p} & = & L_{\rm H\alpha,blue} / (2 \pi F_{\rm blue}). 
\end{eqnarray} 
This equation gives the surface area of the prominence as $1.8\times10^{22}$--$6.3\times10^{22}$ cm$^2$ for flare E1 and $5.3\times10^{20}$--$1.9\times10^{21}$ cm$^2$ for flare E2.
Furthermore, The length scale of the prominence $L_{\rm p}$ is derived as follows, assuming a simple hypothesis that the prominence is roughly square ($A_{\rm p}= L_{\rm p}^2$, the assumption of ``cubic" erupted prominences) for the lower limit and an aspect ratio of the length and width of the prominence of 1:10 for the upper limit ($A_{\rm p}= L_{\rm p} \cdot L_{\rm p}/10$, the assumption of ``cylinder"-like erupted prominences):
\begin{eqnarray} \label{eq:aspect}
L_{\rm p} & = & (A_{\rm p})^{0.5} \sim 10\cdot (A_{\rm p})^{0.5}. 
\end{eqnarray} 
The estimated length scale of the prominence is $(8.3-30)\times 10^{10}$ cm (1.3--4.5 $R_{\rm star}$) $\sim$ $(83-300)\times 10^{10}$ cm (13--45 $R_{\rm star}$) for flare E1 and $(1.4-5.1)\times 10^{10}$ cm (0.2--0.8 $R_{\rm star}$) $\sim$ $(14-51)\times 10^{10}$ cm (2.2--7.8 $R_{\rm star}$) for flare E2.
Since we used the maximum EW value for calculations, the observed length scale would be the one after the prominence is erupted and expands, so it would be possible that the estimated length scale of prominence is larger than the stellar radius.

How significant the assumption of aspect ratio is?
The correction of the prominence aspect ratio with Equation \ref{eq:aspect} tend to give extremely large length scale.
This may indicate that the prominences could be not necessarily a filamentary structure after the eruptions and could look like a halo-type prominence eruptions \citep[cf.][]{2014LRSP...11....1P}.
Anyway, we do not know at all the shape of the stellar prominence for now, so to include all the possibilities, we use the values including the aspect ratio correction as an upper limit of the prominence length scale, following the methods in \cite{2022NatAs...6..241N}.
Actually, this assumption does not significantly affect the following discussion of length scale because the lower limit value is important, but will be revisited when we discuss the general picture of the relationship between flares, prominences, and starspots in Section \ref{sec:picture}.

\begin{figure}
\epsscale{0.5}
\plotone{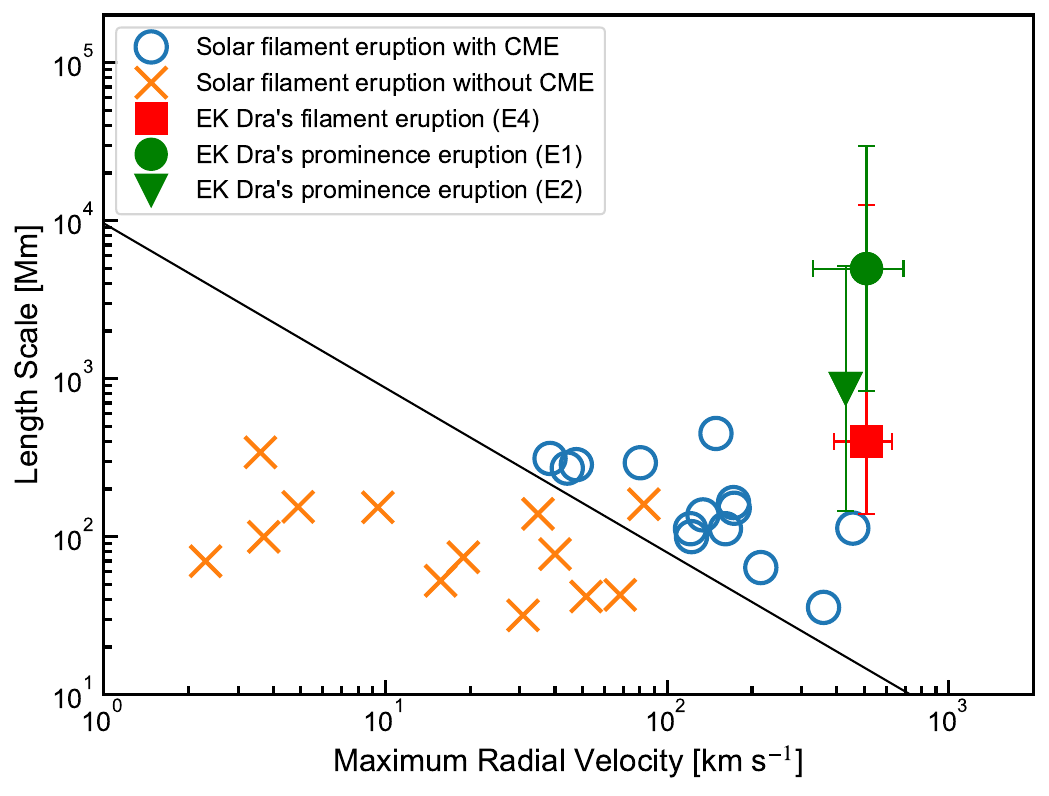}
\caption{Comparison between the velocity ($V_{\rm r\_max}$) and length scale ($L$) of filament/prominence eruptions on the Sun and G-dwarf EK Dra. The blue circles and orange crosses indicate the solar filament eruptions with and without CMEs, respectively. 
The green and red points corresponds to the stellar prominence and filament eruptions on EK Dra, respectively (this study, \citealt{2022NatAs...6..241N}). 
The stellar data are the maximum line-of-sight velocities while the solar data are maximum radial velocity.
The solid black indicates an empirical solar threshold that can roughly distinguish whether a filament eruption is associated with CMEs or not in the formula of  ($V_{\rm r\_max}$/100 km s$^{-1}$)($L$/100 Mm)$^{0.96}$ = 0.8 \citep{2021EP&S...73...58S}. 
}
\label{figs:20}
\end{figure}

Figure \ref{figs:20} shows the plot of the prominence length scale and the maximum radial velocity.
In the figure, the prominence/filament eruptions of EK Dra are compared with solar filament eruptions \citep{2021EP&S...73...58S}. \cite{2021EP&S...73...58S} proposed a criterion, namely 
\begin{eqnarray}
(V_{\rm r\_max}/100 \, {\rm km s^{-1}})(L_{\rm p}/100 \, {\rm Mm})^{0.96} = 0.8,
\end{eqnarray} 
to determine whether a solar prominence eruption is associated with a CME, where $V_{\rm r\_max}$ is the maximum radial velocity of filament eruption. 
In other words, the longer the prominence and the higher the velocity (i.e., the larger the kinetic energy), the higher the likelihood of it leading to a CME. 
This threshold was first applied to stellar filament eruptions on EK Dra by \cite{2022NatAs...6..241N}. 
While this threshold is fundamentally applicable to the Sun, EK Dra has similar surface gravity, surface temperature, and atmospheric stratification as the Sun, suggesting the potential applicability of this threshold. 
As can be seen in the figure, even the lower limit values of prominence eruptions E1 and E2 are positioned well above this criterion. 
This supports the possibility that the prominence eruptions associated with flares E1 and E2 may ultimately lead to stellar CMEs.
Note, however, that the different activity levels between the Sun and EK Dra could potentially affect relative change in this criterion. Future theoretical investigations are required for this threshold.

\subsubsection{Mass and Kinetic Energy}\label{sec:5-2-3}

Here, under the assumption mentioned above, with an optical depth of $\tau=10\sim1000$ for the prominences, we determine the prominence mass $M_{\rm p}$. 
The mass can generally be calculated as $M_{\rm p} \sim A_{\rm p} D n_{\rm{H}} m_{\rm{H}}$, where the spatial LOS depth of the prominence as $D$, the number density of hydrogen as $n_{\rm H}$, and the mass of hydrogen as $m_{\rm H}$. Here, based on the numerical calculations by \citet{1994A&A...292..656H}, a relationship has been derived between the optical depth of the prominence and the emission measure $EM_{\rm blue}$, as follows:
\begin{eqnarray}
EM_{\rm blue} = D n_{\rm e}^2 =   \left\{
\begin{array}{ll}
8.1\times 10^{29}  & ( \tau = 10 ) \\
3.3\times 10^{32}  & ( \tau = 1000 )
\end{array}
\, {\rm [cm^{-5}]} \right. \label{eq:em-ha}
\end{eqnarray}
Here, based on the assumption of ``cubic" or ``cylinder"-like structure introduced in Section \ref{sec:5-2-2}, we assume that the depth of the prominence is approximately equal to $L_{\rm p}$ (cubic) or $0.1\times L_{\rm p}$ (cylinder).
By using the emission measure $EM_{\rm blue}$ value under the assumed optical depth, we can describe the $M_{\rm p}$ as
\begin{eqnarray}
M_{\rm p} &\sim& A_{\rm p} D n_{\mathrm{H}} m_{\mathrm{H}} \\
        &=& A_{\rm p} D \bigg(\frac{EM_{\rm blue}}{D}\bigg)^{0.5} \bigg(\frac{n_{\rm{e}}}{n_{\rm{H}}}\bigg)^{-1} m_{\mathrm{H}}
        \ , 
\end{eqnarray} 

Here we roughly assume the prominence ionization fraction 
$n_{\rm{e}}/n_{\rm{H}}= 0.17 \sim 0.47$ (cf. Table 1 of \citealt{2010SSRv..151..243L}).
Note that this value is modified in \cite{Notsu2023} because the original values used in \cite{2023ApJ...948....9I} was a mistake.
As a result, the mass of prominences are 
$1.3^{+2.9}_{-0.9}\times 10^{20}$ g (i.e., $4.0\times10^{19}$ g -- $4.2\times10^{20}$  g) and 
$1.7^{+3.7}_{-1.1}\times 10^{18}$ g (i.e., $5.2\times10^{17}$ g -- $5.3\times10^{18}$  g) for flare E1 and E2, respectively.
Here, the average of the upper and lower limits in the log scale is denoted as the representative value (the same applies to kinetic energy).


Also, the prominence kinetic energy can be derived from
\begin{eqnarray}
E_{\rm kin} & = & \frac{1}{2} M_{p} V_{\rm typical}^2, 
\end{eqnarray} 
where $V_{\rm typical}$ is the blueshifted velocity at the maximum H$\alpha$ EW.
As a result, the kinetic energy of prominence for E1 and E2 eruptive event are 
$5.8^{+12.8}_{-4.0}\times 10^{34}$ erg (i.e., $1.8\times 10^{34}$ erg -- $1.9\times10^{35}$  erg) 
and $1.2^{+2.7}_{-0.8}\times 10^{33}$ erg (i.e., $3.8\times10^{32}$ erg -- $3.9\times10^{33}$ erg), respectively.
Thus, the kinetic energy of the prominence is about three times larger for the E1 event ($E_{\rm WL,bol}=1.5\times 10^{33}$ erg) and about fifty times smaller for the E2 event ($E_{\rm WL,bol}=12.2\times 10^{33}$ erg) than the radiative energy of the WLF ($E_{\rm WL,bol}$) (see, Figure \ref{figs:20}(b)).
This indicates a diversity depending on the flares; flare E1, which has a smaller energy scale for the WLF than flare E2, is larger in all aspects such as mass, speed, and kinetic energy (see the discussion on diversity in Section \ref{sec:diversity}).

\begin{figure}
\gridline{
\fig{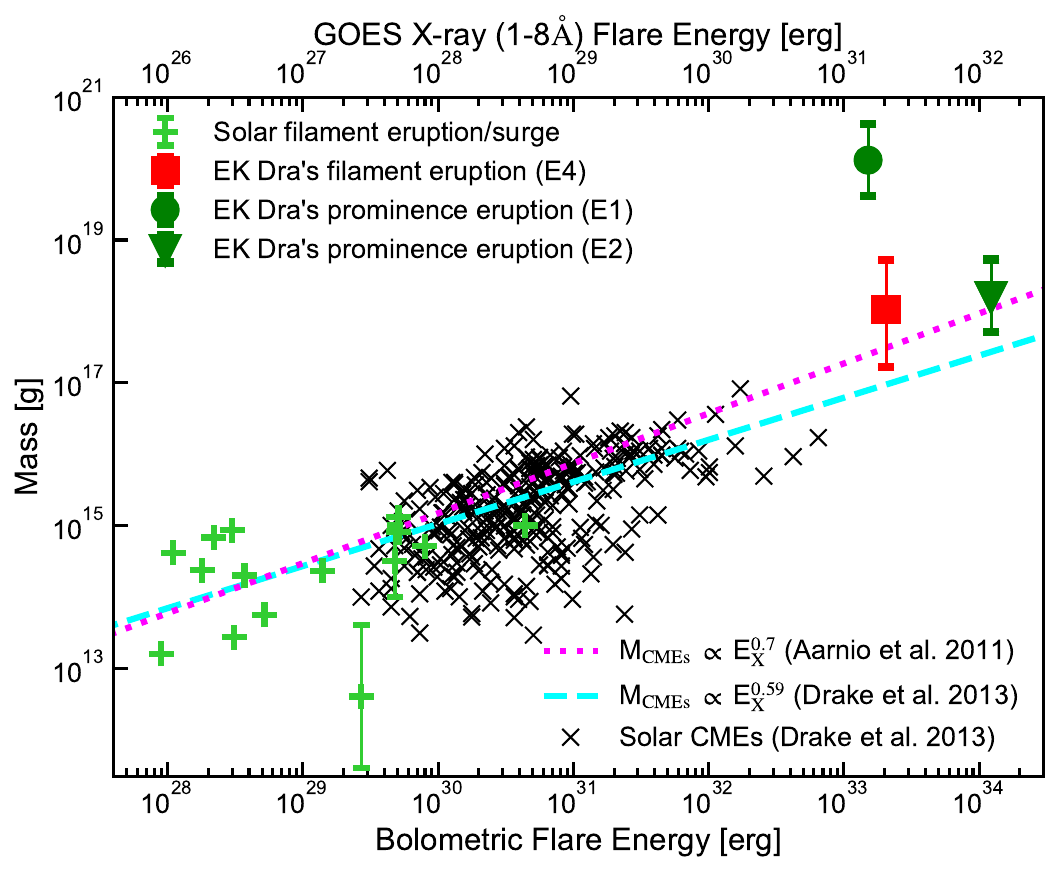}{0.5\textwidth}{\vspace{0mm} (a)}
\fig{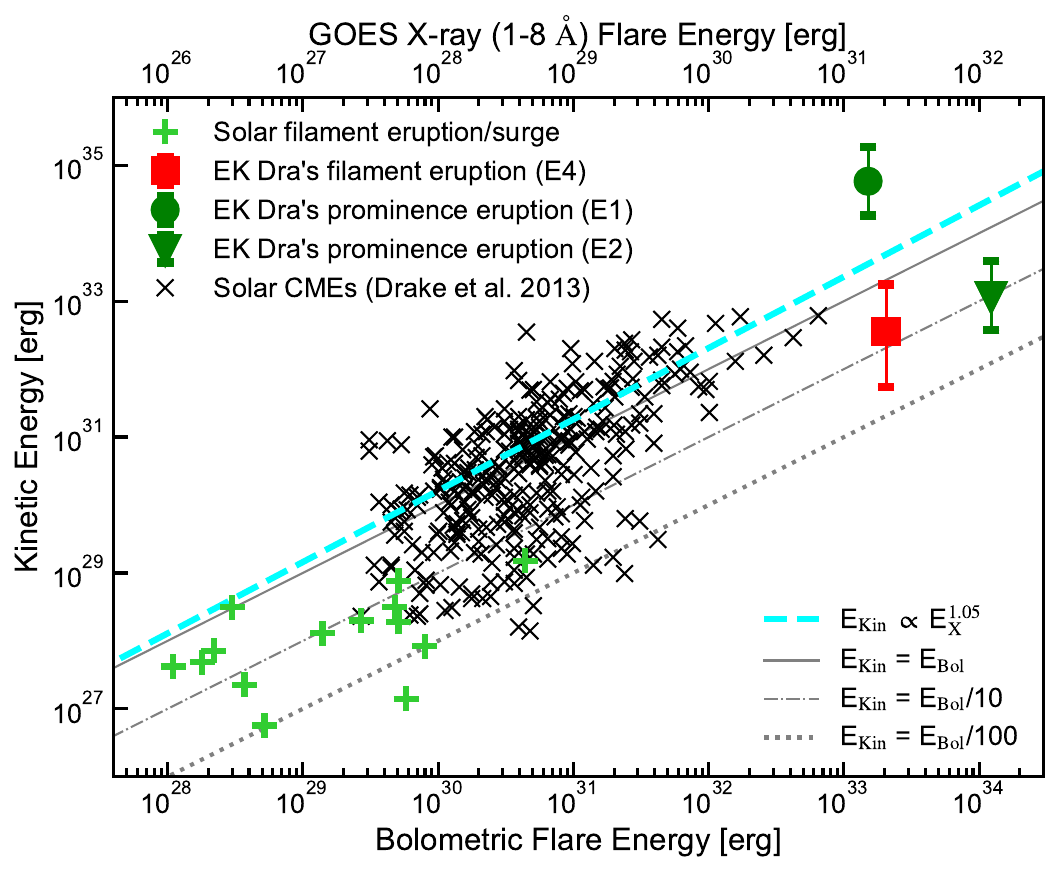}{0.5\textwidth}{\vspace{0mm} (b)}
}
\caption{
Mass and kinetic energy of solar/stellar filament/prominence eruptions and CMEs as a function of bolometric WLF energy or GOES X-ray energy.
The green and red points correspond to the stellar prominence and filament eruptions on EK Dra, respectively (this study, \citealt{2022NatAs...6..241N}). 
The black crosses are solar CME data \citep[][]{2009IAUS..257..233Y,2013ApJ...764..170D} while the green pluses are data for solar prominence/filament eruptions taken from \cite{2022NatAs...6..241N} and \cite{2023ApJ...943..143K}.
The cyan dashed and magenta dotted lines are observational fits for solar CMEs expressed as $M_{\rm CME} \propto E_{\rm bol}^{0.59}$ and $E_{\rm kin} \propto E_{\rm bol}^{1.05}$ \citep[cyan color;][]{2013ApJ...764..170D} and $M_{\rm CME} \propto E_{\rm bol}^{0.7}$ \citep[magenta color;][]{2012ApJ...760....9A}. 
The relations of $E_{\rm kin} = E_{\rm bol}$, $E_{\rm kin} = E_{\rm bol}/10$, and $E_{\rm kin} = E_{\rm bol}/100$ are also plotted with gray color in the right panel.
Here, GOES X-ray energy/flux is related to bolometric energy with the relations of $E_{\rm bol}$ = 100 $E_{\rm GOES}$ \citep{2012ApJ...759...71E} and $E_{\rm bol}$ [erg] = 10$^{35}$ $F_{\rm GOES}$ [W m$^{-2}$] \citep{2013PASJ...65...49S} to connect solar data to stellar data. 
Also, we assume bolometric energy $\approx$ bolometric WLF energy \citep{2010NatPh...6..690K,2011A&A...530A..84K,2015ApJ...809...79O}.
}
\label{figs:21}
\end{figure}

Figure \ref{figs:21} shows the mass and kinetic energy as a function of flare radiated energy for flare E1, E2, and E4 in comparison with solar filament eruptions/CMEs \citep{2022NatAs...6..241N,2023ApJ...943..143K} and their empirical scaling relations \citep{2012ApJ...760....9A,2013ApJ...764..170D}.
Here, we focus comparisons among solar-type stars, the Sun and EK Dra, to show a solar-stellar comparison without much different factors such as surface gravity, atmospheric density, etc, while other studies has already focused on the comparison among different spectral types.
The original figures are taken from \cite{2022NatAs...6..241N}, and here we added two newly discovered stellar prominence eruptions on EK Dra in green filled symbols.
In addition, small-scale solar filament eruptions/surges investigated by \cite{2023ApJ...943..143K} were newly added for increasing samples in small energy regime ($10^{28-30}$ erg).

We found that the mass of the gigantic prominence eruption in flare E1 exceeds the predictions of solar empirical scaling relations between flares and CMEs, while the measurements for E2 and E4 are consistent with these relations. 
Our results suggest that, like solar CMEs/filament eruptions, the ejected mass against a given flare energy scale exhibits considerable scattering in the case of young Sun-like stars, spanning one to two orders of magnitude.
\cite{2023ApJ...943..143K} theoretically suggested that the scattering may be explained by difference in magnetic field strength supporting the prominence.
Hence, our findings in diversity in mass (and kinetic energy below) may be related to the diversity in local magnetic field strength of the star.

Also, the kinetic energy of E1 event aligns with or exceeds the solar flare energy-CME kinetic energy scaling relation. 
In contrast, the kinetic energies of E2, E4 events and solar filament eruptions/surges are smaller by one to two orders of magnitude than the scaling relation.
The smaller kinetic energies associated with solar-stellar prominence/filament eruptions have been explained in the context of two scenarios, as discussed in \cite{2022NatAs...6..241N}. 
The first explanation posits that the velocity of stellar CMEs and prominence/filament eruptions can be suppressed by an overlying robust magnetic field, leading to a smaller allocation of kinetic energy \citep{2018ApJ...862...93A}. 
The second explanation attributes it to the different velocity between CMEs and prominence or filament eruptions \citep{2020PASJ..tmp..253M,2022NatAs...6..241N}.
Solar prominence or filament eruptions are represented by expanding dense and cool lower portion of the self-similarly expanding flux rope. 
Consequently, the velocity of prominence or filament eruption is 4-8 times less than that of overlying CMEs \citep{2003ApJ...586..562G}. 
This perspective naturally accounts for the smaller kinetic energy observed in solar and stellar prominence or filament eruptions than solar CME's scaling relation.
However, given the significant scattering in solar data, the broad range observed in the kinetic energy of stellar prominence/filament eruptions on EK Dra appears plausible by solar analogy. 
The cause of this wide variation of solar and stellar prominences/CMEs warrants further investigation in the future.


\subsubsection{Optical Depth and Electron Density of Prominences}\label{sec:5-2-4}

Then, is the assumption of optical depth $\tau=10-1000$ appropriate? Table \ref{tab:length-scale-prom-E1} summarizes the estimated prominence length scales for different values on the optical depth $\tau$ of 0.1, 1, 10, 100, 1000, and 10000 as references.
Solar prominence eruptions often shows optical depth of $\tau\sim1$ in H$\alpha$ line \citep[e.g.,][]{2018PASJ...70...99S,2022NatAs...6..241N}.
The optical depth $\tau$ = 0.1 and 1 gives extremely large value of the length scale of at least $>$24 $R_{\rm star}$ and $>$8.5 $R_{\rm star}$, respectively, for flare E1.
These values look to be unrealistic considering that the prominence expanding velocity of $\sim$500 km s$^{-1}$ can reach up to only 2.4 stellar radius within 1 hour (although plane-of-sky prominence velocities might be larger than the LOS velocities).
Therefore, it would be reasonable to assume the optical depth of $\tau$=10 as a lower limit.

\begin{deluxetable*}{lccccccccc}
\label{tab:length-scale-prom-E1}
\tablecaption{Length scale of prominences in different prominence parameters for flare E1 and E2.}
\tablewidth{0pt}
\tablehead{
\colhead{} & \multicolumn{4}{c}{Prominence Length} \\
\colhead{} & \multicolumn{2}{c}{E1} & \multicolumn{2}{c}{E2}\\
\colhead{(optical depth)} & [$R_{\rm star}$] & [10$^{10}$ cm] & [$R_{\rm star}$] & [10$^{10}$ cm]
}
\startdata
$\tau = 0.1$ & 24--240  & 160--1600 & 4.1--41  & 27--270\\
$\tau = 1$ & 8.5--85  & 56--560 & 1.5--15  & 9.7--97\\
$\tau = 10$ & 4.5--45  & 30--300 & 0.78--7.8  & 5.1--51\\
$\tau = 100$ & 2.4--24  & 16--160 & 0.41--4.1  & 2.7--27\\
$\tau = 1000$ & 1.3--13  & 8.3--83 & 0.22--2.2  & 1.4--14\\
$\tau = 10000$ & 0.67--6.7  & 4.4--44 & 0.12--1.2  & 0.77--7.7
\enddata
\tablecomments{The lower value is no assumption of the prominence's aspect ratio while the upper value is derived with the assumption of the aspect ratio of the prominence.
}
\end{deluxetable*}

\begin{figure}
\epsscale{0.5}
\plotone{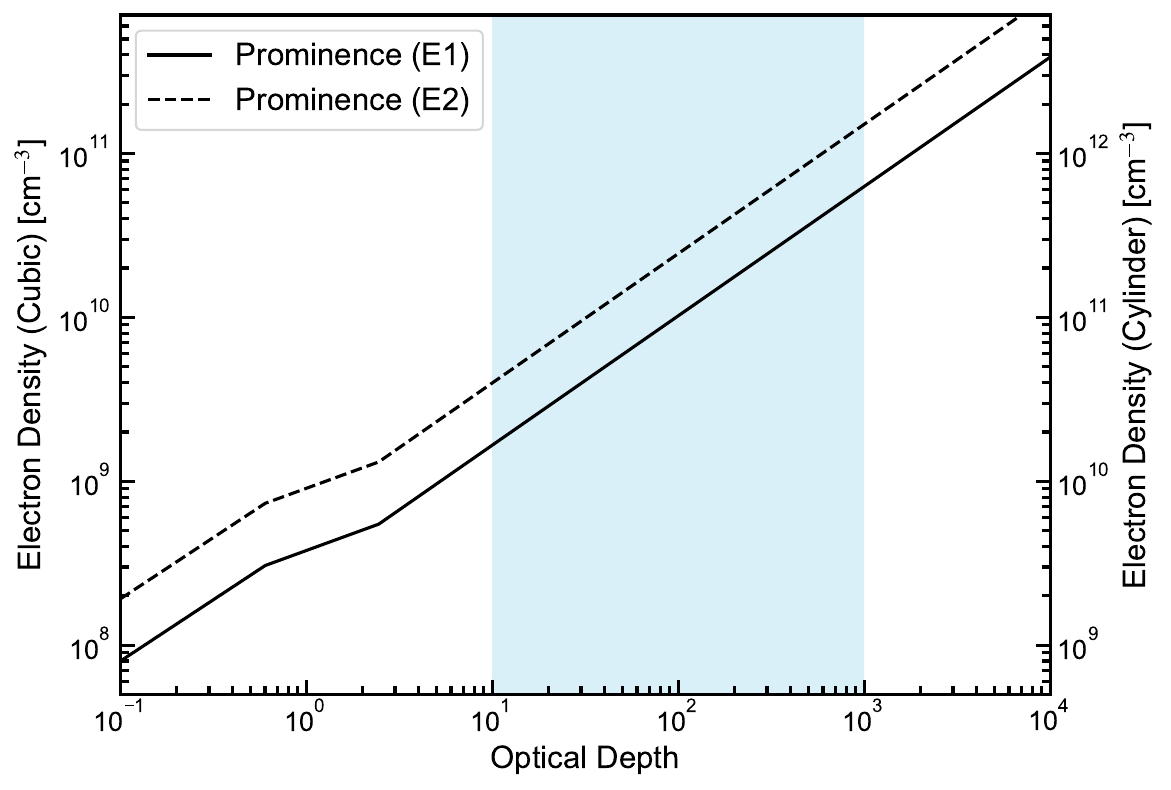}
\caption{
Estimated electron density as a function of optical depth. The left y-axis is the case when the prominence is cubic structure, while right y-axis is the cause when the prominence is cylinder (i.e., filamentary structure). The blue colored region corresponds to the assumed optical depth.
}
\label{figs:new23}
\end{figure}

Next, let us discuss the validity of the assumed optical depth from the consistensy in the electron density.
Figure \ref{figs:new23} shows the estimated electron density as a function of assumed optical depth. The electron densities are estimated from Equation \ref{eq:em-ha}.
As you can see, the optical depth of 10--1000 gives the electron densities of $1.7\times 10^{9}$ -- $6.3\times 10^{10}$ cm$^{-3}$. This is the actually consistent with the solar prominence observations of $10^{9-11}$ cm$^{-3}$ \citep{2014LRSP...11....1P}.
This consistency in electron density supports the assumed range of the optical depth of this study.

Then, what about the electron densities of stellar erupting prominences on active stars?
Past X-ray observations indicate that the electron density in the quiescent corona of EK Dra is at least $4\times 10^{10}$ cm$^{-3}$ with a temperature of $\sim 2\times 10^{6}$ K \citep{1995A&A...301..201G}.
This means that the gas pressure in outside corona is $\sim$40 times larger than solar ones because solar coronal electron density is $\sim$10$^{9}$ cm$^{-3}$ with the similar temperature.
Considering the gas pressure balance between outside corona and prominences, we can estimate that the prominence density in EK Dra can be $\sim 40$ times larger than the solar values ($10^{9-11}$ cm$^{-3}$), i.e., $\sim4\times10^{10-12}$  cm$^{-3}$.
The plasma $\beta$ of solar prominences are roughly $\sim$1, so this could be the upper limit value of the quiescent prominence electron density.
This value ($\sim4\times10^{10-12}$) could be a little bit larger than the estimated prominence electron density ($1.7\times 10^{9}$ -- $6.3\times 10^{10}$ cm$^{-3}$). 
However, since our prominence parameters are estimated from the H$\alpha$ luminosity of $\sim$10 minutes after the prominence initiation for flare E1, 
the prominence could be more dilute than the original value.
Under the assumption of the expansion velocity of 500 km s$^{-1}$ and the timescale of $\sim$10 minutes, the prominence could be expanded into $\sim3\times 10^{10}$ cm, which is $\sim$4 times larger than the flare loop size.
If we assume that the prominence expands in a cubic way, the volume could expand into $\sim$64 larger than the original volume, resulting in $\sim$1/64 of the original electron density, i.e., $\sim6\times10^{8-10}$ cm$^{-3}$. 
This expanded prominence case in active stars is approximately consistent with the electron density estimated from the optical depth and cubic shape (Figure \ref{figs:new23}).


In summary, the optical depth of 0.1$\sim$1 ($\sim$ solar values) have been doubtful considering the unrealistically large prominence size. The upper limit of the optical depth could be difficult to be constrained only from the size, but the assumed optical depth of $<$1000 well cover the realistic prominence electron density.
Table \ref{tab:length-scale-prom-E1} includes estimates with much more extreme optical depth $\tau$=10000 as a reference.
Since Equation \ref{eq:tau-f} shows that the dependence of optical depth on the length scale $L_{\rm p}\propto (\tau)^{\sim -0.25}$ is not so strong, the discussion will not significantly change even though we assume $\tau$=10000.

\color{black}

\subsection{Candidate of Possible X-ray Dimming after Flares and Prominence Eruptions}\label{sec:5-3}

\begin{figure}
\plottwo{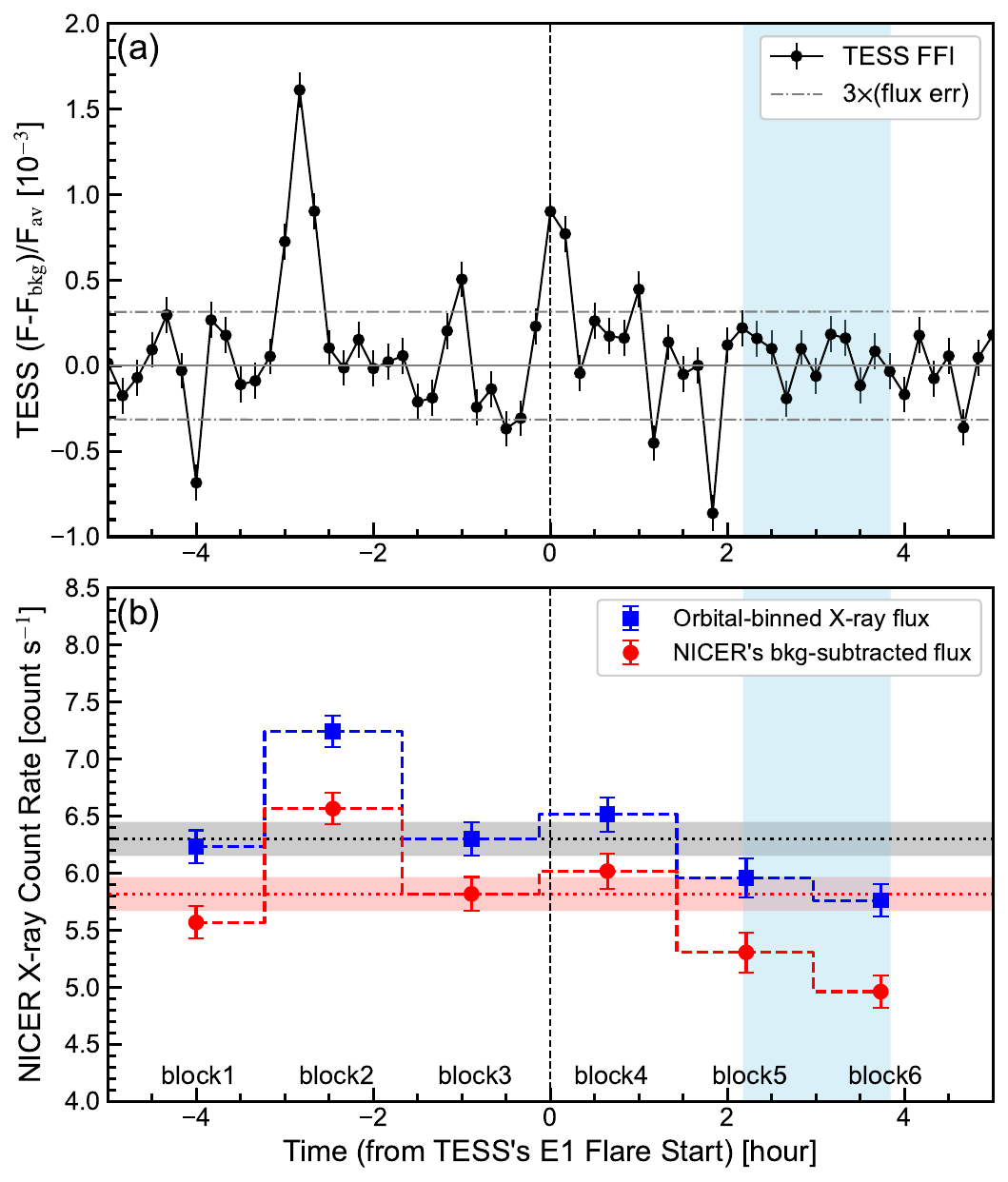}{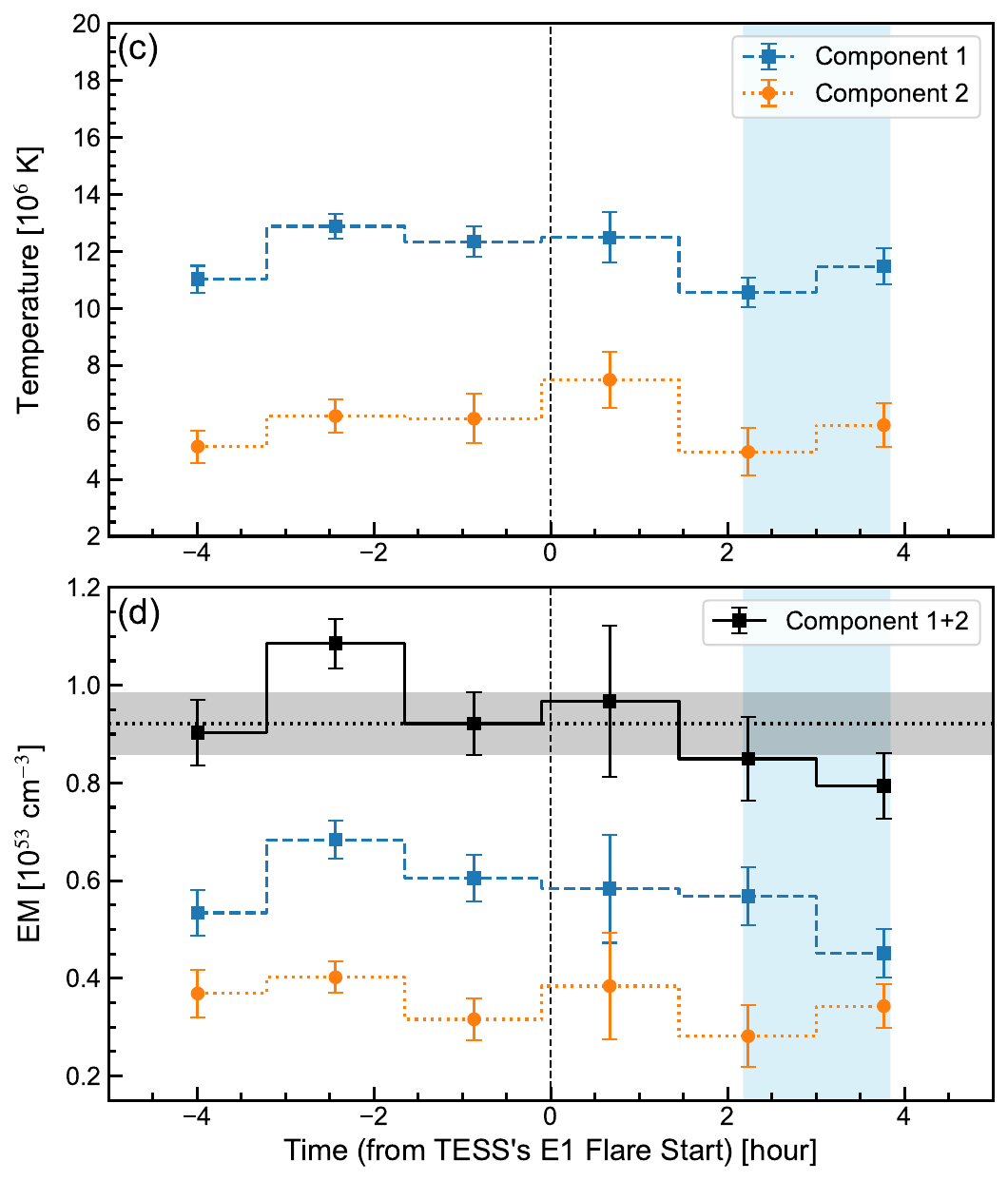}
\caption{Possible X-ray (0.5-3 keV) dimming after the gigantic prominence eruptions associated with the superflare on 2022 April 10 (E1).
(a) Reference TESS light curve.
(b) The X-ray light curve.
The blue points represent the average of the gray data with 2-minute resolution for each orbital period. The red points denote the count rate after NICER's background subtraction. 
The area filled in blue represents the time interval believed to show dimming relative to the pre-flare (block 3).
The area filled with gray and red represent the pre-flare level with error bars for blue and red points data, respectively.
(c), (d) The temporal evolution of the temperature and emission measure of two components. The area filled with gray in panel (d) represents the pre-flare level with error bars. }
\label{figs:22}
\end{figure}

\begin{figure*}[ht!]
\gridline{
\fig{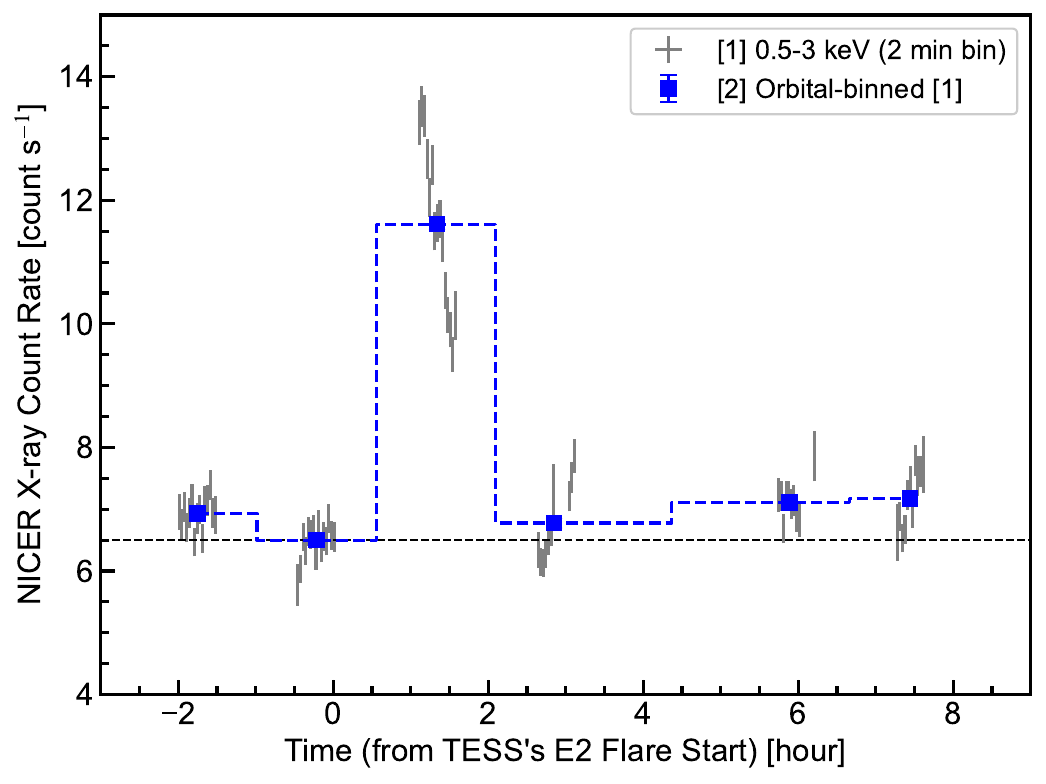}{0.5\textwidth}{\vspace{0mm} (a)}
\fig{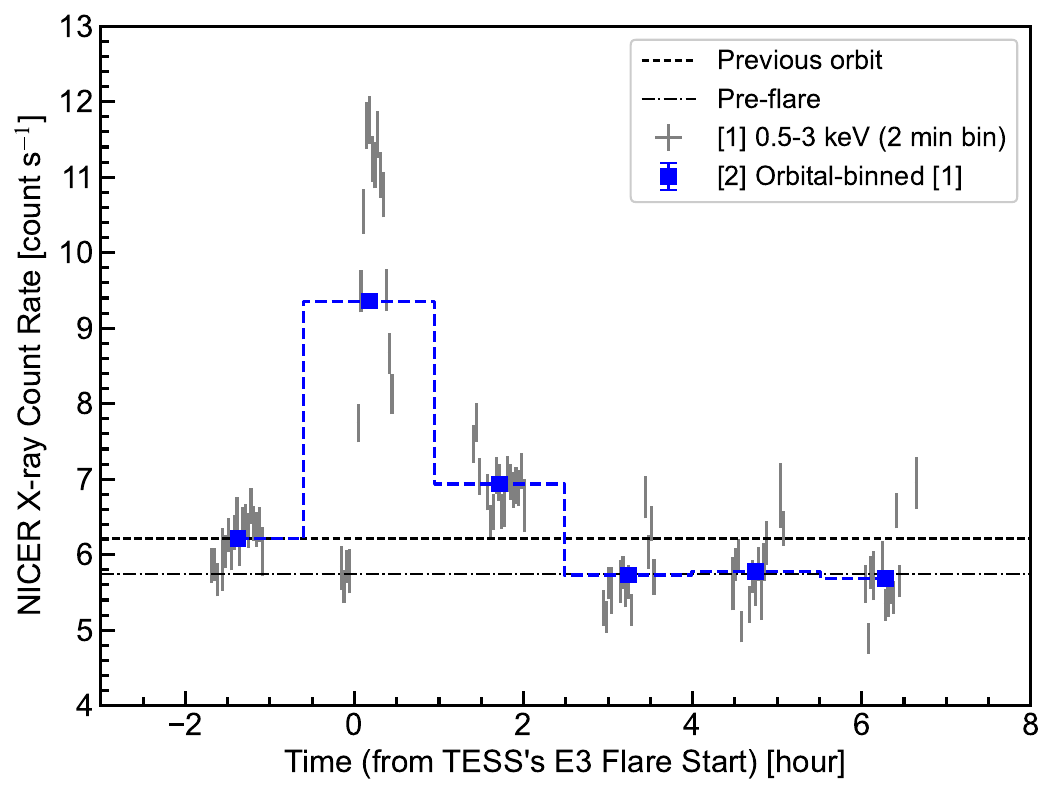}{0.5\textwidth}{\vspace{0mm} (b)}
}
\caption{X-ray (0.5-3 keV) light curves binned for each ISS's orbit around superflares on 2022 April 16 (E2) and 2022 April 17 (E3). The gray points represent the data with a time resolution of 2 minutes, while the blue points correspond to the average of the gray data for each ISS orbit ($\sim$ 90 min period). 
The horizontal lines denote the pre-flare values. 
For flare E3, the average value of the previous orbit and the pre-flare value just before the flare are plotted.}
\label{figs:23}
\end{figure*}

\begin{figure}
\epsscale{0.33}
\plotone{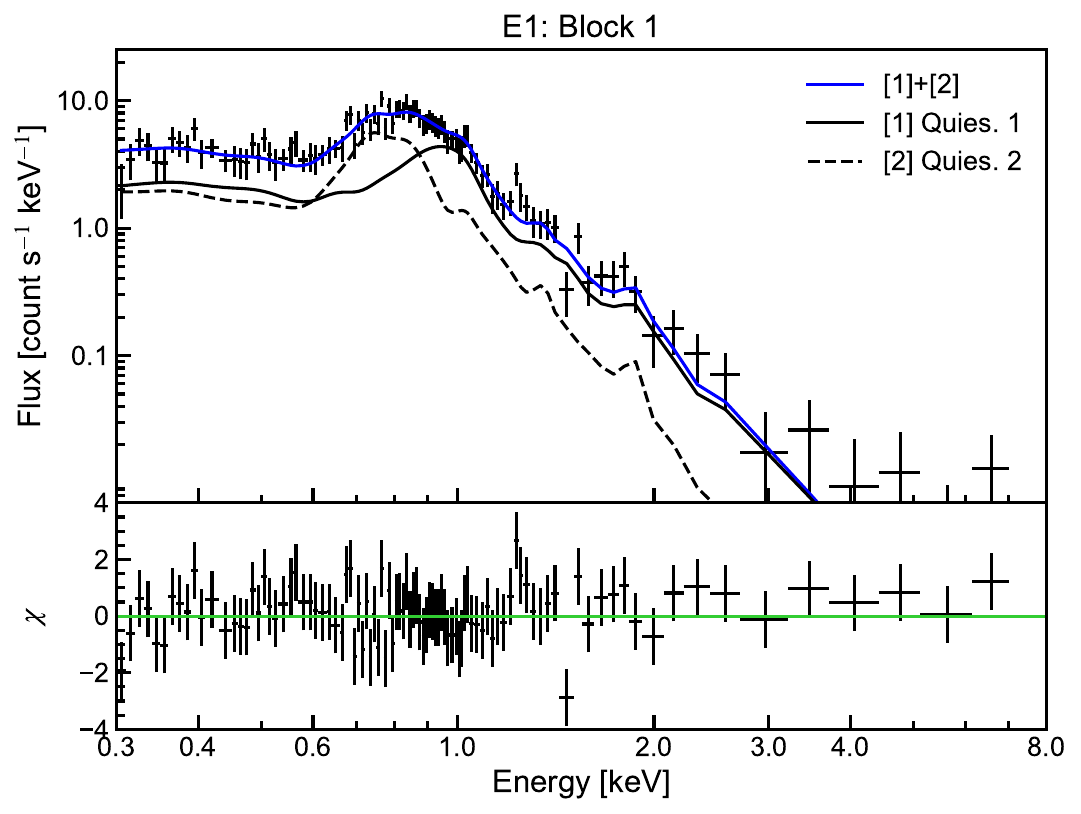}
\plotone{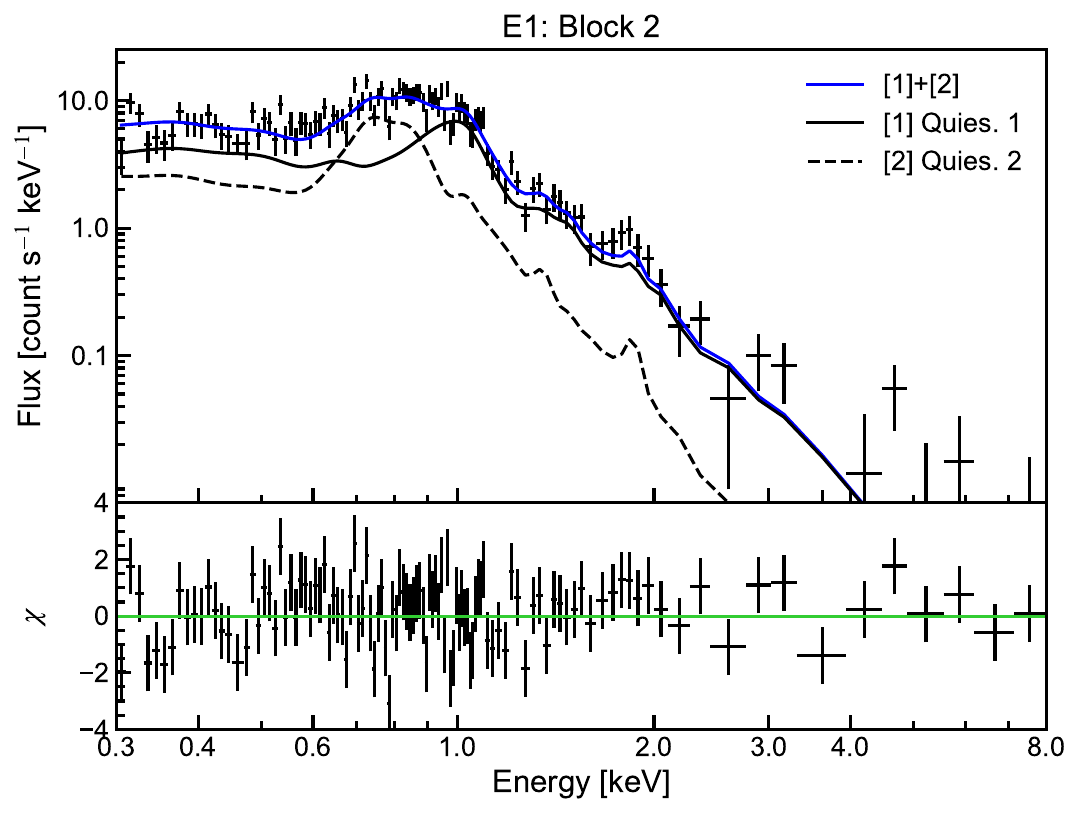}
\plotone{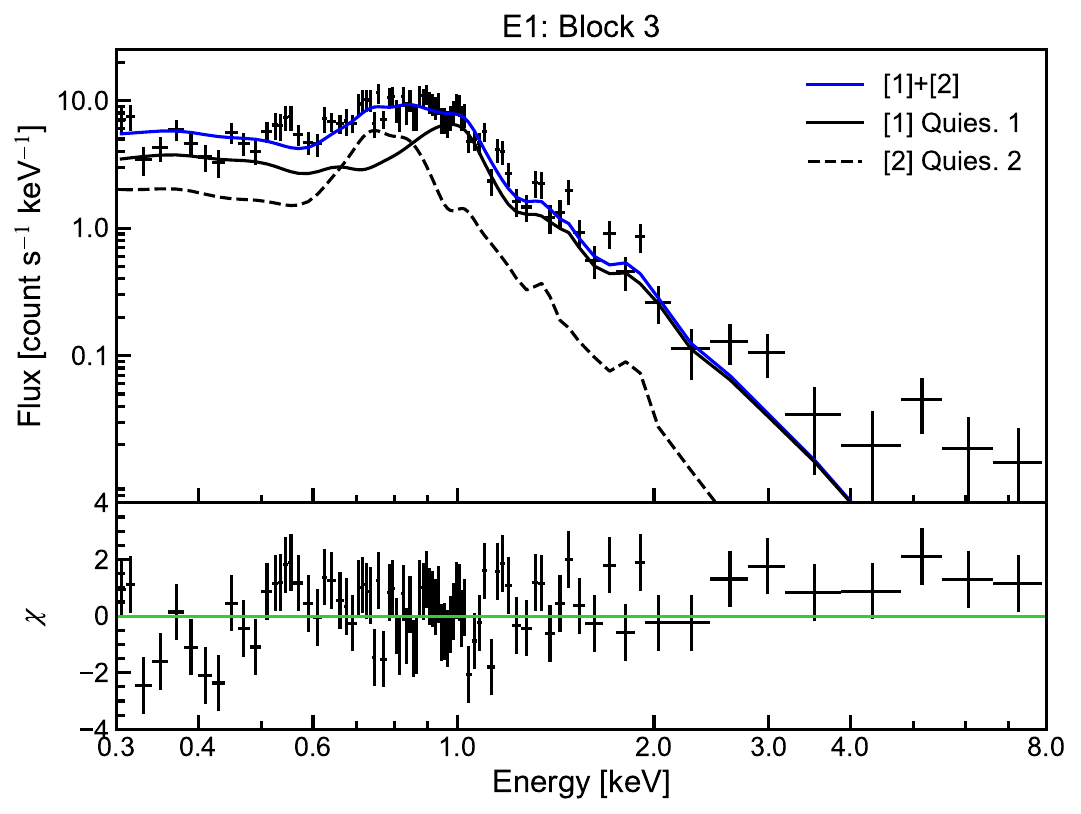}
\plotone{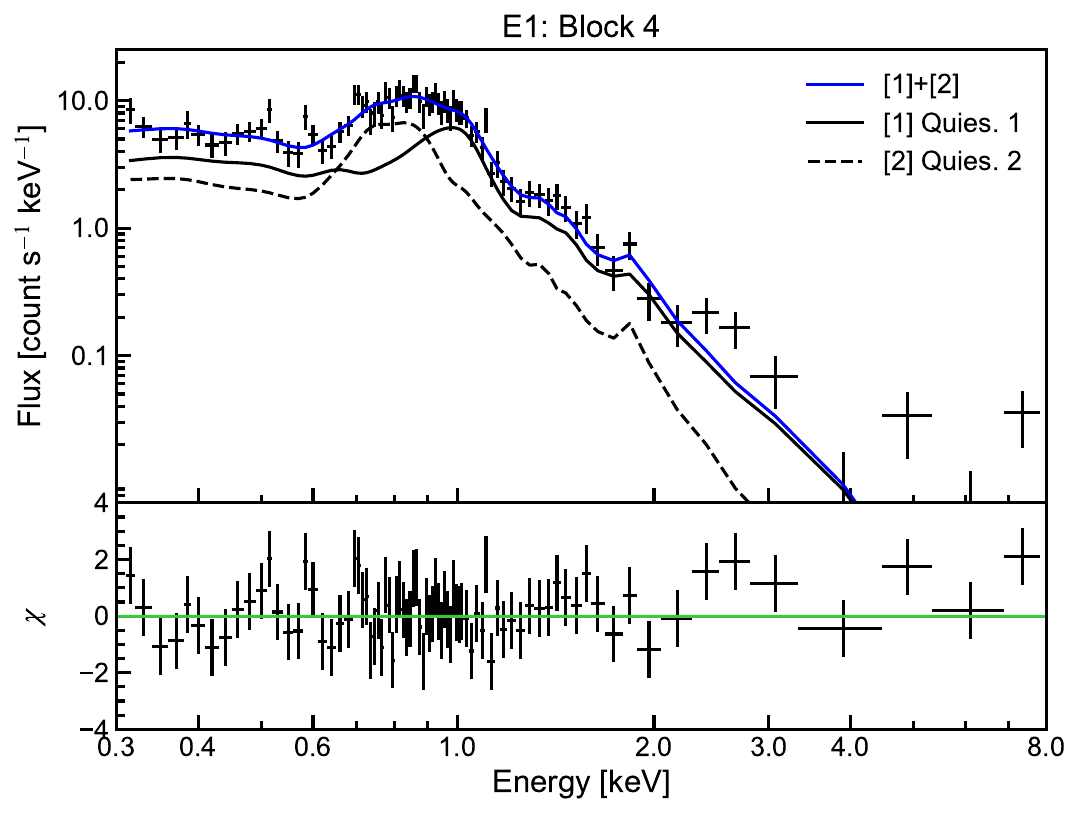}
\plotone{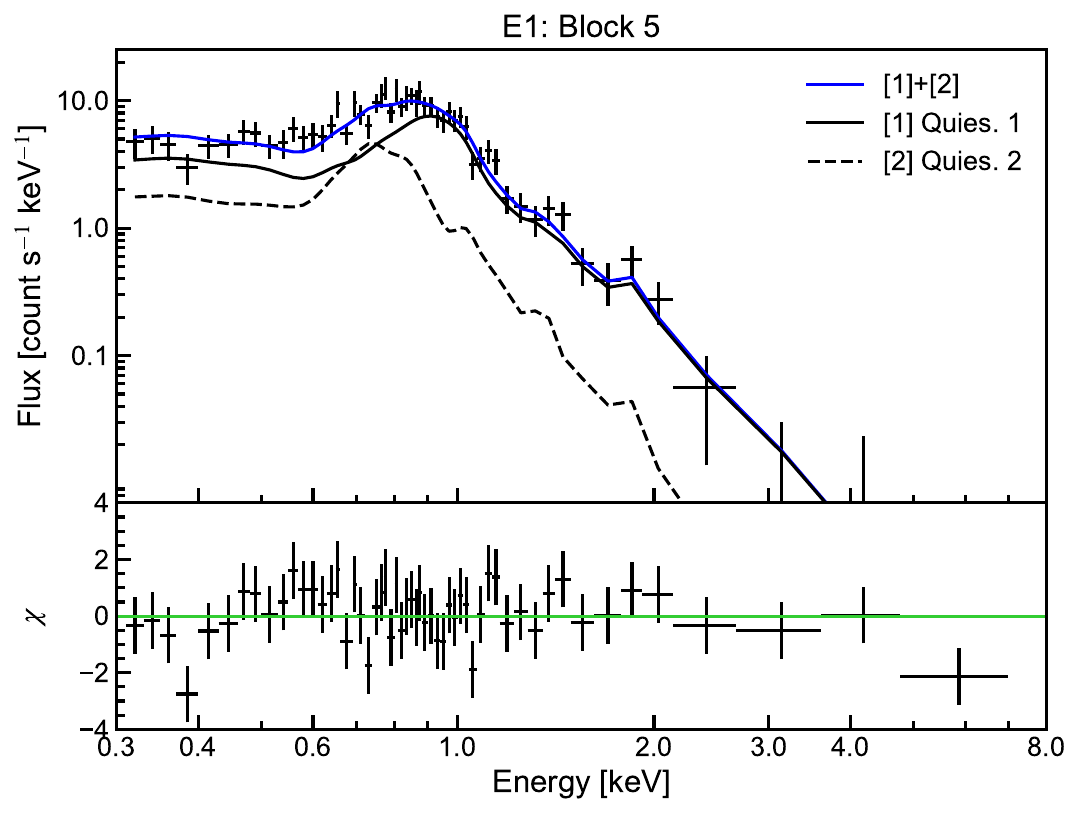}
\plotone{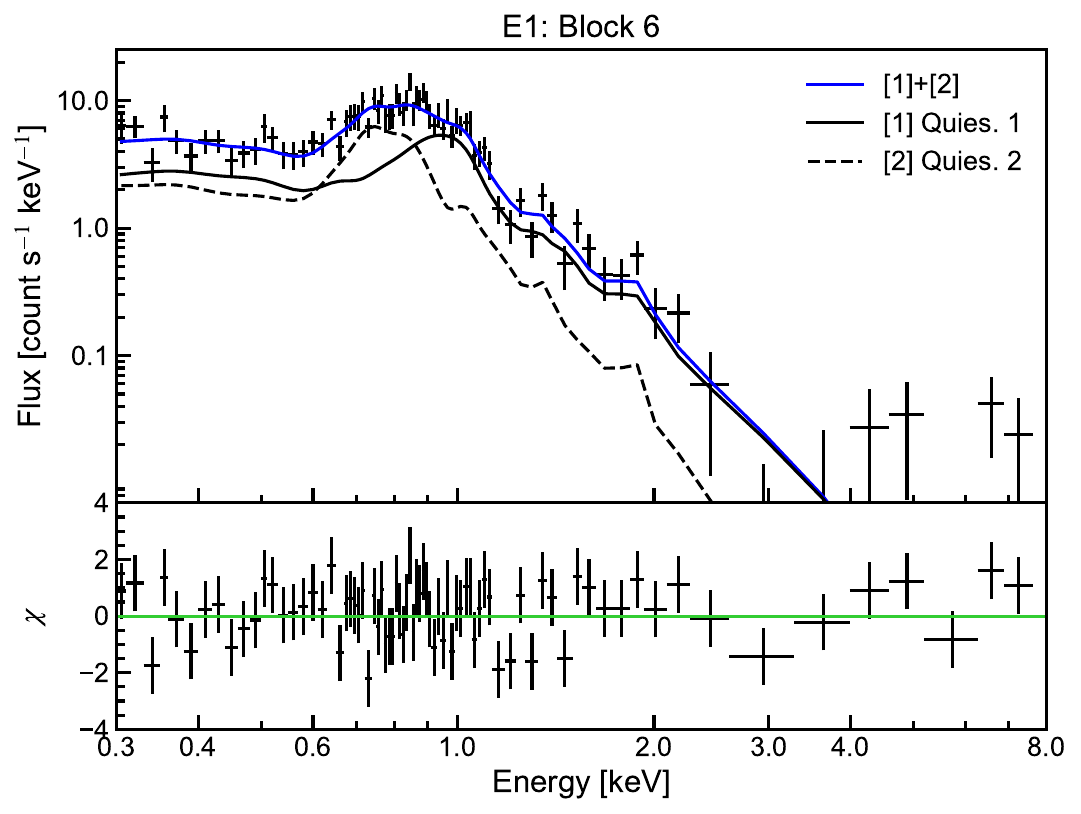}
\caption{The same as Figure \ref{fig:12}, but for the superflare on 2022 April 10 (E1).}
\label{figs:24}
\end{figure}

We investigated whether the prominence eruptions in flares E1 and E2 are associated with X-ray dimmings, as another evidence of CMEs in multi-wavelength \citep[e.g.,][]{2021NatAs...5..697V}.
Figures \ref{figs:22}(b) and \ref{figs:23} show the ISS-orbital-binned X-ray light curves (see also Appendix \ref{app:a} for more details of flare E1).
We found that the X-ray count rates in post flare phase of flare E1 (i.e., block 5 and 6) decrease compared to the possible pre-flare level (block 3) by --5.8$_{\pm5.1}$ \% and --8.5$_{\pm5.6}$ \% for block 5 and 6, respectively (Figures \ref{figs:22}(b)).
The error bars are calculated by considering those in pre-flare phase (block 3) and those in corresponding phases.
The tiny X-ray enhancement in block 2 would be related to a simultaneous \textit{TESS}'s WLF and that in block 4 can include the decay phase in flare E1, so we interpret block 3 (and possibly block 1) is representative of the pre-flare level.
To remove the possible contamination of background X-ray flux (from the \textit{NICER}'s non-imaging field-of-view of 5 arcmin diameter\footnote{\url{https://heasarc.gsfc.nasa.gov/docs/nicer/nicer\_about.html}}; \citealt{2016SPIE.9905E..1HG}), we also calculated a background-subtracted X-ray count rates in Figure \ref{figs:22}(b) using 3C50 background model (\textsf{nibackgen3c50}; \citealt{2022AJ....163..130R}), which produce the dimming amplitude of --8.8$_{\pm6.2}$ \% and --14.7$_{\pm7.1}$ \% for each block. 
On the other hands, Figure \ref{figs:23} shows that there were no significant dimming events for flare E2 and E3.

Then, is this X-ray flux decrease a CME-related X-ray dimming?
Indeed, the \textit{NICER}'s X-ray data is not completely continuous one because \textit{NICER} is on the ISS, so the full variation of the X-ray light curve is not obtained.
This is one of the difficulties to identify whether it could be a dimming.
Also, the net exposure time of this date is relatively short compared to other dates, the data quality is also lower than usual. 
In addition, the definitions of ``per-flare level" is very difficult for finding XUV dimming in active stars \citep{2021NatAs...5..697V,2022ApJ...936..170L} since their ``pre-flare level" are essentially a superposition of frequent small (but solar M/X-class) flares and therefore variable.
Considering these, it may be not possible to conclusively suggest that an X-ray dimming is detected. 
However, simultaneous detection with H$\alpha$ blueshifts in flare E1 is supportive of the possible association of the X-ray dimming, that could be indicative of the occurrence of unknown CMEs.
Therefore, in the following, we call this an X-ray decrease of flare E1 as ``possible X-ray dimming" or just a ``candidate" and investigate its properties and whether they are consistent with flare and prominence eruption sizes, following the previous studies on M/K-dwarfs \citep{2021NatAs...5..697V,2022ApJ...936..170L}.

For estimating the emission measure of the possible X-ray dimming, we performed the spectral fitting of a background-subtracted X-ray spectrum for each ISS orbit on 2022 April 10 with two-component \textsf{apec} models, which are emission spectra from collisionally-ionized diffuse gas calculated from the AtomDB atomic database\footnote{\url{https://heasarc.gsfc.nasa.gov/xanadu/xspec/manual/node134.html}}.
The hydrogen column density is fixed as $N_{\rm H}$ = 3$\times 10^{18}$ cm$^{-2}$, which was the same assumption as \cite{1997ApJ...479..416G}. 
Abundance relative to solar one is fixed as 0.23 in this analysis, which is a median values for two component fitting results for all orbit in this campaign.
Figure \ref{figs:24} shows the X-ray spectra for each orbit and the fitting results whose parameters are summarized in Table \ref{tab:10}.
Figure \ref{figs:22}(c) and (d) show the time evolution of the temperature and emission measure of two plasma component.

\begin{deluxetable*}{ccccccc}
\label{tab:10}
\tabletypesize{\footnotesize}
\tablecaption{Results of NICER X-ray spectral fit for flares E1 X-ray dimming.}
\tablewidth{0pt}
\tablehead{
\colhead{flare E1's block} & \colhead{1} & \colhead{2} & \colhead{3} & \colhead{4} & \colhead{5} & \colhead{6} }
\startdata
T$_1$ [keV] & 0.95$_{\pm 0.04}$ & 1.11$_{\pm 0.04}$ & 1.06$_{\pm 0.05}$ & 1.08$_{\pm 0.08}$ & 0.91$_{\pm 0.05}$ & 0.99$_{\pm 0.05}$ \\
(T$_1$ [10$^7$ K]) & (1.10$_{\pm 0.05}$) & (1.29$_{\pm 0.04}$) & (1.23$_{\pm 0.05}$) & (1.25$_{\pm 0.09}$) & (1.06$_{\pm 0.05}$) & (1.15$_{\pm 0.06}$) \\
T$_2$ [keV] & 0.44$_{\pm 0.05}$ & 0.54$_{\pm 0.05}$ & 0.53$_{\pm 0.07}$ & 0.65$_{\pm 0.09}$ & 0.43$_{\pm 0.07}$ & 0.51$_{\pm 0.07}$ \\
(T$_2$ [10$^7$ K]) & (0.51$_{\pm 0.06}$) & (0.62$_{\pm 0.06}$) & (0.61$_{\pm 0.09}$) & (0.75$_{\pm 0.10}$) & (0.50$_{\pm 0.08}$) & (0.59$_{\pm 0.08}$) \\
EM$_1$ [10$^{52}$ cm$^{-3}$] & 5.34$_{\pm 0.47}$ & 6.84$_{\pm 0.39}$ & 6.05$_{\pm 0.48}$ & 5.83$_{\pm 1.11}$ & 5.68$_{\pm 0.59}$ & 4.51$_{\pm 0.49}$ \\
EM$_2$ [10$^{52}$ cm$^{-3}$] & 3.69$_{\pm 0.49}$ & 4.02$_{\pm 0.32}$ & 3.16$_{\pm 0.42}$ & 3.84$_{\pm 1.09}$ & 2.81$_{\pm 0.63}$ & 3.43$_{\pm 0.45}$ \\
\hline
$\chi^{2}$(dof) & 81(83) & 139(110) & 110(81) & 75(79) & 48(51) & 79(74) \\
reduced $\chi^{2}$ & 0.98 & 1.26 & 1.36 & 0.95 & 0.94 & 1.07 \\
\enddata
\tablecomments{
The hydrogen column density $N_{\rm H}$ in \textsf{TBAbs} is fixed as $3\times10^{18}$ cm$^{-2}$. The abundance in \textsf{apec} is fixed as typical value of 0.23.
}
\end{deluxetable*}

The decrease in total emission measure ($EM_{\rm tot} = EM_{1} + EM_{2}$) of block 5 and 6 compared to the pre-flare orbit (block 3) is 7.2$_{\pm 10.7}\times$10$^{51}$ cm$^{-3}$ (7.8$_{\pm 11.7}$ \%) and 12.7$_{\pm 9.2}\times$10$^{51}$ cm$^{-3}$ (13.8$_{\pm 10.0}$ \%), respectively. 
Here, the error bars are a bit large but this is because the $EM$ was obtained by the spectral fitting while the error bars in the light curves were obtained by energy-integrated count rate. 
This is roughly consistent with the dimming amplitude in their light curve [5.8$_{\pm 5.1}$ (block 5) -- 8.5$_{\pm 5.6}$ (block 6) \% for non-background-subtracted count rates  or 8.8$_{\pm 6.2}$ -- 14.7$_{\pm 7.1}$ \% for background-subtracted count rates].
We can estimate the volume and the length scale of the corona from the equation of $\Delta EM_{\rm tot}$ = $n_{\rm e}n_{\rm H} V$ $\approx$ $n_{\rm e}^{2} V$ = $n_{\rm e}^{2}  L_{\rm dim}^{3}$.
\begin{eqnarray}
V  &\approx& n_{\rm e}^{-2} \Delta EM_{\rm tot} = (7.2_{\pm 11.7}-12.7_{\pm 9.2}) \times 10^{31} \cdot  \left( \frac{n_{\rm e}}{10^{10} {\rm cm^{-3}}} \right)^{-2} \rm cm^3 \\
L_{\rm dim}  &\approx& n_{\rm e}^{-2/3} \Delta EM_{\rm tot}^{1/3} = (4.2_{-4.2}^{+1.6}-5.0_{-1.7}^{+1.0}) \times 10^{10} \cdot  \left( \frac{n_{\rm e}}{10^{10} {\rm cm^{-3}}} \right)^{-2/3} \rm cm
\end{eqnarray}
The estimated mass that escape can be estimated in the formula of
\begin{eqnarray}
M_{\rm dim}  &=& m_{\rm H} n_{\rm H} V \approx m_{\rm H} n_{\rm e} V = m_{\rm H} n_{\rm e}^{-1} \Delta EM_{\rm tot}\\ 
&=& (1.67\times 10^{-24} {\rm g}) \cdot (10^{10} {\rm cm^{-3}})^{-1} \cdot \left( \frac{n_{\rm e}}{10^{10} {\rm cm^{-3}}} \right)^{-1} \cdot ( (7.2_{\pm 11.7}-12.7_{\pm 9.2}) \times 10^{51} {\rm cm}^{-3}) \\ 
&=&  (1.2_{\pm 2.0}-2.1_{\pm 1.5}) \times 10^{18} \cdot \left( \frac{n_{\rm e}}{10^{10} {\rm cm^{-3}}} \right)^{-1} {\rm g} 
\end{eqnarray}
It is reasonable to assume that this dimming region possesses the typical density of the active regions, as numerical simulations by \cite{2005ApJ...634..663S} show that the dimming region can be considered as the immediate adjacent space to the magnetic field lines undergoing reconnection.
Therefore, the assumption of coronal density of 10$^{10-11}$ cm$^{-3}$ would be a reasonable assumption for EK Dra's active region (see, Section \ref{sec:flare-length}; \citealt{1995A&A...301..201G,2004A&A...427..667N,2004A&ARv..12...71G}).
The assumption of a coronal density $n_{\rm e}$ of 10$^{11}$ cm$^{-3}$ gives the estimates of the volume $V$ = $(7.2_{\pm 11.7}-12.7_{\pm 9.2})\times 10^{29}$ cm$^3$, the length scale $L_{\rm dim}$ = $(0.90_{-0.9}^{+0.3}-1.1_{-0.4}^{+0.2})\times 10^{10}$ cm, and the mass $M_{\rm dim}$ = $(1.2_{\pm 2.0}-2.1_{\pm 1.5}) \times 10^{17}$ g.
Likewise, the assumption of a coronal density $n_{\rm e}$ of 10$^{10}$ cm$^{-3}$ gives different estimations as $V$ = $(7.2_{\pm 11.7}-12.7_{\pm 9.2}) \times 10^{31}$ cm$^3$, $L_{\rm dim} = (4.2_{-4.2}^{+1.6}-5.0_{-1.7}^{+1.0})\times 10^{10}$ cm, and $M_{\rm dim}$ = $(1.2_{\pm 2.0}-2.1_{\pm 1.5}) \times 10^{18}$ g.

The length scale of the coronal source of the possible X-ray dimming region ($(0.90_{-0.9}^{+0.3}-1.1_{-0.4}^{+0.2})\times 10^{10}$ cm or $(4.2_{-4.2}^{+1.6}-5.0_{-1.7}^{+1.0})\times 10^{10}$ cm) is roughly consistent with -- or slightly larger than -- the flaring loop length ($0.73\times 10^{10}$ cm) estimated in Section \ref{sec:flare-length}. 
This supports that the observed post-flare decrease in soft X-ray count rate can be related to the escape of hot coronal plasma associated with a CME. 
Now, if this X-ray dimming were indeed the so-called ``dimming", how would it be compare to the properties of the prominence?
The length scale of the dimming region is actually much smaller by one to two orders of magnitude than the estimated length scale of prominence eruption ($(8.3-300)\times 10^{10}$ cm). 
It is speculated that the prominence gradually expands and its area increases as it erupts as in the case of solar prominence/filament eruptions. 
The length scale of the prominence estimated here corresponds to the value when its surface area (i.e., EW) was the largest.
Therefore, it is reasonable that estimated prominence length scale is larger than the length scale of the flare region and dimming region.
Moreover, the escaped coronal mass estimated from the possible X-ray dimming is enough smaller than the estimated mass of prominence ($4.0\times10^{19}$ g -- $4.2\times10^{20}$ g).
Therefore, it is not surprising even if an X-ray dimming associated with the gigantic prominence eruption in E1 is detected.
These comparison in length scale and mass, however, largely depends on the assumed physical quantities of stellar corona and prominences (cf. Section \ref{sec:5-2}). 
More detailed comparisons to ensure that there is no inconsistency between the possible dimming and prominences requires further models \citep[cf.,][]{2022ApJ...936..170L,2022MNRAS.513.6058L}.

\section{Discussion and Conclusion}\label{sec:6}

We report on three superflares (labeled as E1--E3) from the young solar-type star, EK Dra, using our simultaneous optical and X-ray observations taken with Seimei, \textit{TESS}, and \textit{NICER}.
The time-resolved H$\alpha$ spectra were characterized in the relation with simultaneous WL and X-ray flares for the first time on solar-type stars.
A short summary of the preceding sections is as follows: 
Section \ref{sec:4} describes the multi-wavelength time evolution, and energy partitions, and length scales of superflares.
Section \ref{sec:4-1} suggests that occurrence mechanism of superflares on solar-type stars could be common as those of solar and M-dwarf flares, as time delay of X-ray against WL and H$\alpha$ suggest the chromospheric heating followed by chromospheric evaporations.
Section \ref{sec:4-2} provide meaningful empirical laws for flare radiative energy partition into X-ray and WL, linking the well established \textit{Kepler}/\textit{TESS} photometry to assessing their impact on exoplanetary environment.
Section \ref{sec:5} reports on the first discovery and properties of gigantic prominence eruptions (E1 and E2) on a solar-type star.
Section \ref{sec:5:disc} gives a conclusion that the detected blueshifted emissions are stellar prominence eruptions from several observational considerations.
Section \ref{sec:5-3} reports and discusses a possible candidate of the associated post-flare X-ray dimming (E1), which enables the first simultaneous multi-wavelength diagnostics of stellar CMEs on solar-type stars in different signature (i.e., the Balmer line blueshifts and the possible candidate of X-ray dimming).

Here, bringing all information together, we will discuss the following remaining topics entitled in Section \ref{sec:cancme}--\ref{sec:diversity} and give a conclusion for each topic.
In the following, we also mention previous studies reporting two H$\alpha$ superflares on the EK Dra, where one of them shows filament eruptions (labeled as ``E4", \citealt{2022NatAs...6..241N}) and the other do not (labeled as ``E5", \citealt{2022ApJ...926L...5N}), as summarized in Tables \ref{tab:prominence-1}, \ref{tab:prominence-2}, and \ref{tab:length-scale}. 

\subsection{Can the Prominence Eruptions \Add{be Associated with} Stellar CMEs?}\label{sec:cancme}

In Section \ref{sec:5}, we report the discovery of gigantic stellar prominence eruptions associated with superflares (E1 and E2).
The occurrence of the prominence eruption is conclusively suggested in Section \ref{sec:5:disc}.
\Add{The maximum velocity of the prominence eruption in flare E1 was 690 km s$^{-1}$ (conservatively, 330-490 km s$^{-1}$), whereas those in flare E2 was 430 km s$^{-1}$ (Section \ref{sec:5-2-1}).
While the maximum velocity in E1 event is slightly larger than the escape velocity of EK Dra of $\sim$670 km s$^{-1}$, it was gradually decelerated and eventually showed almost zero velocity.
This may suggest that the most of the emitting mass of the prominence stalled and it is unclear if the prominence itself ultimately escaped.
However, we consider that the following points are indicative of the occurrence of the associated CMEs which consist of outer coronal mass and fast prominence components, especially in the flare E1:}
\begin{itemize}
\item \Add{The large velocity dispersion of blueshifted H$\alpha$ component can be interpreted as spatial velocity distribution in the expanding 3D loop structure (Section \ref{sec:5-2-1}), which is also apparent in the Sun-as-a-star analysis (Appendix \ref{app:b}).
Based on this interpretation, the maximum velocity component in E1 event could reach 990-1080 km s$^{-1}$ (conservatively, 630-880 km s$^{-1}$), which is significantly greater than the escape velocity.
In the case of the Sun, prominence eruptions represent the lower-part of CMEs, and some solar events show that the majority of prominence falls back to the Sun while the other part of fast prominence components and outer coronal masses actually lead to CMEs \citep[e.g.,][Appendix \ref{app:b}]{2016ApJ...816...67W}.
These suggest that a part of the fast prominence components might have escaped. 
}
\item Following the scheme suggested by \cite{2022NatAs...6..241N}, comparing solar and stellar eruptions in terms of the length scale and velocity strongly suggests that the prominence eruptions in flare E1 and E2 could be associated with CMEs, considering their gigantic length scales and velocities \Add{(Section \ref{sec:5-2-2})}.
\item \Add{The numerical simulation by \cite{2018ApJ...862...93A} suggests 
that the large-scale magnetic field could suppress CMEs under the average magnetic field strength of 75 G if the kinetic energy of CMEs is below a threshold around $\sim3\times10^{32}$ erg. 
Since the EK Dra's averaged magnetic field strength of 66-89 G is similar to this numerical setup \citep[e.g.,][]{2017MNRAS.465.2076W}, we may be able to apply this threshold to our superflares \citep[cf.][Supplementary Information]{2022NatAs...6..241N}.
The initial kinetic energy of the prominence eruptions from events E1 and E2 are $5.8_{-4.0}^{+12.8}\times10^{34}$ erg and $1.2_{-0.8}^{+2.7}\times10^{33}$ erg, respectively (Section \ref{sec:5-2-4}), which are significantly larger than this threshold. This indicates that the magnetic suppression might not be an efficient process in these cases. 
Indeed, the mentioned model employs a dipolar magnetic field, while spectropolarimetric observations of this star show that large-scale magnetic fields have multi-polar components \citep[e.g.][]{2016A&A...593A..35R}, and we will perform further numerical modelings with the observed data in the future (Section \ref{sec:future}).}

\item After the flare E1, a possible candidate of X-ray dimming were detected, as an indication of escaping the surrounding hot coronal plasma associated with CMEs \Add{(Section \ref{sec:5-3})}. Although the determination of the pre-flare X-ray level is an issue in detecting stellar XUV dimming, the simultaneous detection with H$\alpha$ blueshift event is supportive of the detection of X-ray dimming.
The estimated upper limit of length scale of the dimming region of up to $\sim(9-11)\times10^{9}$ cm is roughly consistent with the length scale of the flares $\sim7.3\times10^{9}$ cm.
The estimated mass of escaping coronal plasma of $1.2\times10^{17}$--$2.1\times10^{18}$ g is enough small that the erupted prominence mass of $4.0\times10^{19}$--$4.2\times10^{20}$ g.
Therefore, it is not surprising that the detected X-ray dimming amplitudes is likely to be interpreted as CME-related dimming (Section \ref{sec:5-3}). However, again we should note that the NICER's X-ray observation is usually sparse due to the ISS orbit, and need to be careful for the treatments of this data (see also Appendix \ref{app:a}). More convincing detections should be required for the further discussions.
\end{itemize}
Given these without any suggestive evidence that is against the occurrence of CMEs in flare E1, we concluded that the flare E1 is associated with the occurrence a stellar CME.
This conclusion is further strengthen than that drawn in \cite{2022NatAs...6..241N}.
Furthermore, solar studies indicates that velocities of outer CMEs are typically 4--8 times larger than the velocities of prominences \citep{2003ApJ...586..562G}, suggesting that the associated CME velocity from flare E1 could reach up to 2800--5500 km s$^{-1}$.

It is important to mention that the center-of-mass velocity of the giant prominence in the E1 evolved from 690 km s$^{-1}$ to near zero in about 20 min, and then disappeared without showing a redshift.
This means that the center-of-mass of the prominence stopped midway and disappeared for some reasons.
The disappearance and terminal zero velocity is puzzling because solar prominence eruptions show redshifted spectra after impulsive blueshifts, as presented in Appendix \ref{app:b}.
By integrating the blueshift velocity over evolution time (up to $t<50$ min), the line-of-sight displacement of the center-of-mass is calculated as $6.0_{\pm 0.3}\times 10^{10}$ cm ($0.91_{\pm 0.04}$ $R_{\rm star}$) for one component fitting and $7.5_{\pm 0.2}\times 10^{10}$ cm ($1.14_{\pm 0.03}$ $R_{\rm star}$) for two component fitting. 
The gravity is expected to be 3.6 times weaker at the height of $0.91$ $R_{\rm star}$ than the surface gravity. 
The time scale of plasma acceleration from the rest to $>$ 50 km s$^{-1}$ from this height is approximately 10 min.
It is likely that before it began falling back to the star, the plasma was heated or became more diffuse and thus became invisible in the H$\alpha$ line.
Another possible explanation is that in the later phase, the emissions in line center come from flare ribbons. 
Also, one may speculate that this zero-velocity erupted prominence could be linked to ``slingshot" prominences \citep[e.g.,][]{2019MNRAS.482.2853J}.
Although there is no observation of slingshot prominences for EK Dra, the E1 event may provide a possiblity that the flare-related prominence could be one source of the slingshot prominences seen in rapidly-rotating stars.


The prominence eruption in flares E2 (this study) and filament eruptions in flare E4 \citep{2022NatAs...6..241N} offer less robust evidence of CMEs than flare E1. 
The prominence eruption associated with E2 flare event was relatively small and short-lived, and it did not exhibit any substantial X-ray dimming. 
The lack of significant X-ray dimming following E2 event could be attributed to its small dimming amplitude, which can be caused by the relatively small prominence area, an order of magnitude smaller than that observed in E1 flare event.
The filament eruptions associated with E4 flare event \citep{2022NatAs...6..241N} exhibited a clear blueshifted absorption, but its maximum velocity of 510 km s$^{-1}$ was lower than the escape velocity, and it eventually showed redshifted absorption in H$\alpha$ at velocities of a few tens of km s$^{-1}$. 
However, these low velocities associated with E2 and E4 flare events, however, do not necessarily rule out the presence of CMEs.
These values could represent a lower limit of velocity along the LOS.
Also, large velocity dispersion in blueshifted components indicates the existence of faster component than the fitting center, which exceeds the escape velocity (see, Section \ref{sec:5-2-1}).
Furthermore, solar observations show that the velocities of CMEs are typically 4--8 times larger than those of prominence eruptions \citep{2003ApJ...586..562G}, as also discussed in Section \ref{sec:5-2-1}. 
Consequently, the low velocity (less than the escape velocity) blueshift plsama motions in E2 and E4 events may suggest that the associated CME velocities could be as high as $\sim$ 1700--3400 km s$^{-1}$ (E2) and 2000--4100 km s$^{-1}$ (E4), respectively. Therefore, following the solar analogy, we speculate that both E2 and E4 eruptive events are likely associated with fast CMEs, as suggested by \citep{2022NatAs...6..241N}.


\subsection{The Overall Picture: Relationship between Starspots, Superflares, and Prominences Eruptions/CMEs}\label{sec:picture}

In this section, we will present an overall picture of the eruptive phenomena observed in the young solar-type star EK Dra. This spans from the occurrence of starspots to the manifestation of superflares and prominence eruptions/CMEs, being compared in terms of length scale, energetics, time evolution, and rotational phase.

\subsubsection{Length Scale}
Table \ref{tab:length-scale} compiles the length scale of starspots $L_{\rm spot}$, superflares (coronal loop length) $L_{\rm flare}$ (Section \ref{sec:flare-length}), and erupting prominences $L_{\rm p}$ (Section \ref{sec:5-2-2}). 
The size and area of the starspots are calculated from the brightness variation amplitude (BVAmp) of \textit{TESS} light curves using the methods of \cite{2017PASJ...69...41M}. Table \ref{tab:prominence-2} also compiles the length scale of possible dimming regions ($L_{\rm dim}$) for E1 flare event (see Section \ref{sec:5-3}). Consequently, the following summarizes the ratios of these length scales to flare loop length:
\begin{eqnarray}
  L_{\rm spot}:L_{\rm flare}:L_{\rm p}:L_{\rm dim} = \left\{
\begin{array}{ll}
 2.9:1:(11\sim410):(1.5\sim6.8), &  \quad\rm for \, (E1) \\
 2.1:1:(1.2\sim44):(-), &  \quad\rm for \, (E2) \\
 5.0:1:(2.9\sim260):(-), &  \quad\rm for \, (E4)
\end{array}
\right. \label{eq:ratio}
\end{eqnarray}
Here we uses the $L_{\rm flare}$ estimated by the method of \cite{2017ApJ...851...91N} since all of these flares have white-light flare observations (see, Section \ref{sec:flare-length}). 
Equation \ref{eq:ratio} data provide an initial picture of eruptive phenomena on young solar-type stars: superflares originating from active regions associated with starspots at length scale of at least 2--5 times larger the flare loop length and producing eruptive prominences/filaments at greater scales (at least 1--10 times of the flare loop length). 
This picture is not unique but looks different depending on events.
These relations are useful when comparing observational data with numerical simulations and solar observations. 
The uncertainty in the length scale of prominences/filaments is currently high as we consider a wide conservative range of optical depths ($\tau=$10--1000). 
The prominence length scale would change depending on the phase of the prominence/filament eruption, as it is expected to expand over time, so we assumed a conservative range of optical depth. 
Also, ionization equilibrium in moving plasma would be different from quiescent prominence \citep{1987SoPh..110..171H}, which can significantly change the relation in Equations \ref{eq:tau-f} and \ref{eq:em-ha}.
Further constraints could be made by combining several spectral lines (e.g., H$\alpha$ and H$\beta$) in future observations or radiative transfer modeling \citep{2020PASJ...72...71O,2022MNRAS.513.6058L}.

\subsubsection{Energy Scales}
Our observations provides an initial picture of flare energetics in eruptive superflares from a young solar-type star. 
The energetics of the eruptive E1, E2, and E4 events can be compiled as follows:
\begin{eqnarray}
  fE_{\rm mag}:E_{\rm WL,bol}:E_{\rm X,bol}:E_{\rm kin} = \left\{
\begin{array}{ll}
 230:1:(-):39^{+85}_{-27}, & \quad \rm for \; (E1) \\
 44:1:(1.2_{\pm0.2}):0.10^{+0.22}_{-0.07}, & \quad \rm for \; (E2) \\
 180:1:(-):0.17^{+0.70}_{-0.02}, & \quad \rm for \; (E4)
\end{array}
\right. \label{eq:ratio2}
\end{eqnarray}
where $fE_{\rm mag}$ is the available magnetic energy stored in active regions associated with starspots, $f$ is the fraction of the non-potential magnetic energy (here, $f=0.1$), $E_{\rm mag}$ is calculated as $B^2L_{\rm spot}^3/8\pi$, and B is assumed as 1000 G.
Both the kinetic and radiation energies are much smaller than the stored magnetic energy of the active regions/starspots. 
This aligns well with our general expectations that giant starspots can produce superflares and prominence/filament eruptions.
The kinetic energies of prominence/filament eruptions is much larger than X-rays and WL radiative energies for flare E1, but the relation is opposite for flare E2 and E4. 
This diversity may require some theoretical explanations, which is beyoud the scope of this paper.
 

Our flare observations of EK Dra shows the energy partition into X-rays and WL appeared to be roughly equal. 
However, when incorporating data from solar flares and flares from other stars, and deriving the energy distribution rule to white light and X-rays via H$\alpha$, we noticed a trend where higher energy flares tend to be more dominated by X-rays (Section \ref{sec:4-2}). 
The derived Equation \ref{eq:WL-x} or \ref{eq:WL-x-ratio} could be useful when estimating the total amount of X-rays irradiated onto the atmospheres of young exoplanets by superflares.
Also, we found that the large prominence eruptions of flare E1 emits the H$\alpha$ flare radiation energy of $\sim$30 \% of WLF energy, which is an extremely large portion compared to usual EK Dra's and M-dwarf superflares of 1 \% (see, Section \ref{sec:5:disc}).
Hence, E1 is one of the outliers in the derived flare energy scaling relationship relationship. 
This may indicate that the flare statistics based on WLFs with \textit{Kepler} and \textit{TESS} can be affected to some extent by the contribution of H$\alpha$ line emission from prominence eruptions.

\subsubsection{Time Evolution}
Table \ref{tab:prominence-2} catalogs the timescales and timing of blueshift events and potential X-ray dimming events relative to WLFs. E1 flare event offers a probable sequence of events, as discussed in Section \ref{sec:cancme}, whereby a stellar prominence begins to erupt almost concurrently with the brightenings of WLFs, potentially leading to the appearance of X-ray dimming approximately 2 hours after WLFs. On the other hand, E2/E4 events have different timescales and timings for the emergence of blueshift events, as seen in Table \ref{tab:prominence-2}, though they lack detections or observations of X-ray dimming. This suggests that the appearances and formation/acceleration processes of stellar prominences/filaments can have a diverse nature (cf. Section \ref{sec:diversity}).

\subsubsection{Rotational Phase}
Additionally, all five detected superflares (E1$\sim$E5) on EK Dra have simultaneous \textit{TESS}'s long-term photometric data, allowing us to explore the relationship between flares and the stellar rotational phase. The rotational phase at which the flares occur is summarized in Table \ref{tab:length-scale}. By assuming the existence of a single dominant active region causing the rotational variation, we can infer that the flare E1 with prominence eruptions occurred when the dominant active region was near the disk center due to rotation. This seems inconsistent with the occurrence of prominence eruptions outside the stellar limb, suggesting that the source of the prominence eruption is not the dominant active region or that there may be multiple sizeble active regions forming the complex rotational modulation. The E2 flare occurred as the dominant active regions neared the limb, which aligns with the occurrence of a prominence eruption. Finally, the E4 flare event occurred when the dominant spots were on the opposite side of the star, suggesting that the associated filament eruption was produced over the relatively small active regions \citep{2022NatAs...6..241N}. 
These conclusions are not yet definitive, given that observations of other solar-type stars suggest there may be multiple starspots on the surface even in single-spot-like optical light curves \citep{2017ApJ...846...99M,2018ApJ...865..142B,2020ApJ...891..103N}. 
The Doppler Imaging of EK Dra, performed in the different epoch from our campaign, shows the gigantic starspots, changing its latitude in time ranging from the equator up to high latitude close to the pole \citep[e.g.,][]{1998A&A...330..685S,2016A&A...593A..35R,2017MNRAS.465.2076W,2018A&A...620A.162J}.
The near-pole spots usually do not appear in optical light curves but could be the sites of the filament/prominence eruptions.
Furthermore, the dark spot causing photometric variability may be not necessarily co-located with regions of strong large-scale magnetic fields \citep{2017MNRAS.465.2076W}, so we also requires a magnetic-field mapping for further understandings.
In our future work, we plan further investigations into the origins of these eruptive phenomena by performing multi-spot and/or magnetic-field mapping using the same observing periods (see Section \ref{sec:future}).


\subsection{Diversity and Frequency of Prominence/Filament Eruptions from Young Solar-type Stars}\label{sec:diversity}

Tables \ref{tab:prominence-1}, \ref{tab:prominence-2}, and \ref{tab:length-scale} provide preliminary insights into the diversity and frequency of eruptive flares on young solar-type stars. 
First, we demonstrate that blueshifted H$\alpha$ components can appear both in emission and absorption from the young solar-type star, EK Dra. This discovery reveals the well-known signatures of filament (absorption) and prominence (emission) observed on the Sun can also hold true for young solar-type stars.
While it might be expected that young solar-type stars emit intense UV radiation, including Ly$\alpha$, that can alter the excitation states of hydrogen within the prominence \citep{2022MNRAS.513.6058L}, our observations indicate that the radiative signatures of filaments and prominences does not significantly change even in young, active stars. 
In other words, the H$\alpha$ source function, $S_{\lambda}$ , of the prominence relative to the background intensity $I_{\lambda}$ seems to remain below unity ($S_{\lambda}/I_{\lambda}<1$).
This could be a very strong constraint in stellar prominence whose emissivity is unknown.

While the occurrence frequency of prominence/filament eruptions (and CMEs) is a crucial factor for habitability of (exo-)planetary environments, its value is poorly understood and is usually estimated based on flare frequency by assuming one-to-one relationship between stellar flares and CMEs.
For example, it is known that large M/X-class solar flares do not necessarily produce CMEs as they could be supressed by overlying coronal magnetic field of active regions \citep{2022MNRAS.509.5075S,2017ApJ...834...56T}.
Our limited sample of eruption events shows that the association rate of stellar prominence/filament eruptions against superflares is 3/5 (60 \%) on the young solar-type stars. 
Blueshifts can be detected with our spectrograph (${\rm R}\sim2000$, $\delta v \sim$ 150 km s$^{-1}$) only when the eruption has relatively high ($>$150 km s$^{-1}$) LOS velocity. 
Considering this limitation on the LOS velocity, we note that the absence of blueshifts in flare E3 and E5 does not necessarily imply the absence of eruptions. 
Therefore, the expected association rate of prominence/filament eruptions (and possibly CMEs) could be higher than 3/5. 
This evidence indicates that the strong suppression by local magnetic fields of active region may not play a significant role in the case of a gigantic eruption associated with the superflare releasing large portions of the magnetic energy stored in an active region  (see, Section \ref{sec:picture}).

Recent numerical simulations of large-sclae photospheric motion driven magnetic flux rope energization suggest an ejection of a CME from a young solar-type star, $\kappa^1$ Ceti  \citep{2019ApJ...880...97L}. Such scenarios can be applicable to global multipole magnetic field environments of young solar-type stars.
Indeed, the magnetic fields observed on EK Dra and young solar-type stars are relatively non-axisymmetric compared to active M dwarfs \citep{2009ARA&A..47..333D}, and the magnetic fields on solar-type stars are organized on smaller spatial scales than fully-convective M dwarfs \citep{2019ApJ...886..120S}, and, therefore, the fall-off with radius is much faster.
Thus, we speculate that young solar-type stars can be relatively more CME-productive, at least than M-dwarfs.
On the other hand, it is expected that ejection of CMEs from M dwarfs with a dipolar magnetic field could be suppressed by the strong global dipole magnetic fields of the star \citep{2018ApJ...862...93A}.
This may explain why recent M-dwarf flare survey shows that M dwarfs blueshift association rate is not so much high (e.g., 7 events among 41 flares on M-dwarfs, YZ CMi, EV Lac, and AD Leo; \citealt{Notsu2023}), compared to our estimate of expected CMEs from EK Dra.



Presently, all flares E1--E5 exhibit different properties in prominence/filament properties in terms of  the radiation signatures, mass, duration, timing, and X-ray counterpart, as summarized in Table \ref{tab:prominence-1}, \ref{tab:prominence-2}, and \ref{tab:length-scale}, and also mentioned in Section \ref{sec:cancme} and \ref{sec:picture}.
This indicates that there could be even more diversity if we increase a sample size.
We will present and discuss further observational results in our forthcoming papers (see Section \ref{sec:future}).


\subsection{Future Studies}\label{sec:future}

Our series of studies, including \cite{2022ApJ...926L...5N,2022NatAs...6..241N}, has offered numerous observational constraints on the eruptions of prominences and filaments from young solar-type stars. This section outlines our future research plans for refining our understanding of these eruptive phenomena.

First, we plan to determine the maps of starspots and large-scale magnetic fields to address the questions such as ``where do prominence and filament eruptions originate?" (refer to Section \ref{sec:picture}). We will model the rotational modulation in the optical light curves from \textit{TESS} using our previously established starspot mapping method \citep{2020ApJ...902...73I,2023ApJ...948...64I}. 
Additionally, at the almost similar period of the campaign periods in April 2022, we conducted high-dispersion optical spectroscopy of EK Dra using the High Dispersion Echelle Spectrograph (HIDES; \citealt{1999oaaf.conf...77I}) onboard the 1.88 m reflector at Okayama Observatory in Japan. We have also utilized the high-resolution echelle spectropolarimeter NARVAL, fiber-fed from the Cassegrain focus of the 2-m Bernard Lyot Telescope at the Pic du Midi observatory in France. These data will be analyzed using Doppler Imaging and Zeeman Doppler Imaging techniques to estimate the surface maps on EK Dra. The resulting maps will be compared with each other, with the occurrence of prominence eruptions (discussed in paper II), and with multi-wavelength rotational modulations observed in H$\alpha$, X-ray, and ground-based B-band photometric observations (Ikuta et al. in prep.).

We will also perform further modeling of the prominence eruptions from flare E1. Specifically, the H$\alpha$ data offers several key constraints on the locations and trajectories of the prominence eruptions (cf. Section \ref{sec:picture}). For instance, an emission H$\alpha$ profile without any absorption signal implies that a prominence's location is consistently outside the stellar limb from the beginning to the end of the event. The occurrence of a WLF indicates that the eruptive flare's footpoints are not -- or at least not entirely -- rooted behind the stellar disk. 
Moreover, the nearly consistent deceleration with surface gravity suggests the eruptions could occur in directions close to LOS. These constraints limit the conditions under which prominence eruptions can occur, thus enabling us to estimate the eruptions' kinematics and evolution. 
In paper II, we will model these processes using a simple model and reveal the probable inclination angle relative to the LoS and its physical mechanism.
By using the model, we plan to apply the radiative transfer to estimate the changes in the surface emission of the prominence as its size/height increases \citep[cf.][]{2022MNRAS.513.6058L}.
Ultimately, we will carry out 3D numerical simulations of prominence eruptions/CMEs, incorporating the data from surface starspots and Zeeman Doppler Imaging (ZDI) maps \citep[e.g.,][]{2019ApJ...880...97L,2021ApJ...916...96A,2022ApJ...928..154J}.


Moreover, our observational campaign of two nearby young solar-type stars, including EK Dra, has been in progress since 2020 with the 3.8m Seimei telescope, 2m Nayuta telescope, Okayama 1.88m telescope, and Bulgarian 2m RCC telescope at Rozhen National Astronomical Observatory. We aim to provide further statistical results of the H$\alpha$ eruptive phenomena by increasing the superflare sample size to more than 10. This expanded statistical study will further constrain the frequency and diversity discussed in Section \ref{sec:diversity}.
The likelihood of viewing the eruptive events as prominences or filaments will be modeled and compared with observed diversity with more samples.
We will later extend this approach to solar-mass stars of various ages: from weak-line/classical T-tauri stars \citep[c.f.,][]{2021ApJ...920..154G,2021ApJ...916...32G,2023A&A...672A...5B} to main sequences, e.g., DS Tuc A and $\kappa^{1}$ Ceti, to eventually understand the Sun in time.

\section*{Acknowledgment}

The authors thank Dr. Shin Toriumi for fruitful comments. 
We thank Dr. Kenji Hamaguchi, Dr. Teruaki Enoto, and the \textit{NICER} team for the helpful suggestions and comments on the \textit{NICER} data analysis.
We thank Dr. Wataru Iwakiri for the helpful comments on the correction of the data of \cite{2019AA...622A.210G}.
We thank Dr. Daisaku Nogami, Dr. Satoshi Honda, and staff at Okayama Observatory for contributions to the Seimei telescope project.
This work was supported by JSPS (Japan Society for the Promotion of Science) KAKENHI Grant Numbers 21J00316 (K.N.), 20K04032, 20H05643 (H.M.), 21J00106 (Y.N.), and 21H01131 (H.M., K.I., and K.S.), and JST CREST Grant Number JPMJCR1761 (K.I.). 
V.S.A. was supported by the GSFC Sellers Exoplanet Environments Collaboration (SEEC), which is funded by the NASA Planetary Science Division’s Internal Scientist Funding Model (ISFM), NASA NNH21ZDA001N-XRP F.3 Exoplanets Research Program grants and NICER Cycle 2 project funds and NICER DDT program.
Y.N. was supported from the NASA ADAP award program Number 80NSSC21K0632.
A.A.A. acknowledges Bulgarian NSF grant No.KP-06-N58/3 (2021).
S.V.J. acknowledges the support of the DFG priority program SPP 1992 ``Exploring the Diversity of Extrasolar Planets" (JE 701/5-1).
A.A.V. acknowledges funding from the European Research Council (ERC) under the European Union's Horizon 2020 research and innovation programme (grant agreement No 817540, ASTROFLOW).
The spectroscopic data used in this paper were obtained through the program 22A-N-CN06 (PI: K.N.) with the 3.8m Seimei telescope, which is located at Okayama Observatory of Kyoto University.
This paper includes data collected with the \textit{TESS} mission, obtained from the MAST data archive at the Space Telescope Science Institute (STScI). Funding for the \textit{TESS} mission is provided by the NASA Explorer Program. STScI is operated by the Association of Universities for Research in Astronomy, Inc., under NASA contract NAS 5-26555. 
Some of the data presented in this paper were obtained from the Mikulski Archive for Space Telescopes (MAST) at the Space Telescope Science Institute. The specific observations analyzed can be accessed via \dataset[10.17909/xgds-j146]{https://doi.org/10.17909/xgds-j146}.
\textit{NICER} analysis software and data calibration were provided by the NASA \textit{NICER} mission and the Astrophysics Explorers Program. 

%


\facilities{3.8m Seimei telescope, \textit{TESS}, \textit{NICER}}

\software{\textsf{astropy} \citep{2018AJ....156..123A} , \textsf{IRAF} \citep{Tody1986}, \textsf{PyRAF} \citep{2012ascl.soft07011S}, \textsf{HEASoft 6.31} (HEASARC 2022), \textsf{xspec} \citep{1996ASPC..101...17A}}







\appendix

\section{Detailed light curves of \textit{NICER} X-ray on 2022 April 10}\label{app:a}

\begin{figure}[h]
\epsscale{1.0}
\plotone{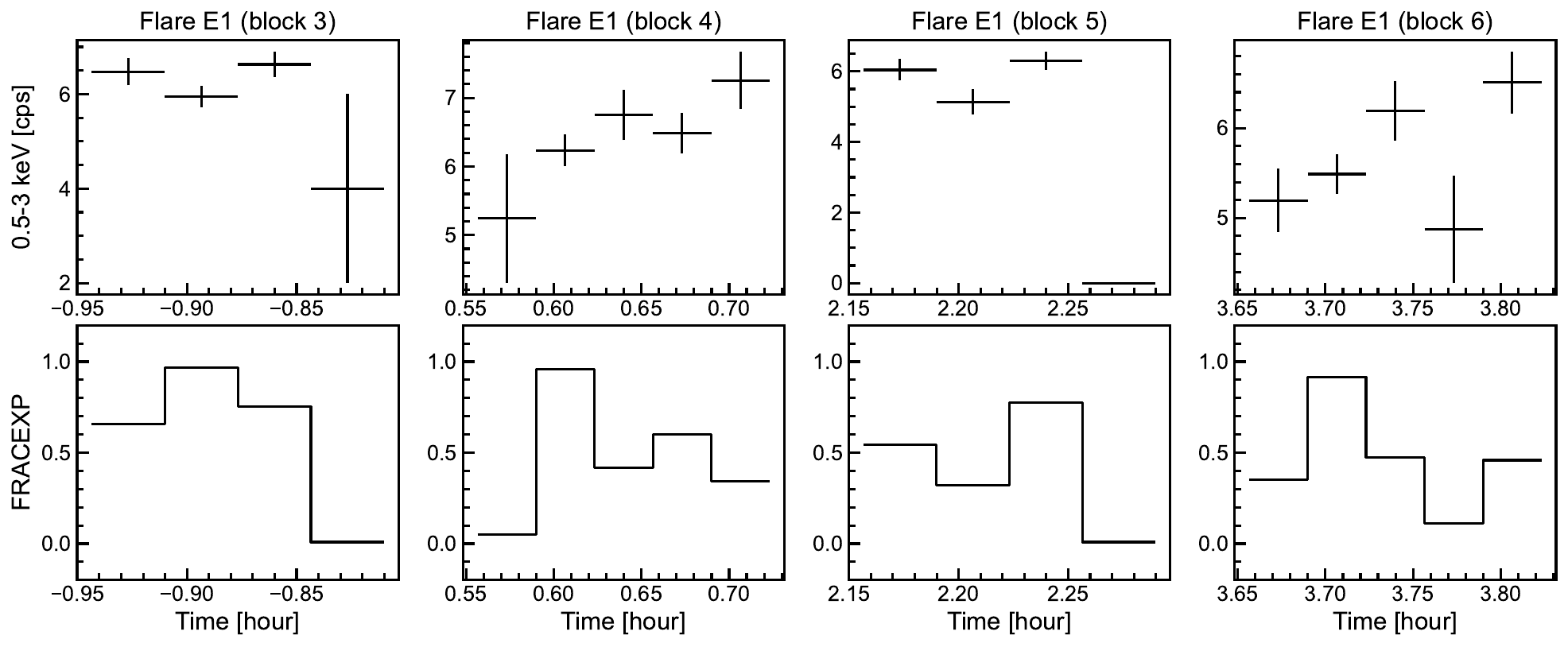}
\caption{(upper) NICER 0.5--3 keV X-ray count rates and (lower) fractional exposure rate for each orbit of block 3--6 on 2022 April 10.
Fractional exposure rate is the ``FRACEXP" column of the \textsf{lcurve} outputs.
The value is the live-time fraction for each time-bin and takes values between zero and one after the deadtime correction with ``LDEADTIME". 
The corrected rate and errors in upper panels are calculated by dividing original values by ``FRACEXP".}
\label{fig:app:xdim}
\end{figure}

Figure \ref{fig:app:xdim} presents the expanded X-ray count rates and exposure fractions of \textit{NICER} for the ISS-orbital block 3 to 6 before and after the flare E1. 
The \textsf{nicerl2}, \textsf{xspec}, and \textsf{lcurve} in HEASoft was used for extracting the data in good time intervals of \textit{NICER} observations, which results in the exposure fractions of each exposure bin not necessarily being unity. 
The \textsf{nicerl2} task functions screens out exposures under unfavorable conditions, notably in situations with a high level of particle background or optical light contamination. When NICER encounters these suboptimal conditions, the Good Time Intervals (GTI) are fragmented. 
Consequently, this fragmentation leads to a decrease in the fractional exposures of the corresponding light curve bins. 
From an analysis of our dataset \textsf{mkf} file, it is observed that the \textsf{TOT\_OVER\_COUNT} is relatively high, nearing a value of $\sim$200. This raises the possibility that the \textsf{nicerl2} task may have excised certain segments from the exposures in each light curve bin, especially those not meeting the standard screening criteria.
This implies that the relatively high noise level of \textit{NICER} X-ray count rates in short time cadence for the data on 2022 April 10 is attributable to the shortened exposure time resulting from the bad time corrections.

\section{Sun-as-a-star Analysis}\label{app:b}

The main text includes some discussion of solar prominences, particularly in relation to CMEs.
Here we show ``Sun-as-a-star" datasets of H$\alpha$ emission during prominence eruptions that could be compared side-by-side with the data from EK Dra.
The first Sun-as-a-star analysis of H$\alpha$ line from solar filament eruptions has been carried out by \cite{2022NatAs...6..241N} using the Solar Dynamics Doppler Imager (SDDI) onboard the Solar Magnetic Activity Research Telescope (SMART) at Hida Observatory. 
Later, \cite{2022ApJ...939...98O} extend the samples to two prominence eruptions outside the solar limb, but they did not perform the spectral analysis of the obtained Sun-as-a-star spectra.
Then, we re-analyzed the two prominence eruptions shown in the paper by \cite{2022ApJ...939...98O} for direct comparison with the EK Dra's data.

\begin{figure*}[ht!]
\gridline{
\fig{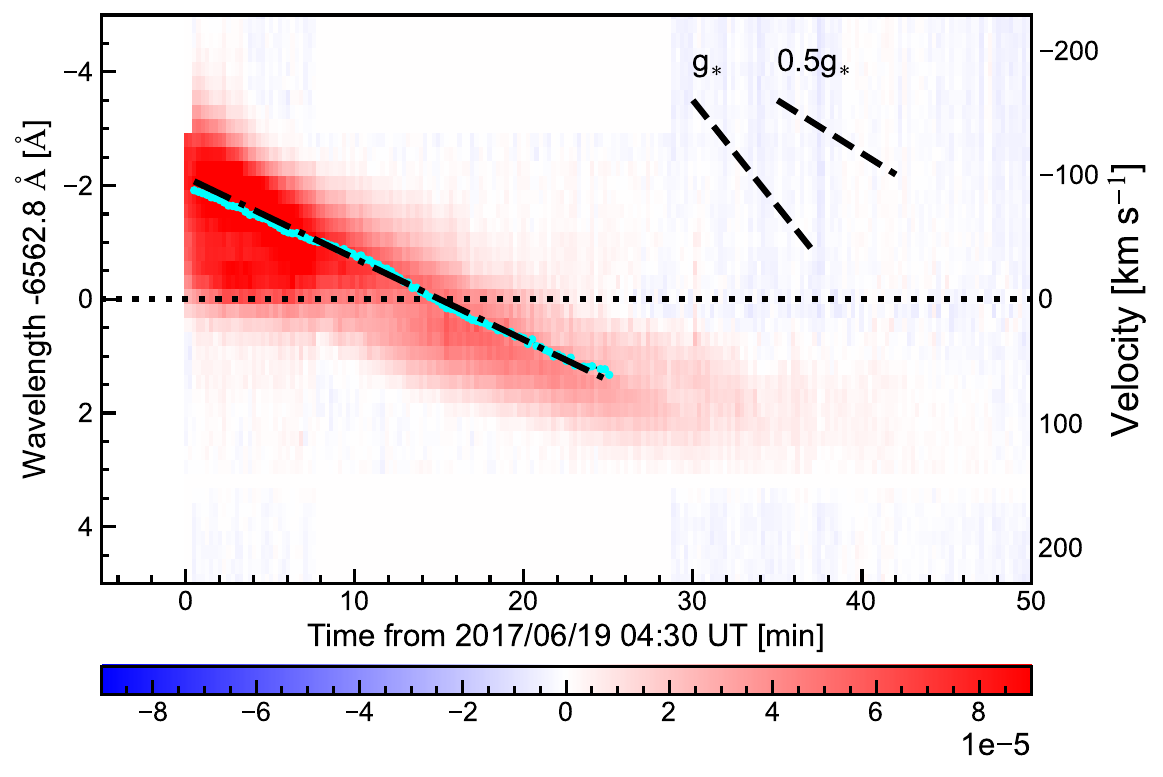}{0.5\textwidth}{\vspace{0mm} (a)}
\fig{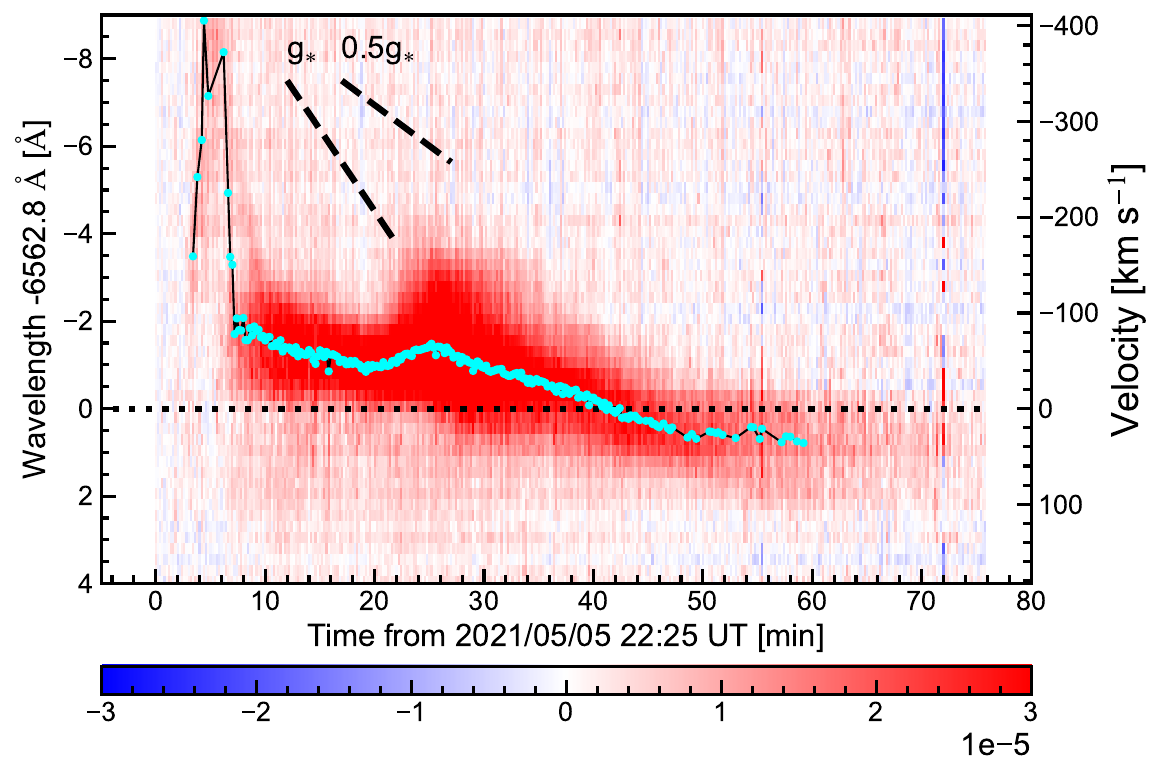}{0.5\textwidth}{\vspace{0mm} (b)}
}
\caption{Dynamic spectra of the pre-flare-subtracted H-alpha spectra of the Sun-as-a-star prominence eruptions. 
(a) The prominence eruption on 2017 June 19 (event ``S1"). (b) The prominence eruption on 2021 May 5 (event ``S2"). 
The data of the dynamic spectra were taken from \cite{2022ApJ...939...98O}, and the spectral fitting is newly done in this paper.  In both events, the emissions from flare ribbons are behind the limb and not visible from the Earth, so the emission almost comes from the erupted prominence.
Some of the white colored region is the unavailable wavelength at that time.
The cyan points are the central wavelength of the emission component.
For the event ``S1", the velocity evolution is fitted by a linear function, and the deceleration is derived as 0.1089$\pm$ 0.0002 km s$^{-2}$.
}
\label{figs:app-sun-1}

\end{figure*}

Figure \ref{figs:app-sun-1} shows the pre-flare-subtracted Sun-as-a-star spectra of H$\alpha$ line of the two prominence eruptions, which are taken from \cite{2022ApJ...939...98O}.
Figure \ref{figs:app-sun-1}(a) is the data for a prominence eruption at around 04:30 UT on 2017 June 19 (event ``S1").
Figure \ref{figs:app-sun-1}(b) is the one for a prominence eruption at around 22:25 UT on 2021 May 5 (event ``S2").
The dynamic spectrum of both S1 and S2 shows that blueshifted emission component of 100-400 km s$^{-1}$ is initially prominent and its velocity decreases in time and finally becomes redshifts of 0-100 km s$^{-1}$.
This represents that the significant protion of the erupted prominences fall down to the solar surface.
However, we noticed that the event S2 is associated with high-speed prominence eruption ($>400$ km s$^{-1}$), and a portion of it moved beyond the field of view (FoV) of the SDDI, which potentially erupted much higher and longer time than seen in the dynamic spectrum. 
Furthermore, \cite{2022ApJ...939...98O} reported that both events are associated with CMEs.
We can learn from these solar dataset that, even though prominence eruptions ultimately show zero or redshift velocity, they can lead to CMEs, as discussed in \ref{sec:cancme}. 

Let us qualitatively compare them with the E1 event on EK Dra. 
The speed of EK Dra's prominence  eventually approaches near 0 km s$^{-1}$. Its maximum velocity is up to 690 km s$^{-1}$ and its duration is about 30 min. 
On the other hand, the solar S1 and S2 events finally show a redshift after decelerating. 
Furthermore, their blueshift speeds range from 10 to 400 km s$^{-1}$, which are typical values for solar prominence ejections, and the duration with blueshift speeds exceeding $\sim$100 km s$^{-1}$, detectable in stellar observations, is also less than about 10 minutes. 
While there are differences between the Sun's events and EK Dra's event in terms of whether they ultimately show a redshift, it can be interpreted that the smaller scale and velocity of the solar prominence eruption compared to stellar events has led to the differences.
At present, large-scale prominence eruptions like the EK Dra's eruption, which eventually reach a velocity of 0 km s$^{-1}$, have not been observed with SDDI. However, by increasing the number of samples in the future, there is a possibility to observe similar events.

\begin{figure*}[ht!]
\gridline{
\fig{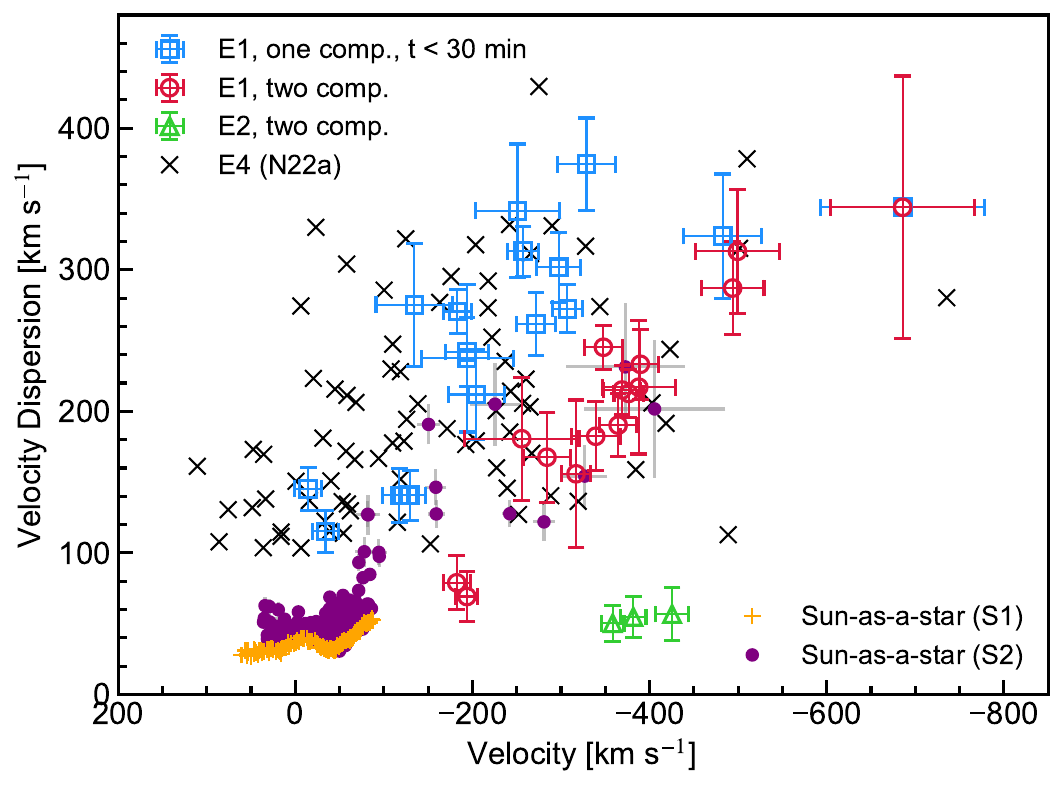}{0.5\textwidth}{\vspace{0mm} (a)}
\fig{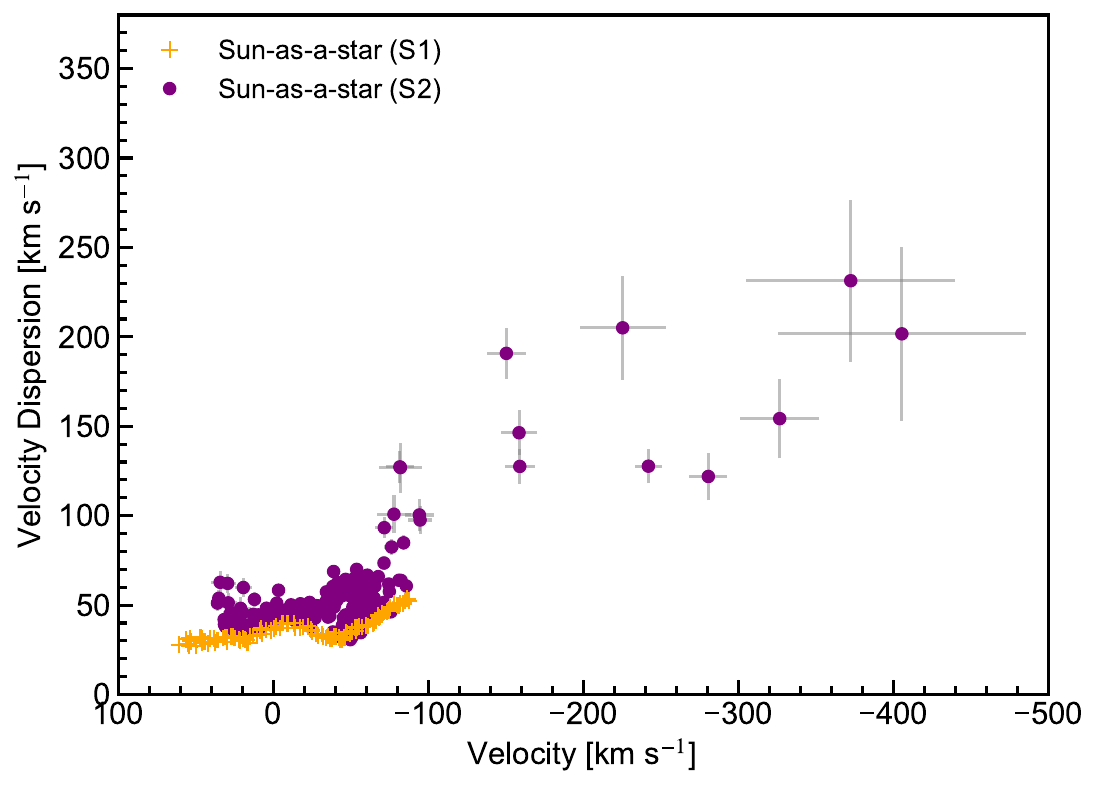}{0.5\textwidth}{\vspace{0mm} (b)}
}
\caption{The same as Figure \ref{fig:11}. The Sun-as-a-star data for event ``S1" and ``S2" is added with orange pluses and purple circles, respectively. (a) The comparison between solar and stellar data. (b) Enlarged panel only for solar data. 
}
\label{figs:app-sun-2}
\end{figure*}

We fitted the H$\alpha$ line profiles of S1 and S2 events with Gaussian function, as performed in stellar data, and derived the velocity and velocity dispersion.
Figure \ref{figs:app-sun-2} is the relation between velocity and velicity dispersion of the H$\alpha$ line profile.
As you can see, they exhibit velocity dispersion that scales with the velocity of the prominence eruption, and its relation is very consistent with stellar data.
This support our suggestion in Section \ref{sec:5-2-1} that the velocity dispersion represents superposition of different velocity components from spatially different regions.

\color{black}



\clearpage
\bibliography{namekata_EKDra_paper1_ver1}{}
\bibliographystyle{aasjournal}



\end{document}